\pdfoutput=1
\documentclass[11pt,twoside,a4paper,cmspaper,final,collab]{cms-tdr}
\def\svnVersion{1bc8a4d}\def\svnDate{2025/02/27}\def\cmsCernNoTag{CERN-EP-2024-251}\def\cmsCernDate{\today}\def\cmsMessage{Published in the Journal of High Energy Physics as \href{http://dx.doi.org/10.1007/JHEP02(2025)177}{\doi{10.1007/JHEP02(2025)177}.}}
\begin{document}\cmsNoteHeader{TOP-23-004}

\newlength\cmsTabSkip\setlength{\cmsTabSkip}{1ex}

\newcommand{\PellW}{{\HepParticle{\Pell}{\PW}{}}\xspace}

\newcommand{\pp}{\ensuremath{\Pp\Pp}\xspace}
\newcommand{\WZ}{\ensuremath{\PW\PZ}\xspace}
\newcommand{\ZZ}{\ensuremath{\PZ\PZ}\xspace}
\newcommand{\ttZ}{\ensuremath{\ttbar\PZ}\xspace}
\newcommand{\tWZ}{\ensuremath{\PQt\PW\PZ}\xspace}
\newcommand{\ttZplustWZ}{\ensuremath{\ttZ\text{+}\tWZ}\xspace}
\newcommand{\tZq}{\ensuremath{\PQt\PZ\PQq}\xspace}
\newcommand{\ttW}{\ensuremath{\ttbar\PW}\xspace}
\newcommand{\ttH}{\ensuremath{\ttbar\PH}\xspace}

\newcommand{\mll}{\ensuremath{m(\Pellp\Pellm)}\xspace}
\newcommand{\mZ}{\ensuremath{m(\PZ)}\xspace}

\newcommand{\abseta}{\ensuremath{\abs{\eta}}\xspace}
\newcommand{\pTZ}{\ensuremath{\pt(\PZ)}\xspace}
\newcommand{\cosThetaStar}{\ensuremath{\cos\theta^\ast_{\PZ}}\xspace}

\newcommand{\ptlW}{\ensuremath{\pt(\PellW)}\xspace}
\newcommand{\DRZlW}{\ensuremath{\DR(\PZ,\PellW)}\xspace}
\newcommand{\Dphill}{\ensuremath{\Delta\phi(\Pellp,\Pellm)}\xspace}
\newcommand{\chisq}{\ensuremath{\chi^2}\xspace}

\newcommand{\sttZtWZ}{\ensuremath{\sigma(\ttZplustWZ)}\xspace}
\newcommand{\stZq}{\ensuremath{\sigma(\tZq)}\xspace}
\newcommand{\mttZtWZ}{\ensuremath{\mu_{\ttZplustWZ}}\xspace}
\newcommand{\mtZq}{\ensuremath{\mu_{\tZq}}\xspace}

\cmsNoteHeader{TOP-23-004}

\title{Measurements of inclusive and differential cross sections for top quark production in association with a \texorpdfstring{\PZ}{Z} boson in proton-proton collisions at \texorpdfstring{$\sqrt{s}=13\TeV$}{sqrt(s) = 13 TeV}}

\date{\today}

\abstract{Measurements are presented of inclusive and differential cross sections for \PZ boson associated production of top quark pairs (\ttZ) and single top quarks (\tZq or \tWZ). The data were recorded in proton-proton collisions at a center-of-mass energy of 13\TeV, corresponding to an integrated luminosity of 138\fbinv. Events with three or more leptons, electrons or muons, are selected and a multiclass deep neural network is used to separate three event categories, the \ttZ and \tWZ processes, the \tZq process, and the backgrounds. A profile likelihood approach is used to unfold the differential cross sections, to account for systematic uncertainties, and to determine the correlations between the two signal categories in one global fit. The inclusive cross sections for a dilepton invariant mass between 70 and 110\GeV are measured to be $1.14\pm0.07\unit{pb}$ for the sum of \ttZ and \tWZ, and $0.81\pm0.10\unit{pb}$ for \tZq, in good agreement with theoretical predictions.}

\hypersetup{%
pdfauthor={CMS Collaboration},%
pdftitle={Measurements of inclusive and differential cross sections for top quark production in association with a Z boson in proton-proton collisions at sqrt(s) = 13 TeV},%
pdfsubject={CMS},%
pdfkeywords={CMS, top quarks, top+X}}

\maketitle

\section{Introduction}
\label{sec:intro}

At the CERN LHC, precise inclusive and differential measurements of top quark production in association with a \PZ boson have made it possible to directly probe the couplings between the top quark and the \PZ boson. Possible physics beyond the standard model (SM) of particle physics could induce measurable deviations from SM expectations, as described, \eg, in Ref.~\cite{Rontsch:2014cca}. Furthermore, processes involving top quarks and \PZ bosons are significant backgrounds to other important SM processes, such as Higgs boson production in association with a top quark~\cite{LHCHiggsCrossSectionWorkingGroup:2016ypw}. These measurements also allow for tests of theory calculations in perturbative quantum chromodynamics (pQCD) and studies of the modeling in Monte Carlo (MC) generators.

Measurements of \PZ boson production in association with top quark pairs (\ttZ) and single top quarks with an additional light-flavor quark \PQq (\tZq) have been performed by the ATLAS~\cite{ATLAS:2015qtq,ATLAS:2016wgc,ATLAS:2017dsm,ATLAS:2019fwo,ATLAS:2020bhu,ATLAS:2021fzm,ATLAS:2023eld} and CMS~\cite{CMS:2015uvn,CMS:2017ugv,CMS:2017wrv,CMS:2018sgc,CMS:2019too,CMS:2021ugv,CMS:2021aly}
experiments. The cross sections for \ttZ and \tZq are expected to be of similar order of magnitude, despite their different orders in the strong coupling \alpS~\cite{Campbell:2013yla}.
Generally good agreement with the SM expectations was found~\cite{LHCHiggsCrossSectionWorkingGroup:2016ypw,Frixione:2015zaa,Frederix:2018nkq,Kulesza:2019adl,Pagani:2020mov}.
The CMS Collaboration recently reported the first evidence for single top quark production in association with both a \PW and a \PZ boson (\tWZ)~\cite{CMS:2023krq}. The inclusive \tWZ cross section was measured to be $\sigma_{\tWZ}=354\pm54\stat\pm95\syst\unit{fb}$, about two standard deviations above the SM expectation of $136\,^{+9}_{-8}\unit{fb}$~\cite{ref:madgraph}.

\begin{figure}[!ht]
\centering
\includegraphics[width=0.3\textwidth]{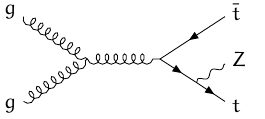}%
\hfill%
\includegraphics[width=0.3\textwidth]{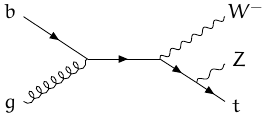}%
\hfill%
\includegraphics[width=0.3\textwidth]{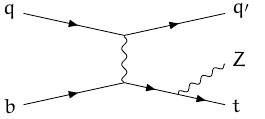}
\caption{Leading order Feynman diagrams for the \ttZ (left), \tWZ (middle), and \tZq (right) processes.}
\label{fig:feyn}
\end{figure}

Leading order (LO) Feynman diagrams for the processes \ttZ, \tWZ, and \tZq are shown in Fig.~\ref{fig:feyn}. Experimentally, the \PZ boson is reconstructed in its leptonic decays into a pair of leptons (electrons or muons, in the following collectively denoted by the \Pell symbol).
For signal events with three leptons in the final state, at LO, the three signal
processes differ in the jet multiplicity distributions. For \ttZ, in which the \ttbar system decays in the lepton+jets channel, four central jets are expected of which two originate from \PQb quarks. For \tWZ, three central jets including one \PQb jet are expected. For \tZq, there is one \PQb jet and one light quark jet, and the light-flavor recoil jet often has a high absolute pseudorapidity \abseta and transverse momentum \pt. Higher-order effects and limitations in detector acceptance lead to a large overlap between the measured jet multiplicity distributions. In addition, at next-to-leading order (NLO) in pQCD, interference terms between \ttZ and \tWZ appear~\cite{Faham:2021zet}, making the individual measurements more difficult. Therefore, in this analysis, the two processes \ttZ and \tWZ are measured together.

In this paper, the cross sections for single and pair production of top quarks in association with a \PZ boson are simultaneously measured for the first time. Both inclusive and differential measurements are reported for \ttZplustWZ and \tZq. In previous publications, separate analyses for the \ttZ and \tZq processes were reported, assuming SM expectations for one process while measuring the other. In the simultaneous analysis presented here, information about the correlation between the processes is obtained directly. This information can be useful, \eg, in global fits using effective field theory (EFT)~\cite{BUCHMULLER1986621,Burgess:2007pt} in which the combined sensitivity of data from different measurements can be used to obtain optimal constraints~\cite{Ethier:2021bye,Castro:2022zpq,Degrande:2018fog}.

The analysis makes use of proton-proton (\pp) collision data recorded at the CMS experiment in the years 2016--2018 at a center-of-mass energy of 13\TeV, corresponding to an integrated luminosity of 138\fbinv. For the differential measurement, events are selected with exactly three isolated leptons, of which two originate from the \PZ boson, while the third comes from the decay of a \PW boson. For the inclusive measurement, events with four leptons are also included, as described in more detail in Section~\ref{sec:results}. In the cases of \ttZ and \tZq, the \PW boson originates from the top quark decay, while in the \tWZ process it can also be produced directly. A multiclass deep neural network (DNN) is used to determine event likelihood scores for each of the three categories: 1)~the signal category \ttZplustWZ, used to measure the combined cross section of \ttZ and \tWZ; 2)~the \tZq signal category; and 3)~the background category. An important background process is \WZ production. Another important background, referred to as ``nonprompt leptons'', arises in processes with large cross sections, such as {\PZ}+jets and inclusive \ttbar production, if particles not originating from the decay of a real vector boson mimic the signature of leptons in the detector.
In the fit, a given event is assigned to the category with the highest score. Both inclusive and differential cross sections for the two signal processes are measured simultaneously using a profile likelihood fit approach, in which nuisance parameters are implemented to constrain the systematic uncertainties, and the free parameters of interest are determined simultaneously, for each signal process and in each bin of the measurement, together with their correlation matrix.

The paper is structured as follows. In Section~\ref{sec:cms}, an overview of the CMS detector is given. Subsequently, in Section~\ref{sec:samples}, the data and MC simulation samples are described.
The event selection, the resulting composition of signal and background events and their classification into two signal categories and the background are presented in Section~\ref{sec:evtsel}. The systematic uncertainties and the results of the measurement are reported in Sections~\ref{sec:uncs} and~\ref{sec:results}. Finally, a summary is given in Section~\ref{sec:summary}. Tabulated results can be found in HEPData~\cite{hepdata}.

\section{The CMS detector and event reconstruction}
\label{sec:cms}

The central feature of the CMS apparatus is a superconducting solenoid of 6\unit{m} internal diameter, providing a magnetic field of 3.8\unit{T}. Within the solenoid volume are a silicon pixel and strip tracker, a lead tungstate crystal electromagnetic calorimeter (ECAL), and a brass and scintillator hadron calorimeter (HCAL), each composed of a barrel and two endcap sections. Forward calorimeters extend the pseudorapidity coverage provided by the barrel and endcap detectors. Muons are measured in gas-ionization detectors embedded in the steel flux-return yoke outside the solenoid. A more detailed description of the CMS detector, together with a definition of the coordinate system used and the relevant kinematic variables, can be found in Refs.~\cite{CMS:2008xjf,CMS:2023gfb}.

The particle-flow (PF) event reconstruction~\cite{CMS:2017yfk} aims to reconstruct and identify each individual particle in an event, with an optimized combination of information from all CMS detector components. In the PF algorithm, the identification of the particle type (photon, electron, muon, charged or neutral hadron) plays an important role in the determination of the particle direction and energy. Additional \pp interactions within the same or nearby bunch crossings (pileup) can contribute additional tracks and calorimetric energy deposits to the jet momentum. To mitigate this effect, charged particles identified as originating from pileup vertices are discarded and an offset correction is applied to correct for the remaining contributions~\cite{CMS:2020ebo}.
The primary vertex is taken to be the vertex corresponding to the hardest scattering in the event, evaluated using tracking information alone, as described in Section 9.4.1 of Ref.~\cite{CMS-TDR-15-02}.

The electron momentum is estimated by combining the energy measurement in the ECAL with the momentum measurement in the tracker.
Muons are identified as tracks in the central tracker consistent with either a track or several hits in the muon system, and associated with calorimeter deposits compatible with the muon hypothesis.
Charged hadrons are identified as charged-particle tracks that are neither identified as electrons nor as muons. The energy of charged hadrons is determined from a combination of the track momentum and the corresponding ECAL and HCAL energies, corrected for the response function of the calorimeters to hadronic showers. Neutral hadrons are identified as HCAL energy clusters not linked to any charged-hadron trajectory, or as a combined ECAL and HCAL energy excess with respect to the expected charged-hadron energy deposit. The energy of neutral hadrons is obtained from the corresponding corrected ECAL and HCAL energies.

For each event, hadronic jets are clustered from these reconstructed particles using the anti-\kt algorithm~\cite{Cacciari:2008gp, Cacciari:2011ma} with a distance parameter of 0.4.
Jet energy corrections (JEC) are derived from simulation studies so that the average measured energy of jets becomes identical to that of particle level jets, and residual differences between data and simulation are determined using momentum balance measurements in dijet, photon+jet, {\PZ}+jet, and multijet events~\cite{CMS:2016lmd}.

To identify (tag) jets originating from \PQb quarks (\PQb jets), information about the energy of particles, tracks, and vertices inside the jet is fed into the \textsc{DeepJet} DNN~\cite{CMS:2017wtu,Bols:2020bkb}. A \PQb tagging requirement is applied in this analysis by using a minimum threshold on the \PQb tagging score, corresponding to a 75\% efficiency (1\% mistagging rate) for \PQb jets (light-flavor jets excluding charm).
This reconstruction method is applied only for jets within the acceptance of the silicon tracker ($\abseta<2.4$ for 2016 and $\abseta<2.5$ for 2017--2018).

The missing transverse momentum vector \ptvecmiss is computed as the negative vector \pt sum of all the PF candidates in an event, and its magnitude is denoted as \ptmiss~\cite{CMS:2019ctu}. The \ptvecmiss is modified to account for corrections to the energy scale of the reconstructed jets in the event.

\section{Event simulation}
\label{sec:samples}

The MC simulation samples for the signal processes \tZq, \ttZ, and \tWZ, as well as those for the backgrounds arising from the production of three electroweak gauge bosons, are generated at NLO with \MGvATNLO~v2.6.5~\cite{ref:madgraph}.
The generator output is passed to \PYTHIA~v8.240~\cite{ref:pythia} for hadronization using the CP5 underlying event tune~\cite{ref:cp5}.
For the production of a \PW boson and a \PZ boson (\WZ), the \ttbar production in association with a \PW boson (\ttW), and the production of a photon in association with a \PW or a \PZ boson ($\PV\PGg$), \MGvATNLO is used in the setup with up to one additional parton included in the matrix element simulation.
To remove double counting with \PYTHIA, the FxFx~\cite{ref:fxfx} matching scheme is used. For the \ttW sample, more details of the specific settings are described in Ref.~\cite{Ferencz:2023fso}.

Other backgrounds, such as \ttbar production in association with a photon or with two electroweak gauge bosons, are simulated with \MGvATNLO at LO accuracy using MLM matching~\cite{Alwall_2007}. Some minor backgrounds, \eg, \ttbar production in association with a Higgs boson (\ttH) and the quark-initiated production of two \PZ bosons (\ZZ), are simulated at NLO accuracy in pQCD using \POWHEG~v2~\cite{ref:powheg}, while the gluon-initiated \ZZ production is simulated at LO with the \textsc{MCFM}~v7.0.1~\cite{CAMPBELL201010} generator. In the following, backgrounds arising from \ttW, \ttH, as well as from \ttbar production in association with a photon or with two vector bosons, are referred to as $\ttbar\PX$, the production of two or three heavy bosons, excluding \WZ production, is referred to as multiboson production.

For the \tZq sample, the four-flavor scheme (4FS) is used. In the 4FS, the \PQb quark in the initial state originates from gluon splitting at the matrix element level. The 4FS is expected to provide a better modeling of the event kinematics than the five-flavor scheme (5FS)~\cite{Pagani:2020mov}, which is employed for all other samples. In the 5FS, the \PQb quark comes directly from the parton distribution function (PDF), and the total cross section prediction is expected to be more precise~\cite{Pagani:2020mov}. The samples are simulated with the NNPDF~v3.1~\cite{ref:nnpdf} PDF set at NLO accuracy.

At NLO, \tWZ production interferes with the LO \ttZ process.
The diagram removal (DR) and diagram subtraction (DS) approaches are used to remove the double counting. In the DR scheme, the overlap is removed at the amplitude level, either excluding (DR1) or including (DR2) the interference term~\cite{Frixione:2008yi,Demartin:2016axk}. In the DS scheme, the overlap is removed at the cross section level.
In this analysis, the nominal \tWZ event samples are generated using the DR1 method, and alternative samples, produced with the DR2 method, are used to estimate the corresponding uncertainty. As reported in Ref.~\cite{Demartin:2016axk}, the result of the DS scheme lies between DR1 and DR2, and it is thus not included as an uncertainty in this measurement.

The CMS detector response is simulated using GEANT4~\cite{Agostinelli:2002hh}. The effect of pileup is taken into account by adding
simulated minimum-bias interactions to the simulated data.

\section{Event selection and background estimation}
\label{sec:evtsel}

The event selection criteria are designed to efficiently select \tZq, \ttZ, and \tWZ signal events while suppressing the background.
Events are selected if they contain three or more leptons, which are reconstructed within $\abseta<2.5$ for electrons and $\abseta<2.4$ for muons. The three leptons, when ordered in decreasing \pt, are required to have $\pt>25$, 15, and 10\GeV. Two of the leptons must be of the same flavor and opposite charge, and have an invariant mass between 70 and 110\GeV; the third one is expected to originate from the \PW boson. A combination of trigger algorithms requiring the presence of either one, two, or three leptons is used to record the events.

To suppress backgrounds, selected leptons are required to fulfill additional ''tight`` selection requirements based on a multivariate classifier~\cite{CMS:2019too,CMS:2021ugv,CMS:2023ftu}, which distinguishes prompt leptons that are created in the decay of a \PW or \PZ boson from nonprompt leptons. The prompt-lepton selection efficiencies in the simulation are corrected to match those in the data using scale factors (SFs) which are evaluated using a ``tag-and-probe'' method in $\PZ\to\Pellp\Pellm$ events~\cite{CMS:EGM-17-001, CMS:MUO-16-001}.

Jets are required to have $\pt>25\GeV$ and $\abseta<5.0$ and to be separated in space from each lepton by $\DR=\sqrt{\smash[b]{(\Delta\eta)^2+(\Delta\phi)^2}}>0.4$, where $\Delta\eta$ and  $\Delta\phi$ are the differences in $\eta$ and azimuthal angle, respectively, between the jet and the lepton. At least two jets, of which at least one must be \PQb tagged, are also required in each event.

Dominant contributions to the nonprompt background come from \ttbar and Drell--Yan production where two genuine prompt leptons are expected; the third lepton candidate then often comes from a decay of a hadron. The nonprompt-lepton background contribution is estimated from data, separately for electrons and muons, following an approach similar to that described in Ref.~\cite{CMS:2021ugv}. In this approach, a lepton misidentification rate is determined from simulated QCD multijet samples. This information is then used to determine a transfer factor that is applied to events in data that are selected similarly to those in the signal region, but with looser lepton requirements. The obtained distribution describes the expectation for the nonprompt background in the signal region.

\begin{figure}[!p]
\centering
\includegraphics[width=0.45\textwidth]{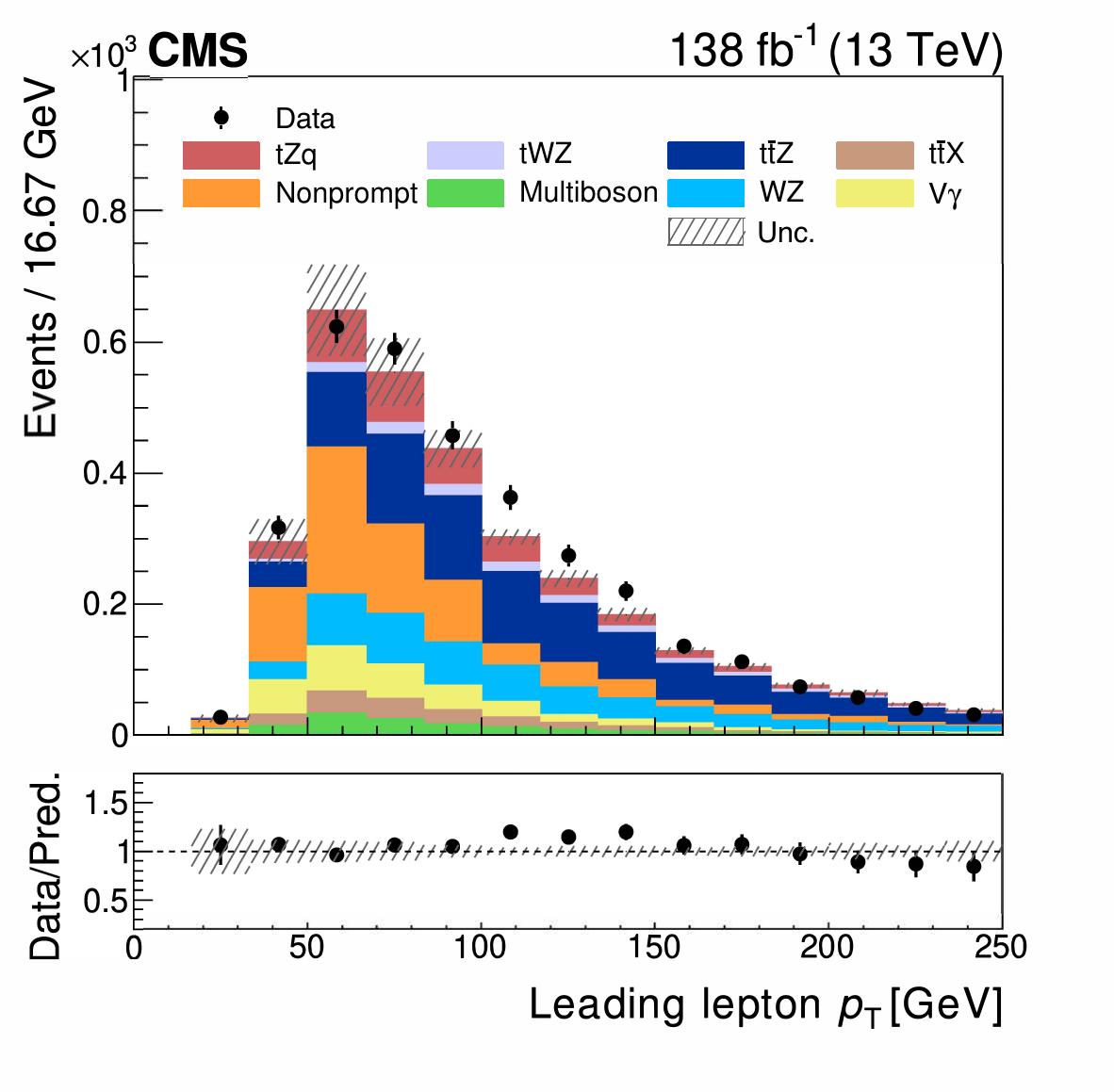}
\includegraphics[width=0.45\textwidth]{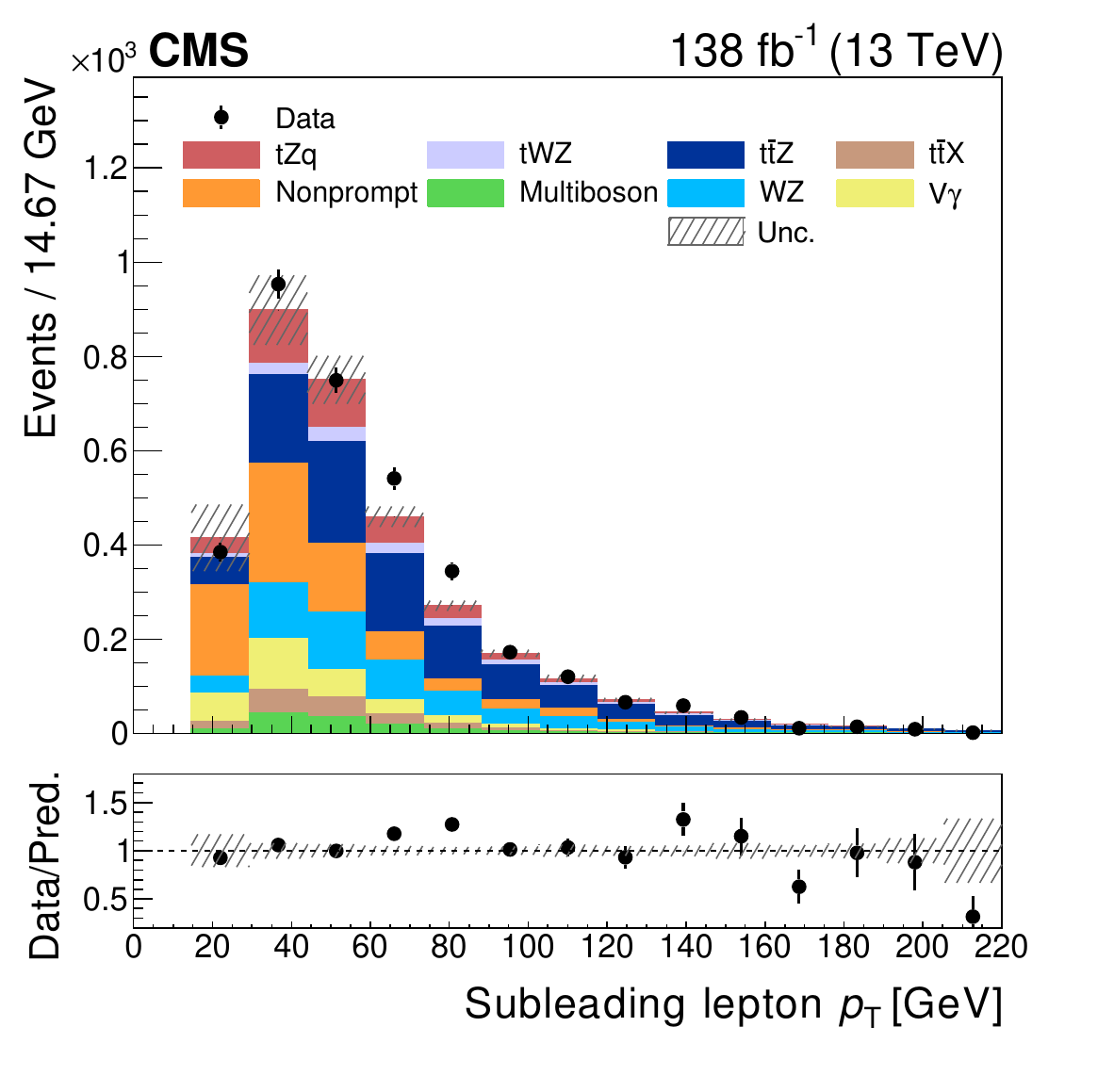} \\
\includegraphics[width=0.45\textwidth]{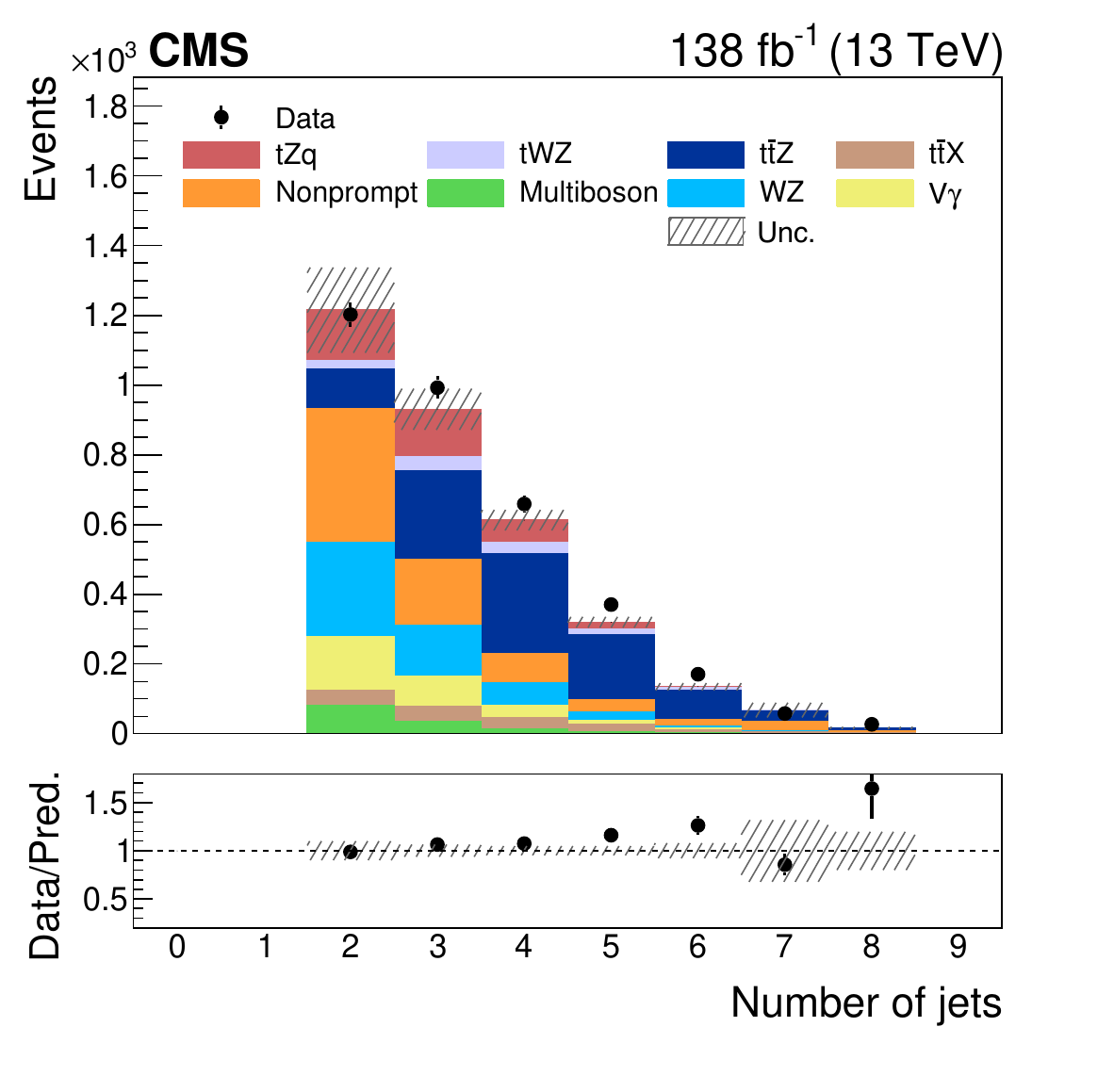}
\includegraphics[width=0.45\textwidth]{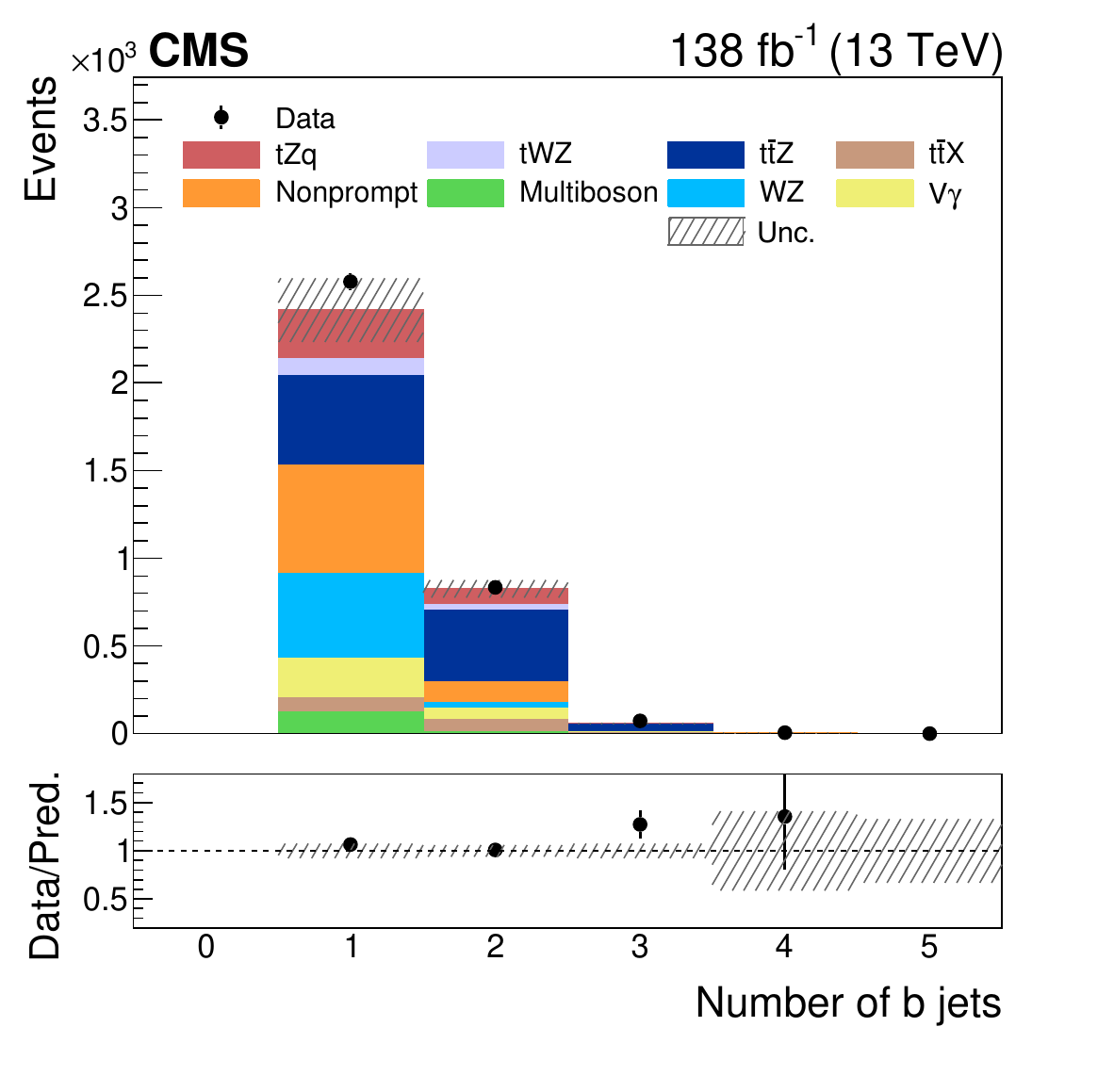} \\
\includegraphics[width=0.45\textwidth]{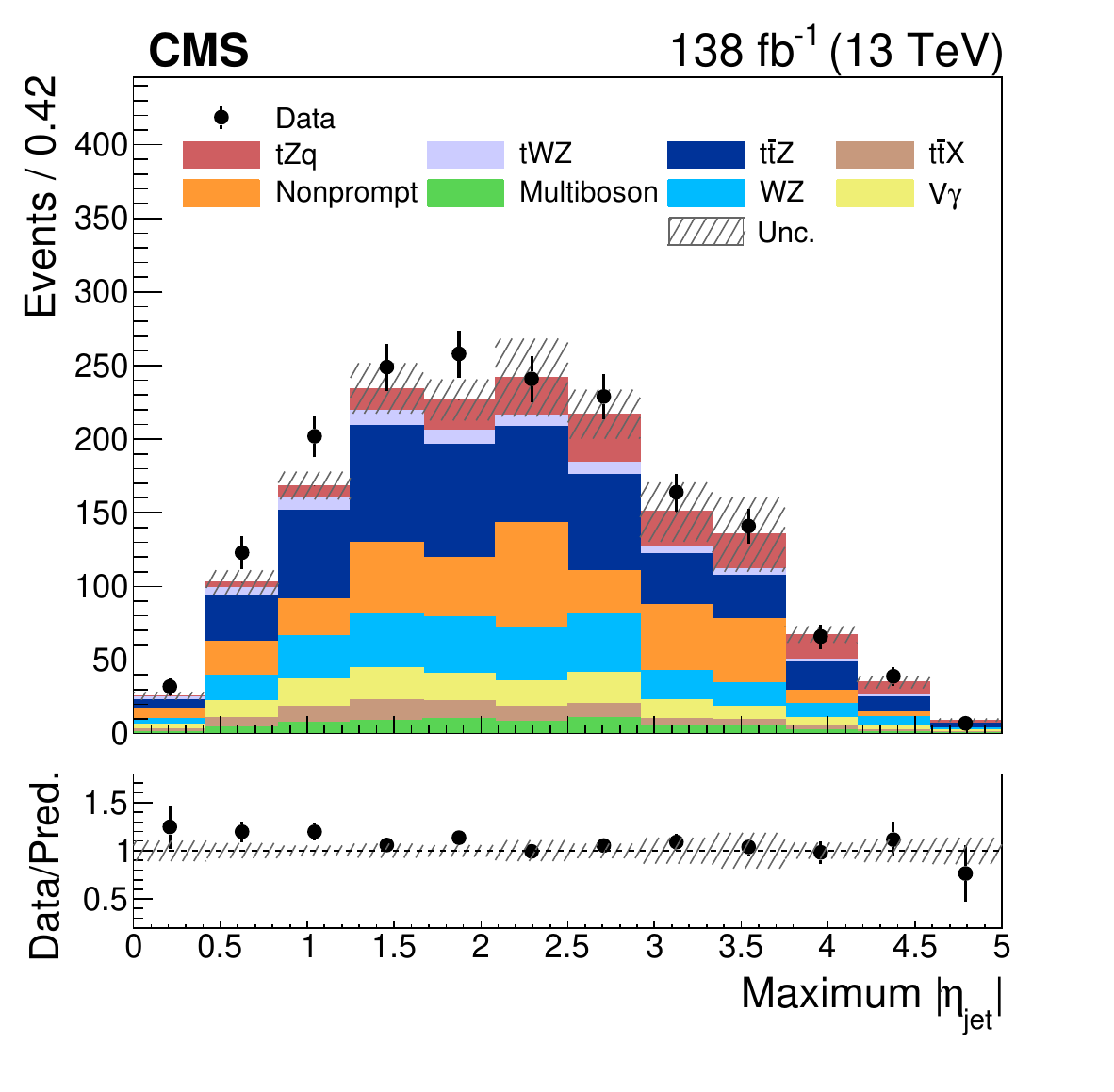}
\caption{Distributions after final event selection: the \pt of the lepton with the highest (upper left) and second highest (upper right) \pt, the number of jets (middle left), the number of \PQb jets (middle right), and the \abseta of the jet with the highest \abseta (lower). The data are displayed as points with statistical error bars and the expectation is shown with a histogram with the systematic uncertainty given by the hatched area.}
\label{fig:controlplots}
\end{figure}

\begin{figure}[!p]
\centering
\includegraphics[width=0.45\textwidth]{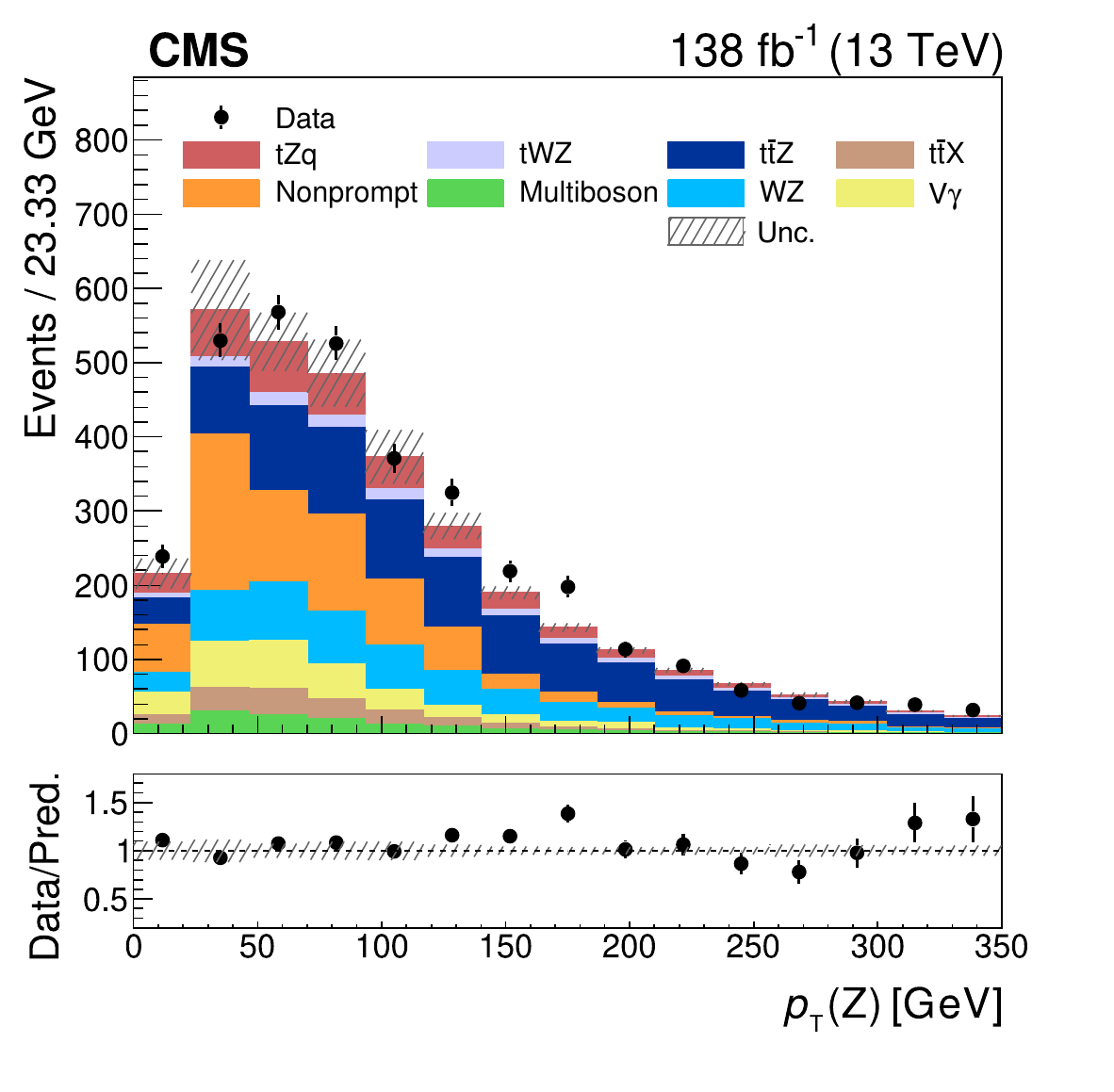}
\includegraphics[width=0.45\textwidth]{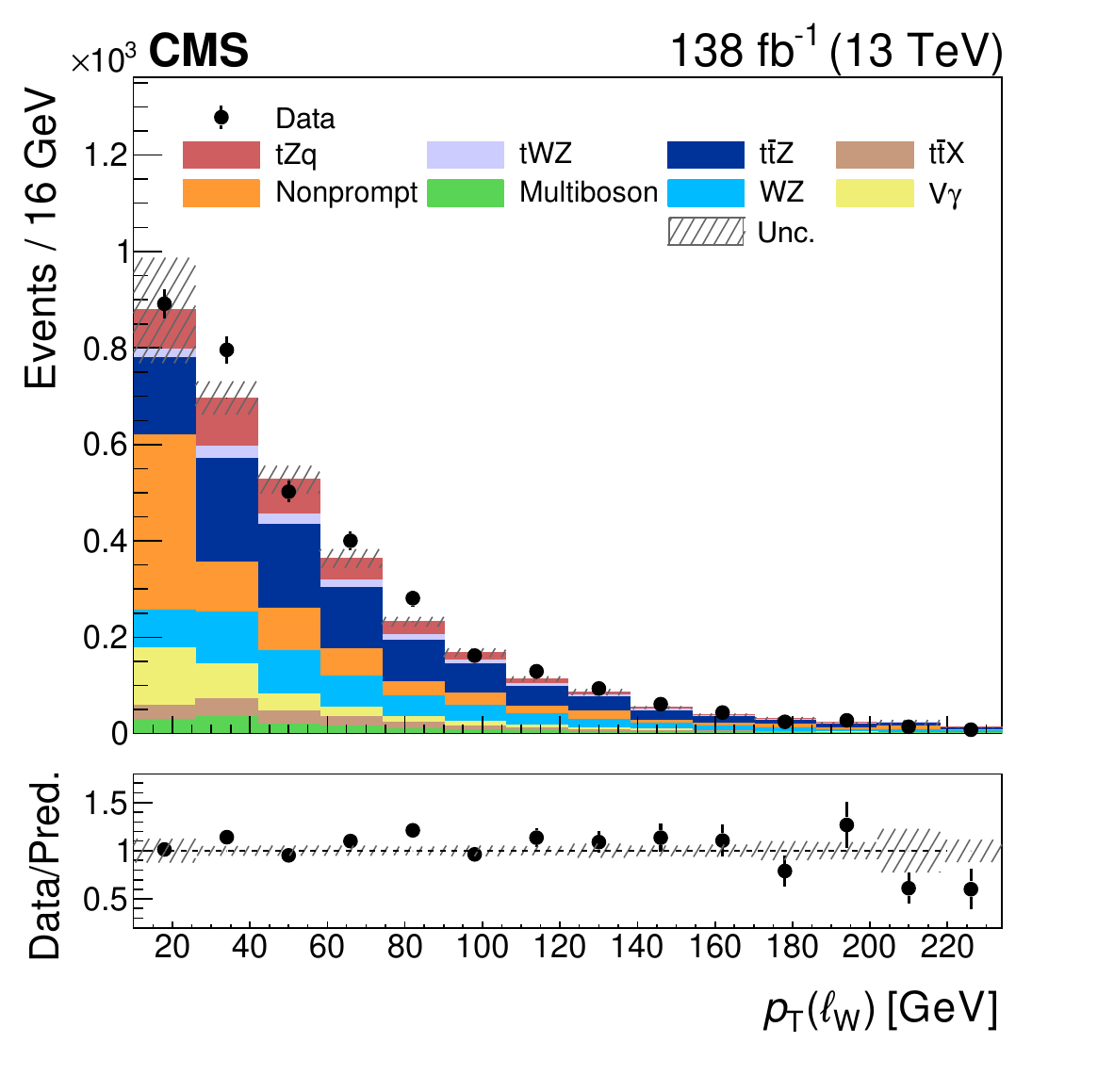} \\
\includegraphics[width=0.45\textwidth]{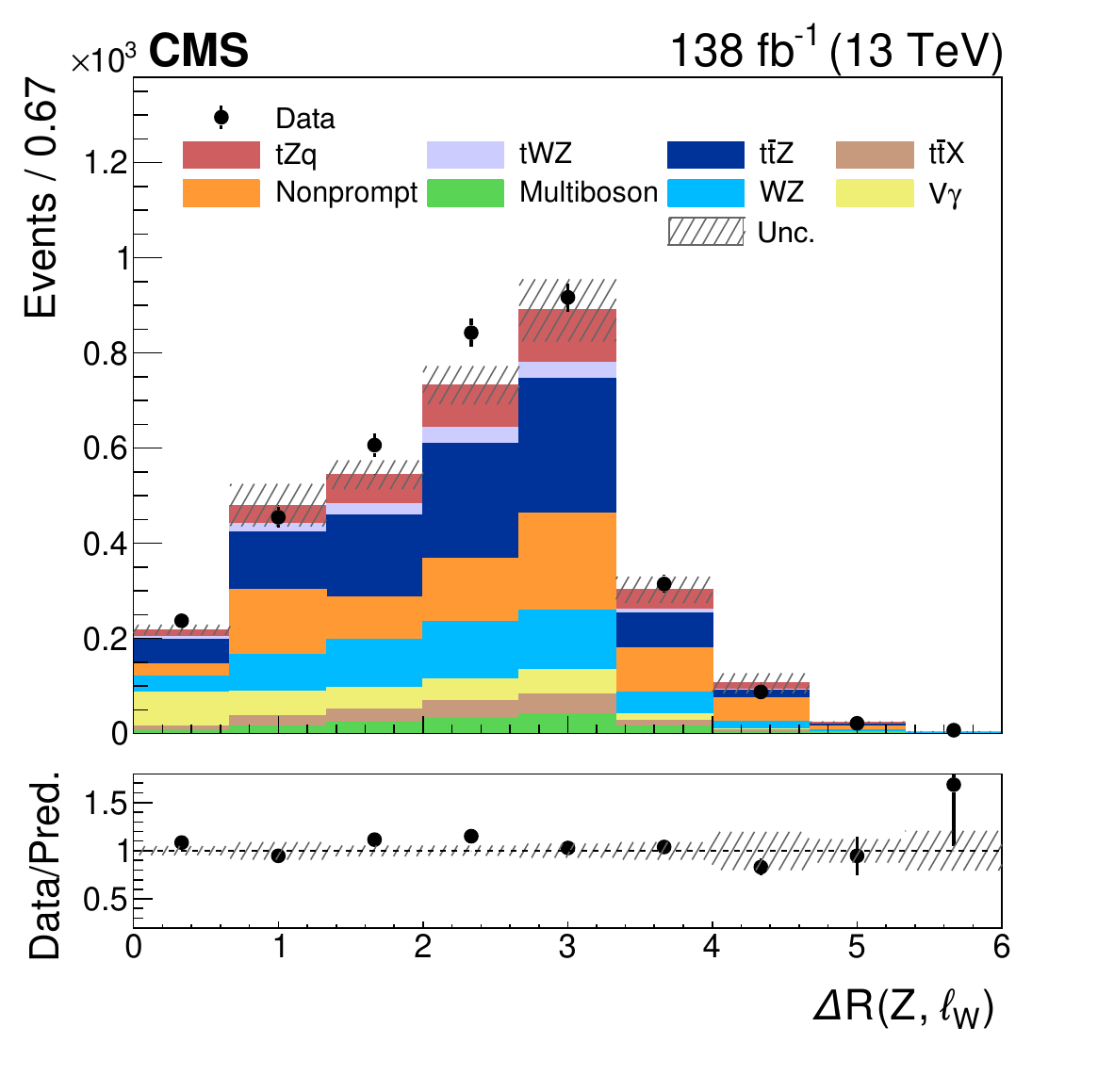}
\includegraphics[width=0.45\textwidth]{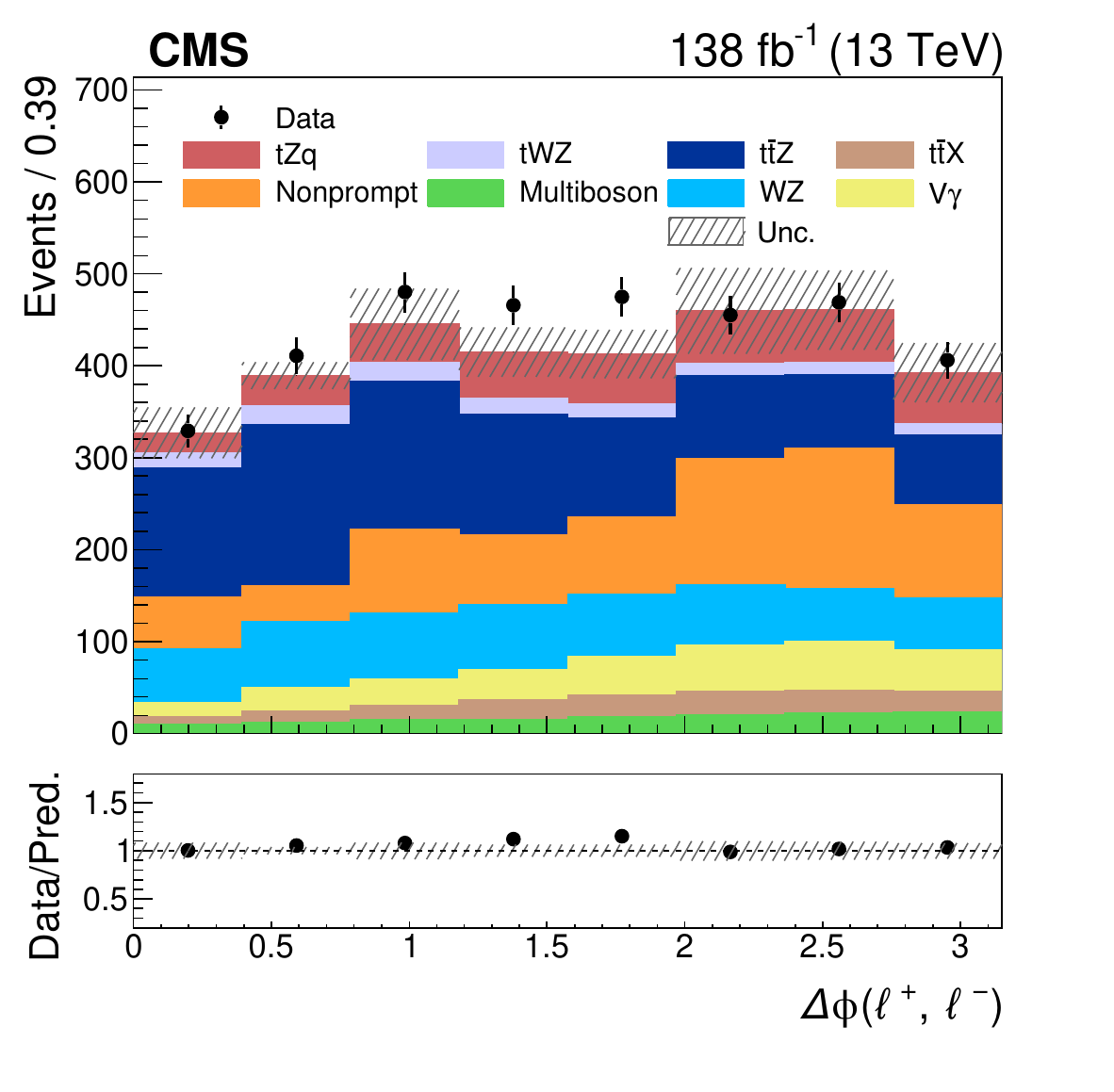} \\
\includegraphics[width=0.45\textwidth]{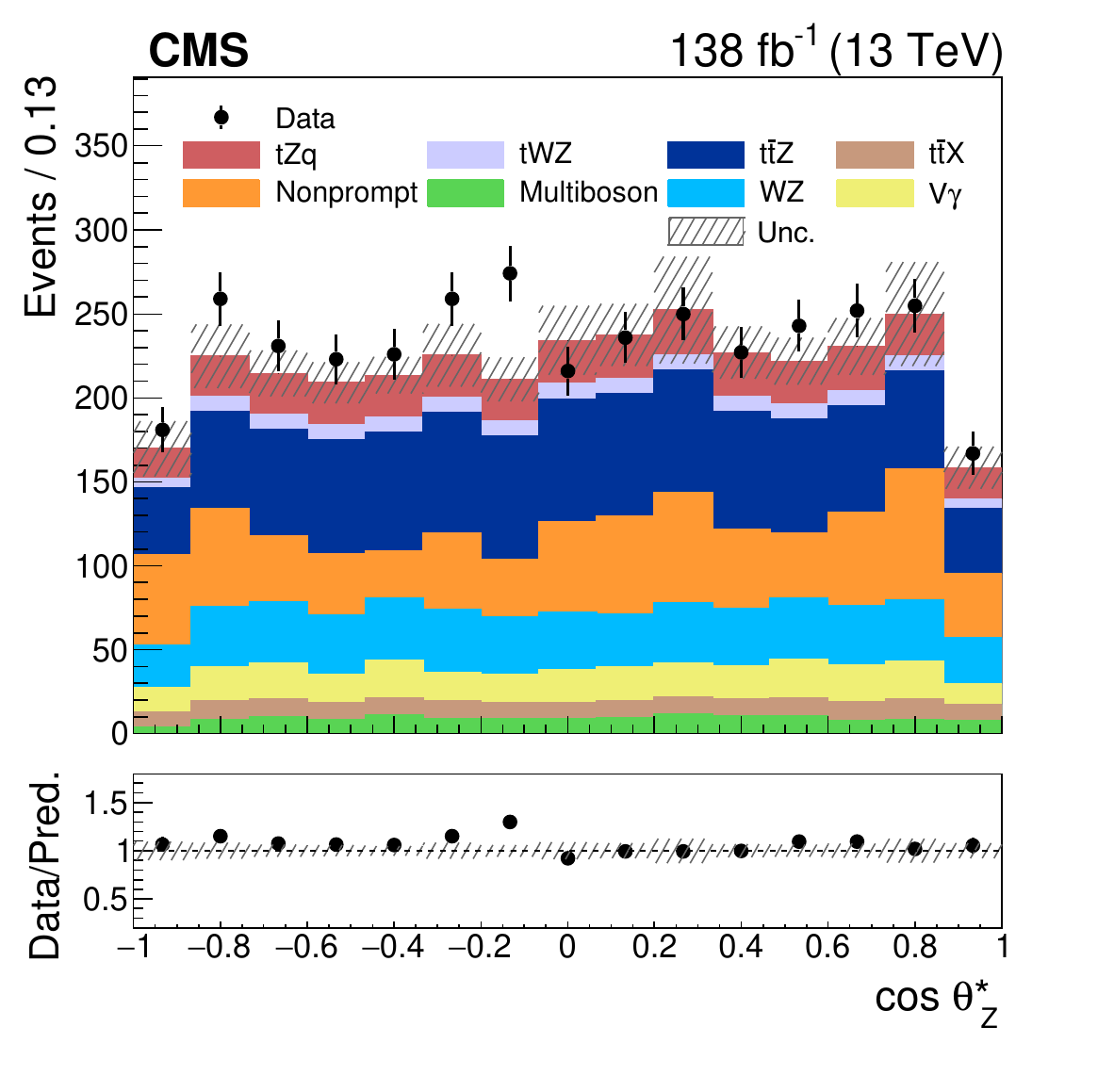}
\caption{Distributions after final event selection: the \pt of the reconstructed \PZ boson (upper left), the \pt of the lepton arising from the \PW boson \ptlW (upper right), \DRZlW (middle left), \Dphill (middle right), and the cosine of the angle \cosThetaStar between the \PZ boson and its negatively charged decay lepton in the \PZ boson rest frame (lower). The data are displayed as points with statistical error bars and the expectation is shown with a histogram with the systematic uncertainty given by the hatched area.}
\label{fig:controlplots_unfolding}
\end{figure}

Figures~\ref{fig:controlplots} and \ref{fig:controlplots_unfolding} show comparisons of measured and expected distributions of the events that fulfill the signal selection criteria.
In Fig.~\ref{fig:controlplots}, the transverse momenta of the leptons with the highest and second highest \pt are presented, the jet and \PQb jet multiplicities, and the \abseta of the highest-\abseta jet. Figure~\ref{fig:controlplots_unfolding} shows distributions of the \pt of the reconstructed \PZ boson, the \pt of the lepton originating from the \PW boson \ptlW, the separation \DRZlW between the \PZ boson and the lepton originating from the \PW boson decay, the azimuthal angle \Dphill between the two leptons originating from the \PZ boson decay in the lab frame, and the cosine of the  angle \cosThetaStar between the \PZ boson and its negatively charged decay lepton in the \PZ boson rest frame. These are the observables for which the differential cross sections are reported in Section~\ref{subsec:diff}.

The \PZ boson candidate is reconstructed from the opposite-sign same-flavor lepton pair with the closest invariant mass to that of the \PZ boson.
The third lepton is then interpreted as originating from the \PW boson.
Reasonable agreement can be seen between the expectation and the data. In Fig.~\ref{fig:controlplots}, a small excess of the data over the expectation is observed in the range of medium lepton \pt (upper row), towards larger number of jets (middle left), and at small values of maximum $\abs{\eta_{\text{jet}}}$. In this region the relative contribution from \ttZplustWZ is large.

\begin{figure}[!t]
\centering
\includegraphics[width=0.45\textwidth]{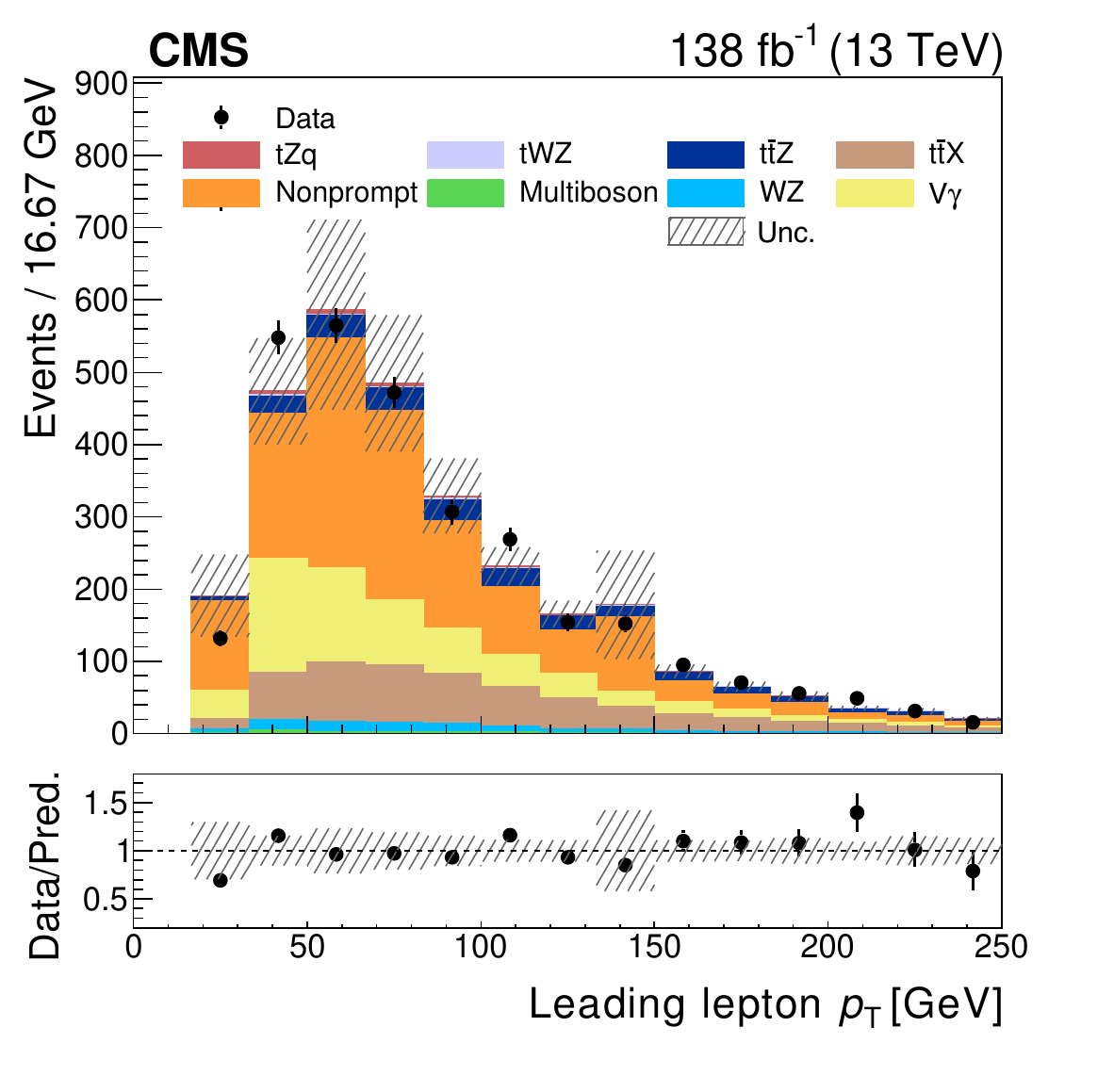}
\includegraphics[width=0.45\textwidth]{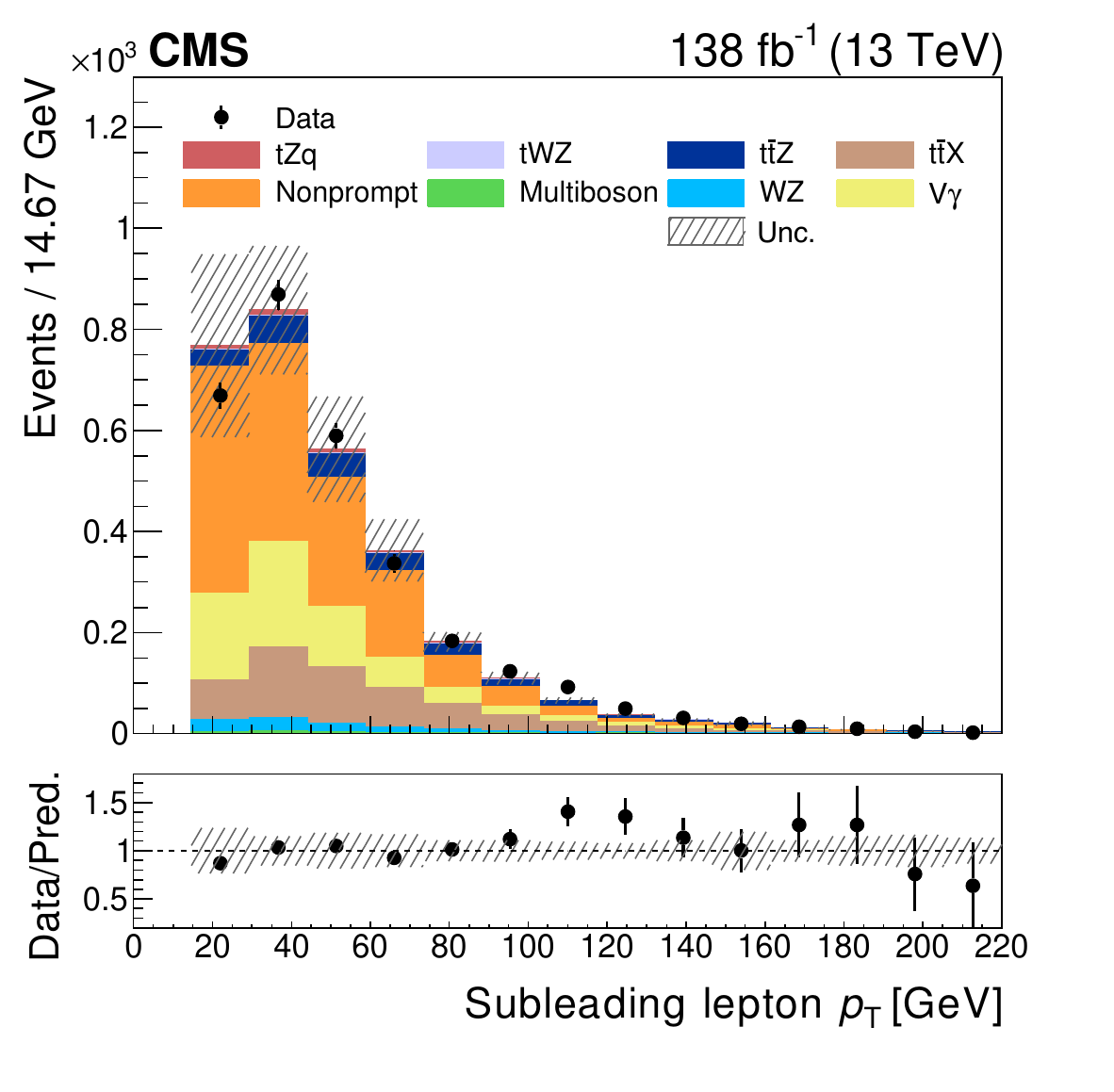} \\
\includegraphics[width=0.45\textwidth]{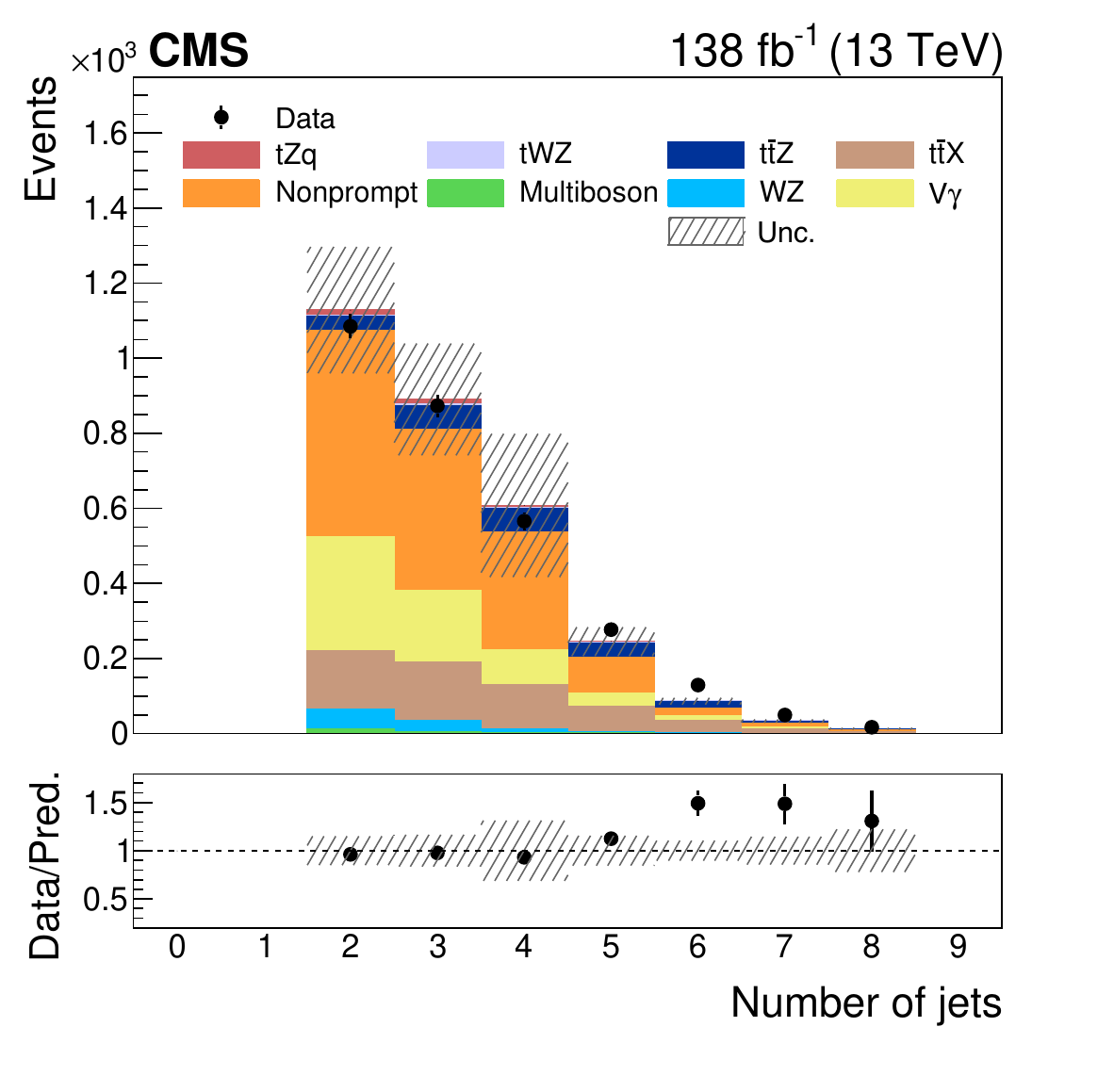}
\includegraphics[width=0.45\textwidth]{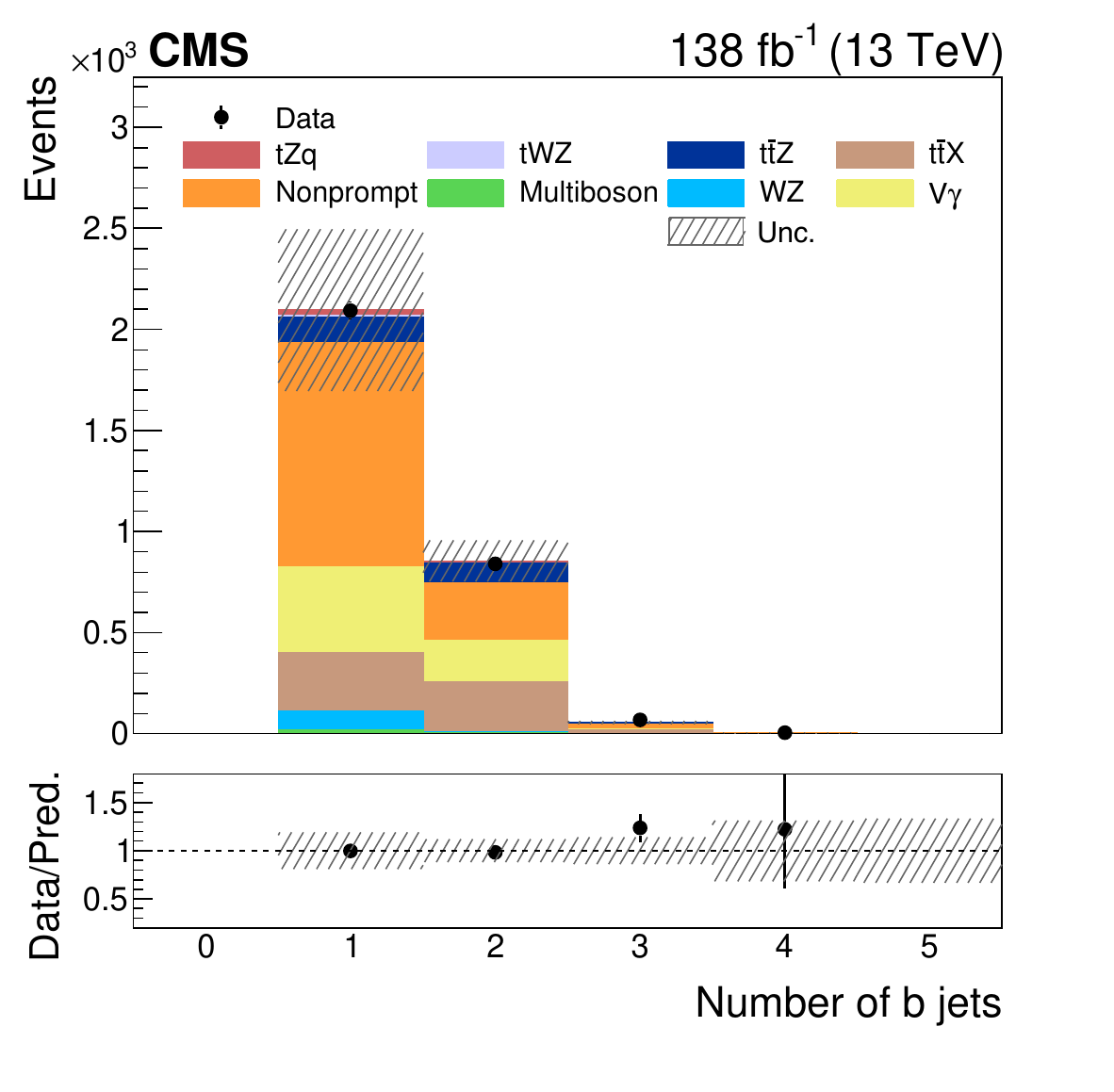}
\caption{Distributions for events selected in the region with $\abs{\mll-\mZ}>20\GeV$ for: the \pt of the lepton with the highest (upper left) and second highest (upper right) \pt, the number of jets (lower left), and the number of \PQb jets (lower right). The data are displayed as points with statistical error bars and the expectation is shown with a histogram with the systematic uncertainty given by the hatched area.}
\label{fig:offZpeak}
\end{figure}

The expectations for signals and most background processes are taken from the MC simulations described in Section~\ref{sec:samples}, while the nonprompt contribution is estimated from the data as described above. To validate the estimation of the nonprompt-lepton contribution, events outside the \PZ boson resonance region are selected, $\abs{\mll-\mZ}>20\GeV$. In this region, the contribution from events with nonprompt leptons is enhanced. Event distributions in this control region are shown in Fig.~\ref{fig:offZpeak}. The data are well described by the sum of the simulations and the contribution from nonprompt leptons as determined from the data.

To obtain an optimal separation between signals and backgrounds, a DNN based on \textsc{tensorflow} (v2)~\cite{ref:tensorflow} interfaced with \textsc{keras}~\cite{chollet2015keras} is used to categorize the events into three output nodes. In total, 26 variables are used as input to the DNN. These include kinematic properties of leptons, jets, and the reconstructed top quark candidates,  as well as jet and \PQb jet multiplicities, and \ptvecmiss.

For top quark candidates with a lepton in the final state, the neutrino kinematics are reconstructed from the measured \ptvecmiss and the charged lepton originating from the \PW boson decay, using the \PW boson mass as a constraint~\cite{PDG2022}. Depending on the jet multiplicity of the event, three different cases are considered: in events with two jets, of which at least one is a \PQb jet, the leptonic top quark candidate is reconstructed from the lepton, the neutrino momentum, and a \PQb jet; in events with three or more jets, of which at least one is a \PQb jet, squared distances \chisq of the reconstructed top quark mass with respect to the expected top quark mass 172.69\GeV~\cite{PDG2022}, weighted by their uncertainties, are calculated for both the {\Pell}+jets and all-hadronic top quark reconstructions; in events with four or more jets, two top quarks can be fully reconstructed and the overall \chisq value is determined as the sum of the two \chisq terms. In all cases, the combination of jets that gives the smallest overall \chisq is taken as the solution.

\begin{figure}[!p]
\centering
\includegraphics[width=0.45\textwidth]{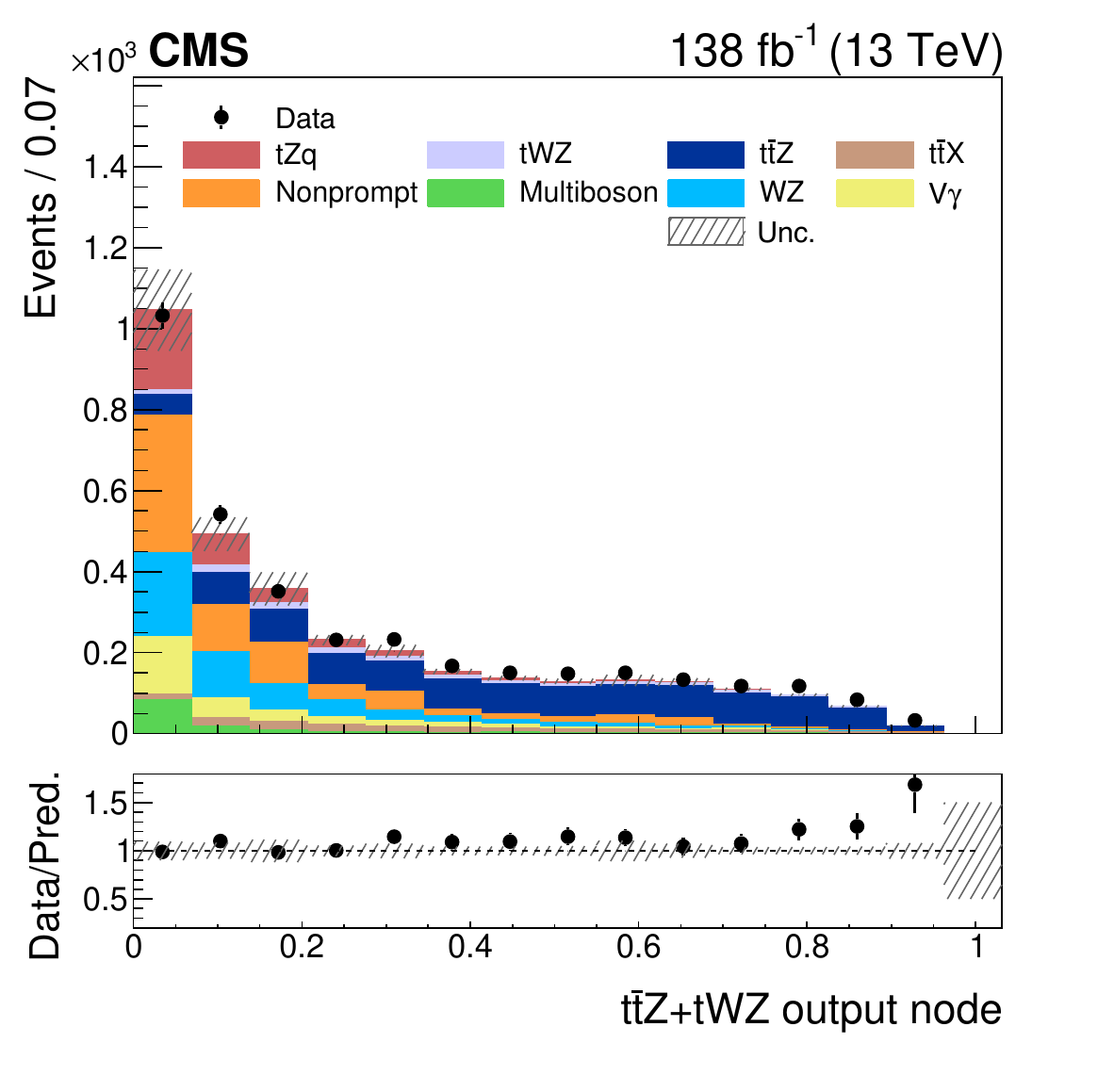}
\includegraphics[width=0.45\textwidth]{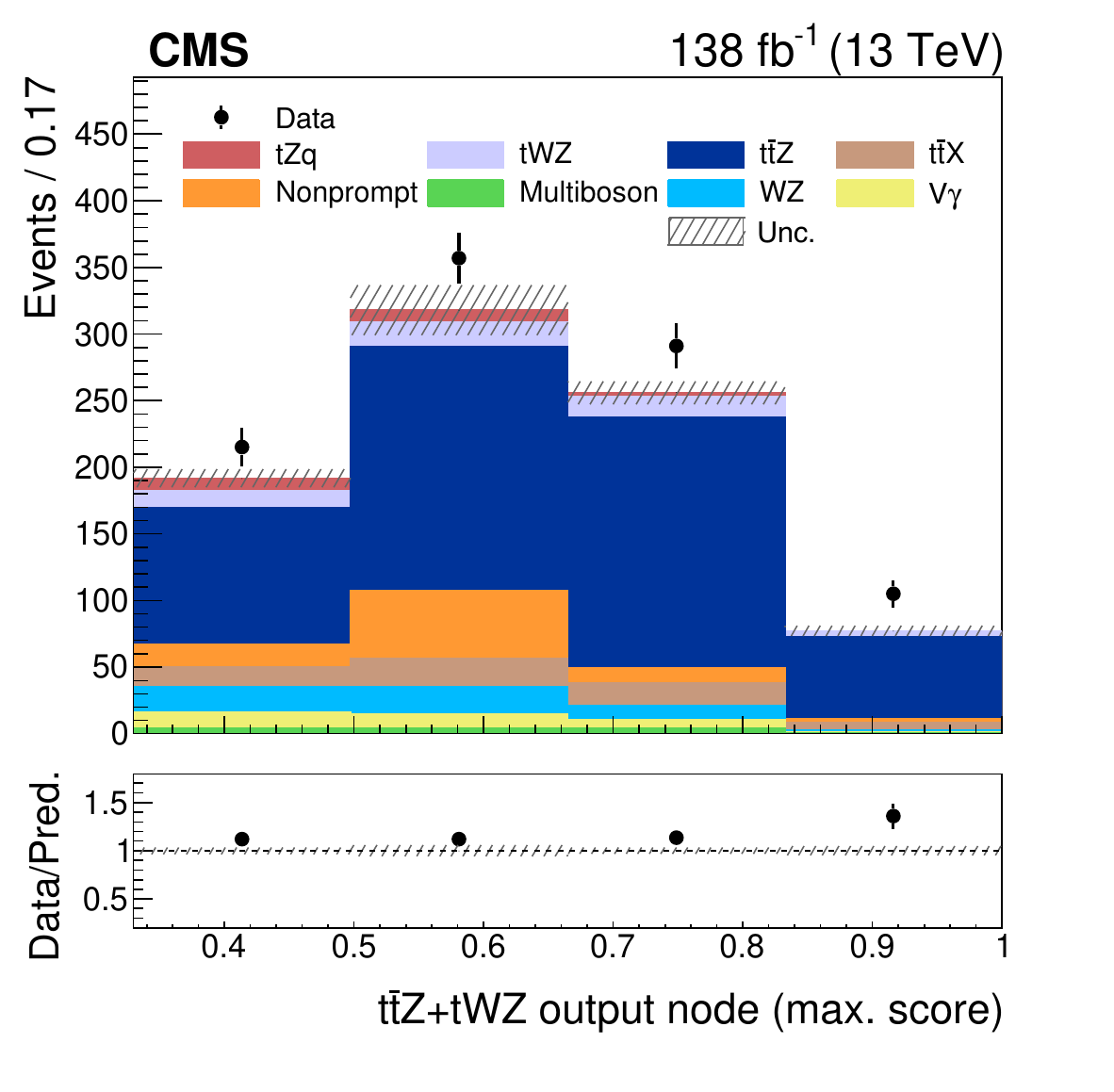} \\
\includegraphics[width=0.45\textwidth]{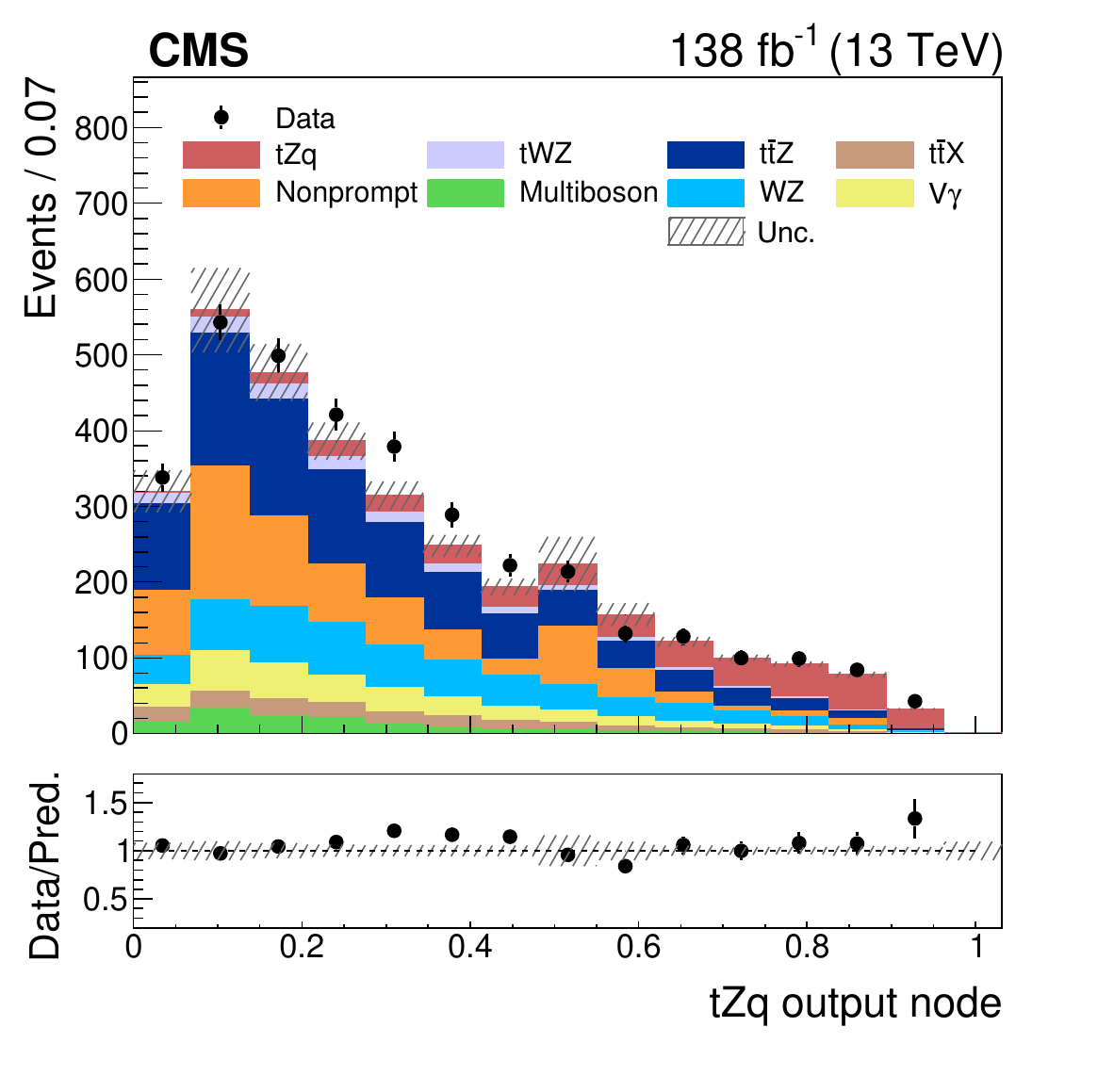}
\includegraphics[width=0.45\textwidth]{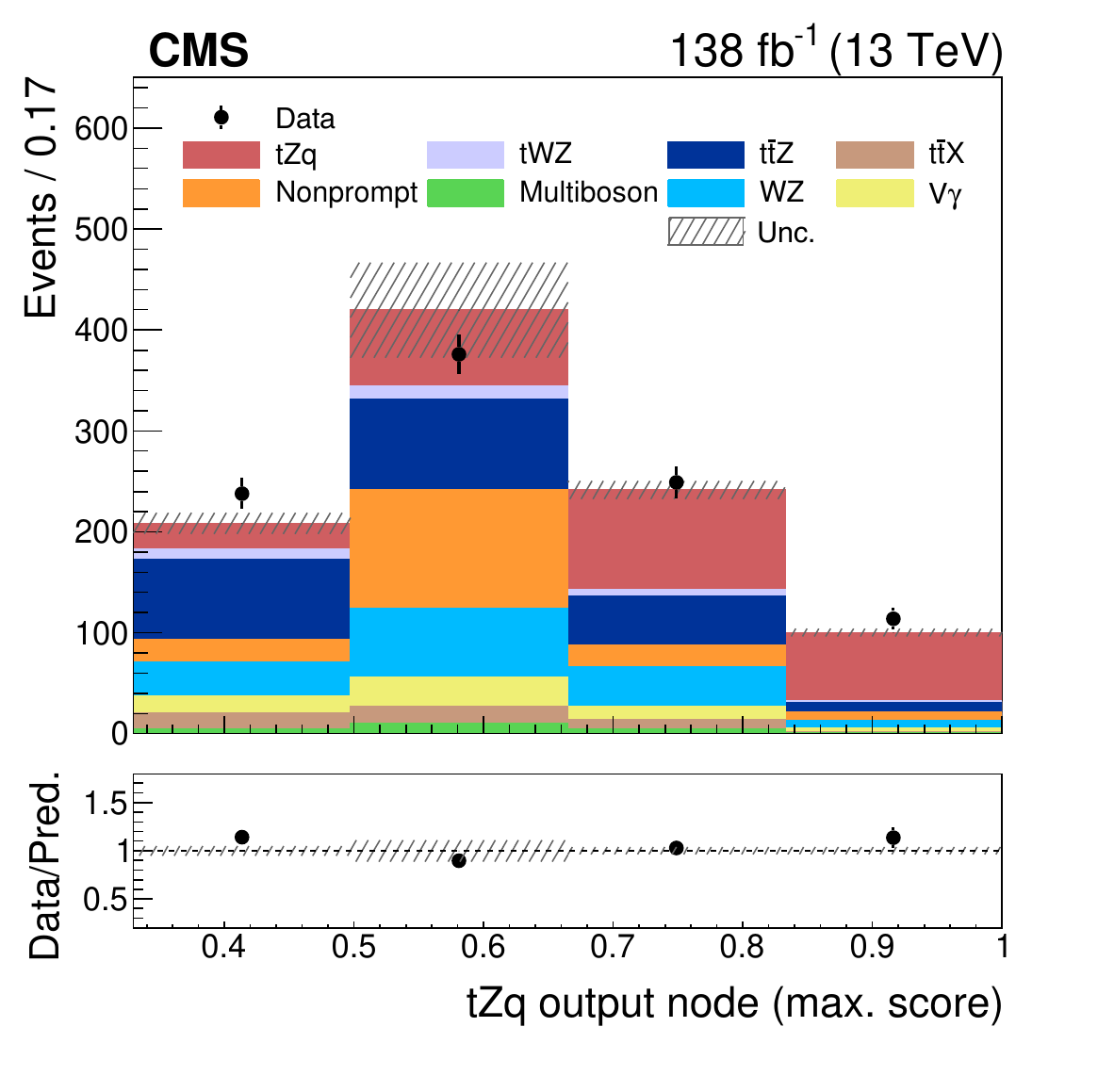} \\
\includegraphics[width=0.45\textwidth]{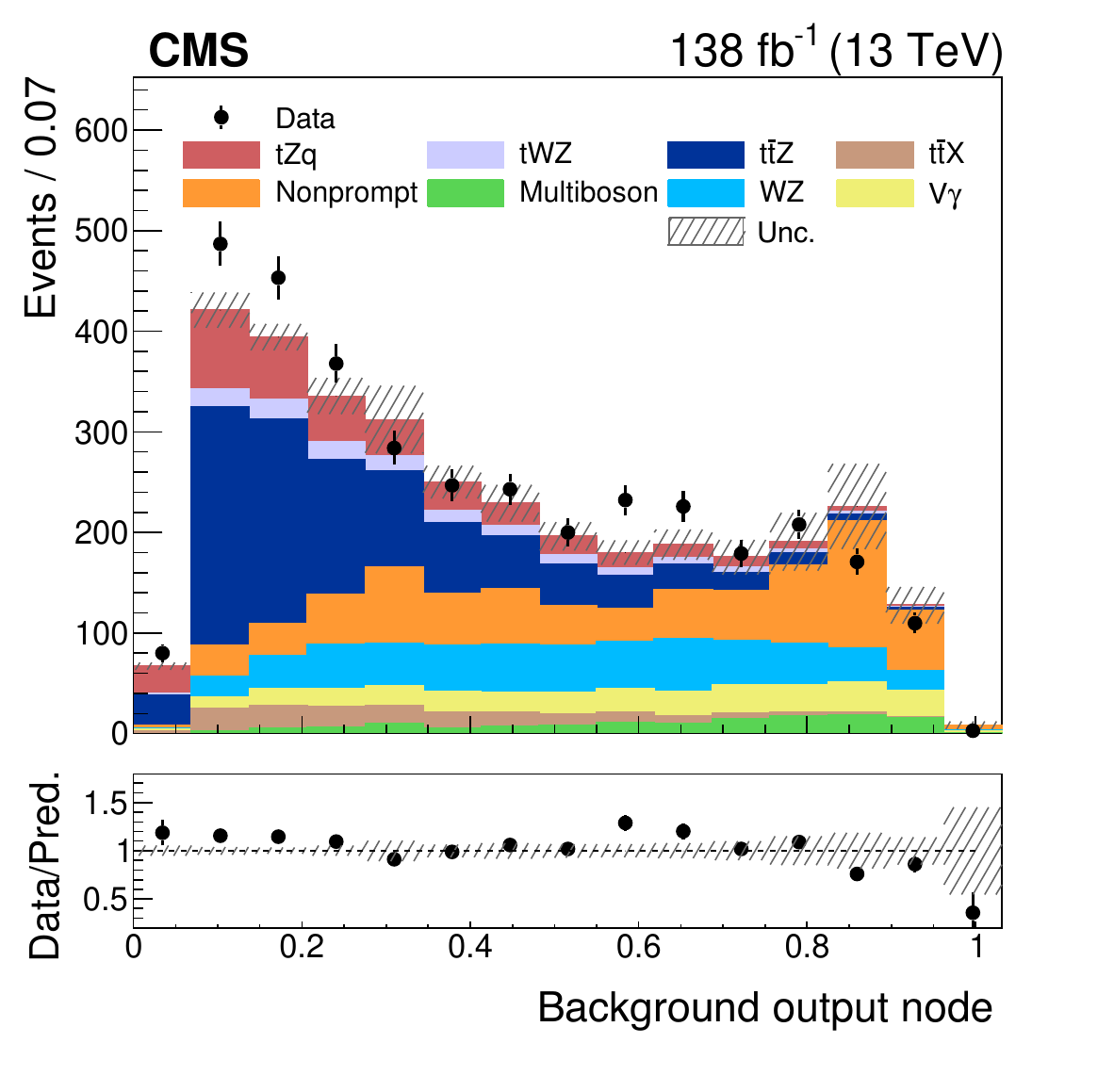}
\includegraphics[width=0.45\textwidth]{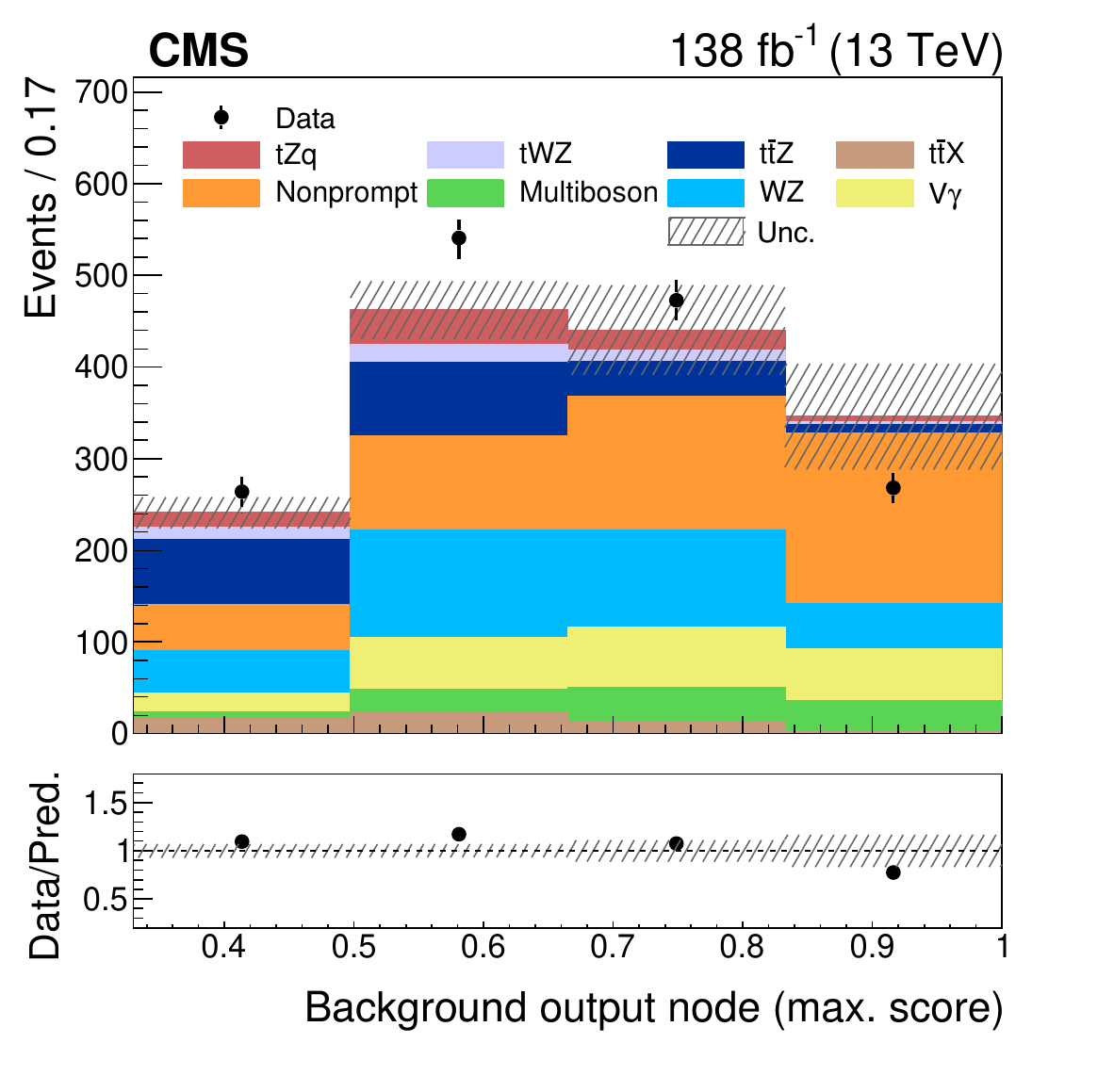}
\caption{Distributions of the output values in the three DNN output nodes for the sum of \ttZ and \tWZ (upper), \tZq (middle), and background (lower).
The data are compared with expectations.
In the left column, the inclusive distributions are shown, \ie, each selected event enters each of the output nodes.
In the right column, each event enters exactly one of the three distributions, namely the one for which the output score is largest. The data are displayed as points with statistical error bars and the expectation is shown with a histogram with the systematic uncertainty given by the hatched area.}
\label{fig:dnn}
\end{figure}

For the training of the DNN, MC simulation samples of signals and backgrounds are used. The samples are divided into two equal parts, based on even and odd event numbers. For each of these two halves, 80\% is used for training, the remaining 20\% is used for testing, and the other half is used to evaluate the model in order to build the output score distributions that will be used in the final stage of the analysis. 
For simulated events with negative weights, the absolute value of the event weight is used.
The model trained on odd-numbered events is used to evaluate the even-numbered subsample, and vice versa, ensuring an unbiased evaluation. In the case of data, the two samples based on even and odd event numbers are assigned to one of the two models each. The three output score values are normalized such that for each event their sum gives unity.
In Fig.~\ref{fig:dnn}, the data are compared to the expectation. Towards large values of the respective scores, high purities in \ttZplustWZ, \tZq, and background are achieved.

\section{Systematic uncertainties}
\label{sec:uncs}

Systematic uncertainties arise from experimental uncertainties in the detector response, such as calibrations, resolutions, and efficiencies, in the integrated luminosity, and from modeling uncertainties, which can affect the efficiency and acceptance.
Depending on the type of uncertainty, different shapes are assumed for the prior probability density functions. A given
systematic uncertainty can affect both the shape and the normalization of the distribution, or only the normalization.
To determine the effect of the uncertainty on the distribution, the analysis is repeated for each source of uncertainty,
shifting the relevant uncertainty parameter by one standard deviation. The systematic uncertainties are described in the following and their quantitative impacts on the inclusive \ttZplustWZ and \tZq cross sections are listed in Table~\ref{tab:syst}.
These are obtained by freezing the corresponding group of nuisance parameters in the fit and then subtracting in quadrature the resulting uncertainty from the uncertainty returned by the nominal fit.

\begin{table}[!ht]
\centering
\topcaption{%
    Systematic uncertainty sources and their relative impact on the inclusive \ttZplustWZ and \tZq cross section measurements.
}
\label{tab:syst}
\renewcommand{\arraystretch}{1.1}
\begin{tabular}{lcc}
    Source & \sttZtWZ & \stZq \\
    \hline
    Trigger &2\% &2\% \\
    Trigger prefiring &$<$1\% &2\% \\
    Lepton identification efficiencies &1\% &2\% \\
    Jet energy scale &1\% &3\% \\
    Jet energy resolution &$<$1\% &1\% \\
    \PQb tagging &1\% &2\% \\
    Missing transverse momentum &$<$1\% &3\% \\
    Nonprompt background &2\% &3\% \\
    Pileup &$<$1\% &1\% \\
    Integrated luminosity &2\% &2\% \\
    Statistical &3.7\% &10\% \\[\cmsTabSkip]
    \WZ background modeling &2\% &4\% \\
    Factorization scale &1\% &1\% \\
    Renormalization scale &1\% &2\% \\
    Parton shower &$<$1\% &2\% \\
    PDF and strong coupling \alpS &$<$1\% &$<$1\% \\
    Underlying event and color reconnection &1\% &2\% \\
    \tWZ modeling &$<$1\% &$<$1\% \\
    MC statistical &$<$1\% &1\% \\[\cmsTabSkip]
    Total &6\% &13\% \\
\end{tabular}
\end{table}

\subsection{Experimental uncertainties}

\begin{description}

\item[The trigger efficiency] is found to be close to unity. No trigger SFs are applied, and a flat 2\% uncertainty, uncorrelated across the years, is applied per event to account for the uncertainty in the trigger efficiency.
During certain periods of the data taking, the so-called \textit{Level-1 trigger prefiring}, \ie, a time jitter of the trigger signal in the muon~\cite{CMS:2021yvr} and ECAL~\cite{CMS:2020cmk} systems, led to a loss of events at the level of 0.5\% or less.

\item[The lepton identification efficiencies] are treated separately for electrons and muons. As described in Section~\ref{sec:evtsel}, the simulation is corrected to the data using SFs. The systematic components of the uncertainties are treated as correlated across the data-taking years.

{\tolerance=800
\item[The jet energy scale and resolution] uncertainties (JES and JER) include several different com\-po\-nents~\cite{CMS:2016lmd}. For each of these components, the four-momenta of the jets, as well as the SF for the \PQb tagging efficiencies, are varied and the analysis is repeated.
\par}

\item[\PQb tagging:] separate SFs and uncertainties are applied for jets originating from bottom and charm quarks, and those from light quarks or gluons~\cite{CMS:2017wtu,Bols:2020bkb}.

\item[Missing transverse momentum:] the observable \ptvecmiss includes contributions from jets and unclustered energy in the calorimeters, with systematic uncertainties determined for both components. Jets are considered only if their uncorrected \pt is larger than 15\GeV and their electromagnetic energy fraction is less than 0.9. The uncertainty in the unclustered \ptmiss is taken as uncorrelated between the years.  The uncertainty due to the jets includes each of the JES and JER variations~\cite{CMS:2019ctu}.

\item[Nonprompt-lepton background estimation:] the uncertainties in the lepton misidentification rates and the statistical uncertainties on the transfer factors are propagated into the final histogram templates, and these are included as shape uncertainties in the fit.
Additionally, a 30\% uncorrelated bin-by-bin normalization uncertainty is added to account for any residual mismodeling.

\item[Pileup:] before applying the event selection, event weights are applied to the MC simulation to match the pileup distributions of the data. Subsequently, to estimate the pileup uncertainty, the \pp inelastic cross section~\cite{CMS:2018mlc} is varied by $\pm$4.6\%, which corresponds to a variation of one standard deviation. This uncertainty is considered as fully correlated across the years.

\item[The integrated luminosities] for the 2016, 2017, and 2018 data-taking periods have individual
uncertainties of 1.2--2.5\%~\cite{CMS-LUM-17-003,CMS-PAS-LUM-17-004,CMS-PAS-LUM-18-002}. The combined uncertainty for the full data taking period is 1.6\%.
\end{description}

\subsection{Modeling uncertainties}

\begin{description}
\item[Backgrounds] other than the nonprompt background are modeled using MC simulation samples and an uncertainty of 15\% is estimated for each source, except for the $\PV\PGg$ process, for which a normalization uncertainty of 40\% is taken. For the \WZ background, the component with \PQc or \PQb induced jets is treated separately from that with light-flavor jets.

\item[Factorization and renormalization scale] uncertainties are considered for the signal samples \ttZ, \tWZ, and \tZq, as well as the background sample \WZ. For each process the renormalization and factorization scales are varied independently and the resulting variations are treated as uncorrelated between the different processes.

\item[Parton shower] uncertainties are computed by varying the renormalization scale for initial- and final-state radiation (ISR and FSR, respectively) independently. These uncertainties are considered for the processes \tZq, \ttZ, \tWZ, and \WZ. The ISR uncertainty is treated as uncorrelated between the processes, while only two nuisances are considered for FSR, one for the signal samples and one for the \WZ process.

{\tolerance=800
\item[PDF and strong coupling \alpS] uncertainties are considered for the signal samples and for the main background \WZ. For each event, 100 eigenvectors and two \alpS variations are provided~\cite{Butterworth:2015oua}, all of them treated as different nuisances in the fit. These uncertainties are considered correlated across the years and between the processes.
\par}

\item[Underlying event tune and color reconnection] uncertainties are estimated with dedicated MC samples~\cite{Argyropoulos:2014zoa,ref:cp5}. The uncertainties are treated as correlated among years and processes.

\item[\tWZ modeling:] two different schemes (DR1 and DR2) are used in the MC generation of \tWZ in order to account for the uncertainty due to interference between the \ttZ and the \tWZ processes.
The interference between \tZq and \tWZ is small and not considered in this analysis.

\item[The statistical uncertainty of the MC simulation samples] is estimated in the Barlow--Beeston lite approach~\cite{BARLOW1993219}.

\end{description}
The relative contribution of the \ttZ and \tWZ processes in their sum was also investigated as a possible source of systematic uncertainty by varying the theoretical assumptions on the cross sections by their maximum uncertainty, approximately 10\unit{fb}. This variation is performed anti-correlated between \ttZ and \tWZ, representing the most conservative estimate. However, the resulting effect, estimated to be 0.3\% for the inclusive measurement and ranging from 0.1 to 0.5\% for differential distributions, is found to be negligible at the current level of precision, and is not included in the fit.

\section{Results}
\label{sec:results}

The cross sections are measured using a profile likelihood ratio, in which for each cross section hypothesis
the optimal nuisance parameters are determined from a fit using the CMS statistical analysis tool \textsc{combine}~\cite{CAT-23-001}.
For the simultaneous measurement of the sum of the inclusive \ttZ and \tWZ cross sections,
and the inclusive \tZq cross section, two parameters of interest are used and their best fit values are the result of the measurement.
Correspondingly, for the simultaneous measurement of the differential cross sections,
the number of parameters of interest in the fit is twice the number of bins of the differential measurement.

\subsection{Inclusive cross sections}

\begin{figure}[!ht]
\centering
\includegraphics[width=0.45\textwidth]{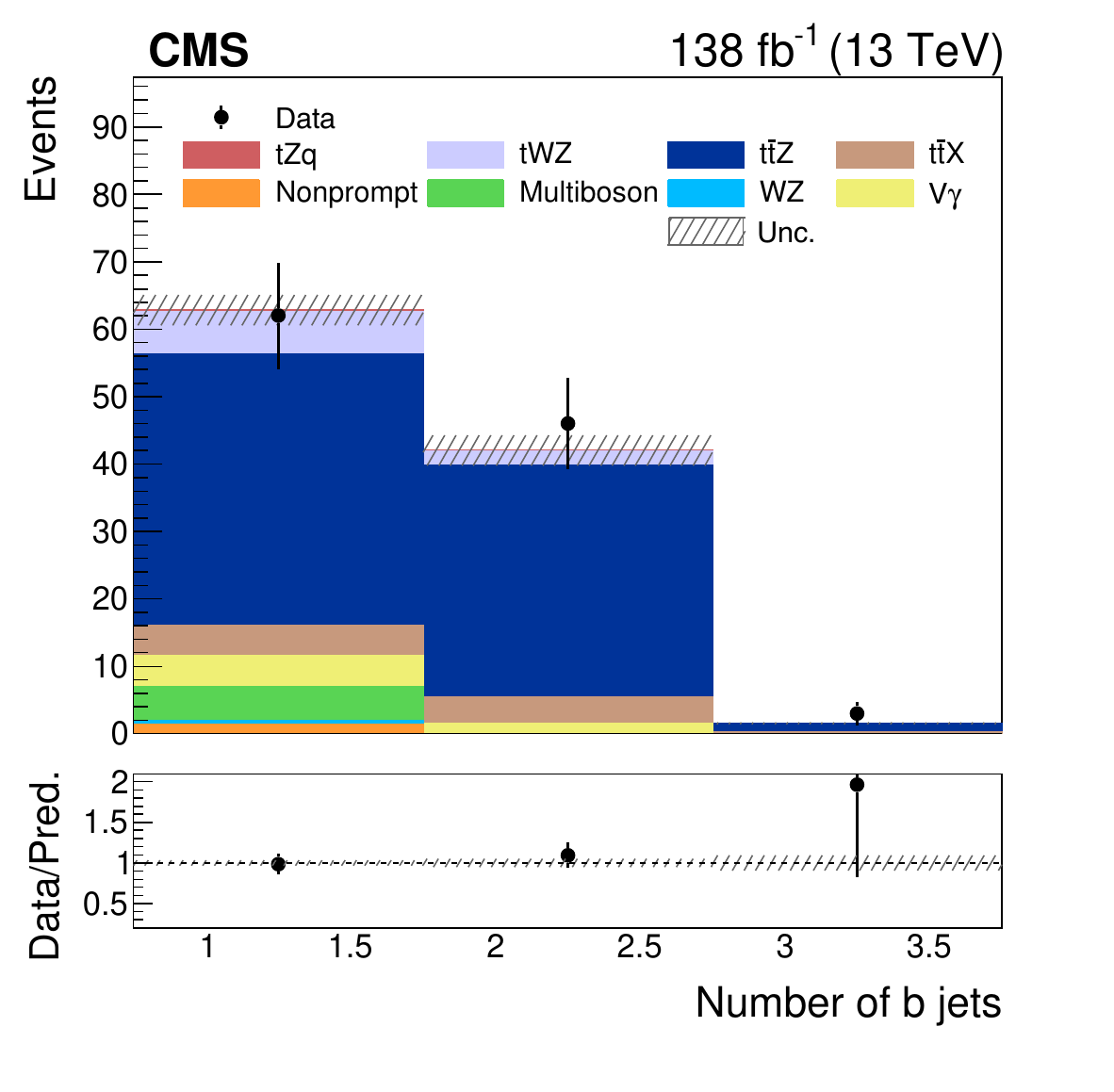}
\includegraphics[width=0.45\textwidth]{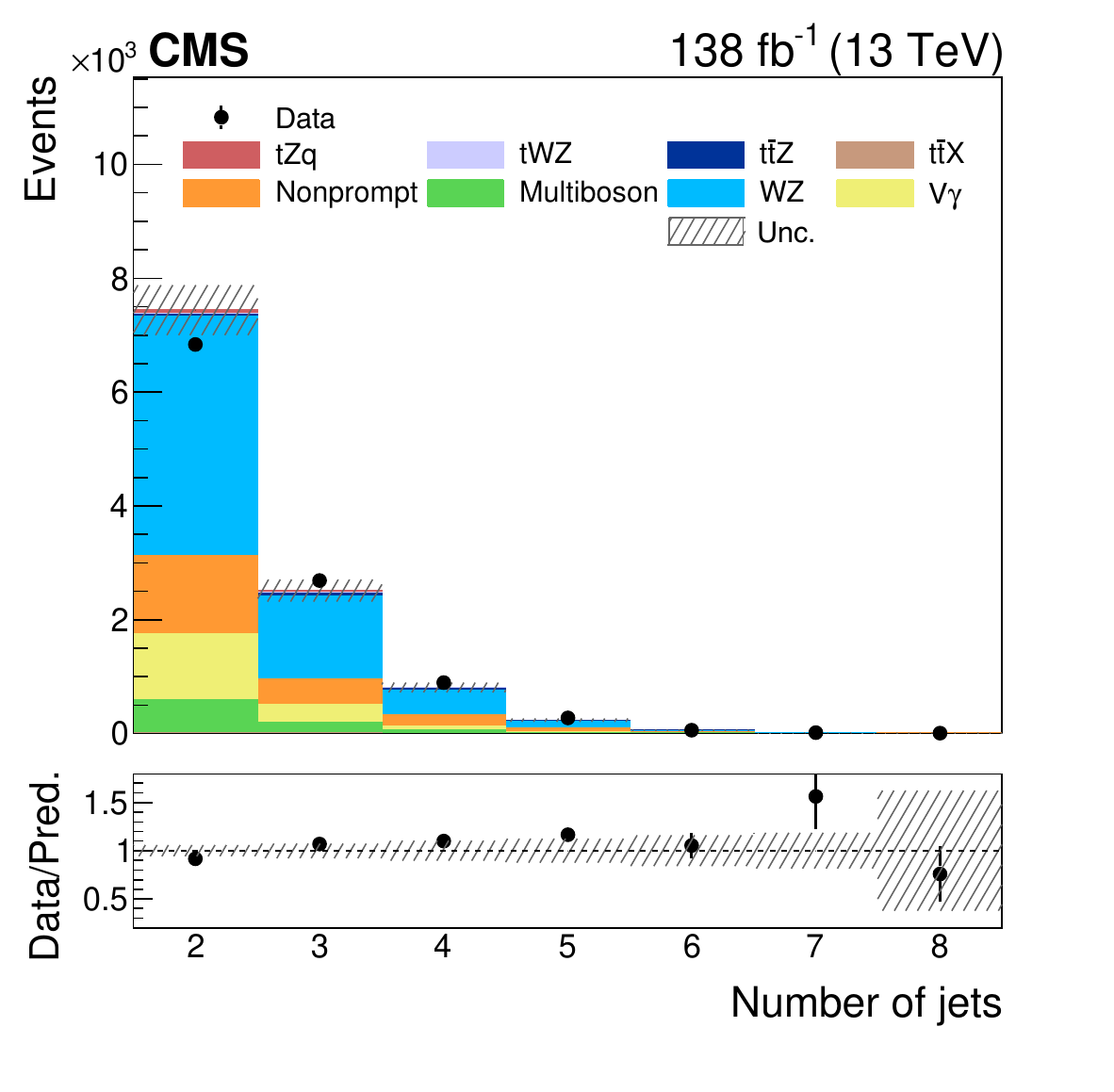}
\caption{Distributions of the \PQb jet multiplicity in the four lepton region (left) and the jet multiplicity in the zero \PQb jet control region (right). The data are compared to the expectation. The data are displayed as points with statistical error bars and the expectation is shown with a histogram with the systematic uncertainty given by the hatched area.}
\label{fig:inclusive}
\end{figure}

For the inclusive cross sections, the three-lepton event selection described above is extended to include a category of events with four leptons, as well as a category of three-lepton events without \PQb jets. The region with four leptons and at least one \PQb jet has a high purity in \ttZ events, as can be seen in Fig.~\ref{fig:inclusive} (left), where the \PQb jet multiplicity distribution for this region is shown.
The region without \PQb jets is enriched in \WZ background events and is useful as a control region. The jet multiplicity distribution for this region is shown in Fig.~\ref{fig:inclusive} (right).

Both distributions shown in Fig.~\ref{fig:inclusive}, as well as the three distributions in the right column of Fig.~\ref{fig:dnn}, are included in the profile likelihood fit for the extraction of the inclusive cross sections. The \PQb jet multiplicity distribution in the region with four leptons adds sensitivity to the \ttZ cross section, and the jet multiplicity in the region without \PQb jets improves the modeling of the \WZ background.

We define the inclusive cross sections for top quark production in association with a \PZ boson in the phase space including resonant and nonresonant production of opposite-sign and same-flavor lepton pairs with an invariant mass $70<m_{\Pellp\Pellm}<110\GeV$. The predicted cross sections in this phase space, as evaluated from the signal generator \MGvATNLO, are $0.84\pm0.08\unit{pb}$, $0.14\pm0.01\unit{pb}$, and $0.82\pm0.05\unit{pb}$, for the \ttZ, \tWZ, and \tZq processes, respectively.

\begin{figure}[!p]
\centering
\includegraphics[width=0.45\textwidth]{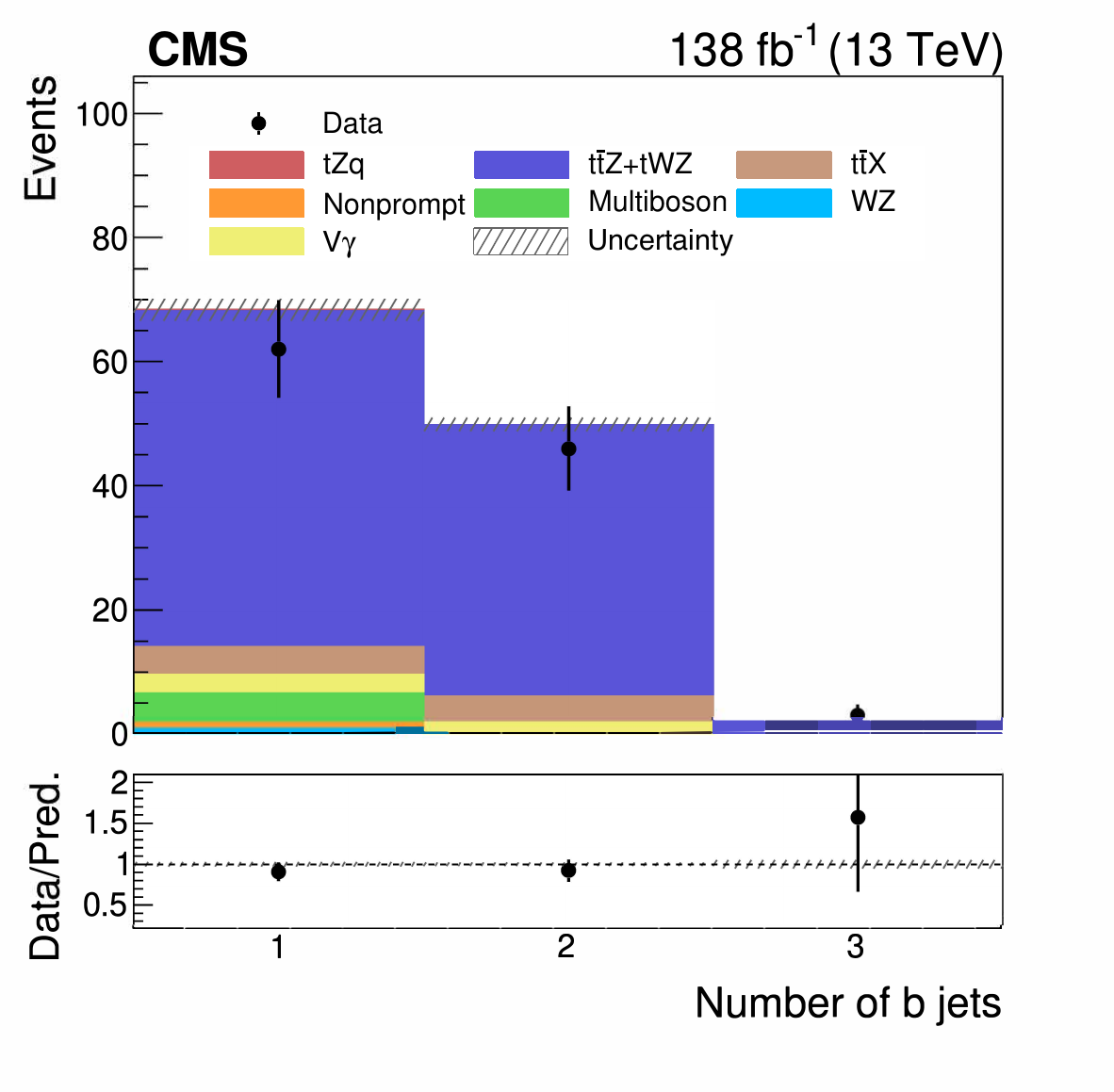}
\includegraphics[width=0.45\textwidth]{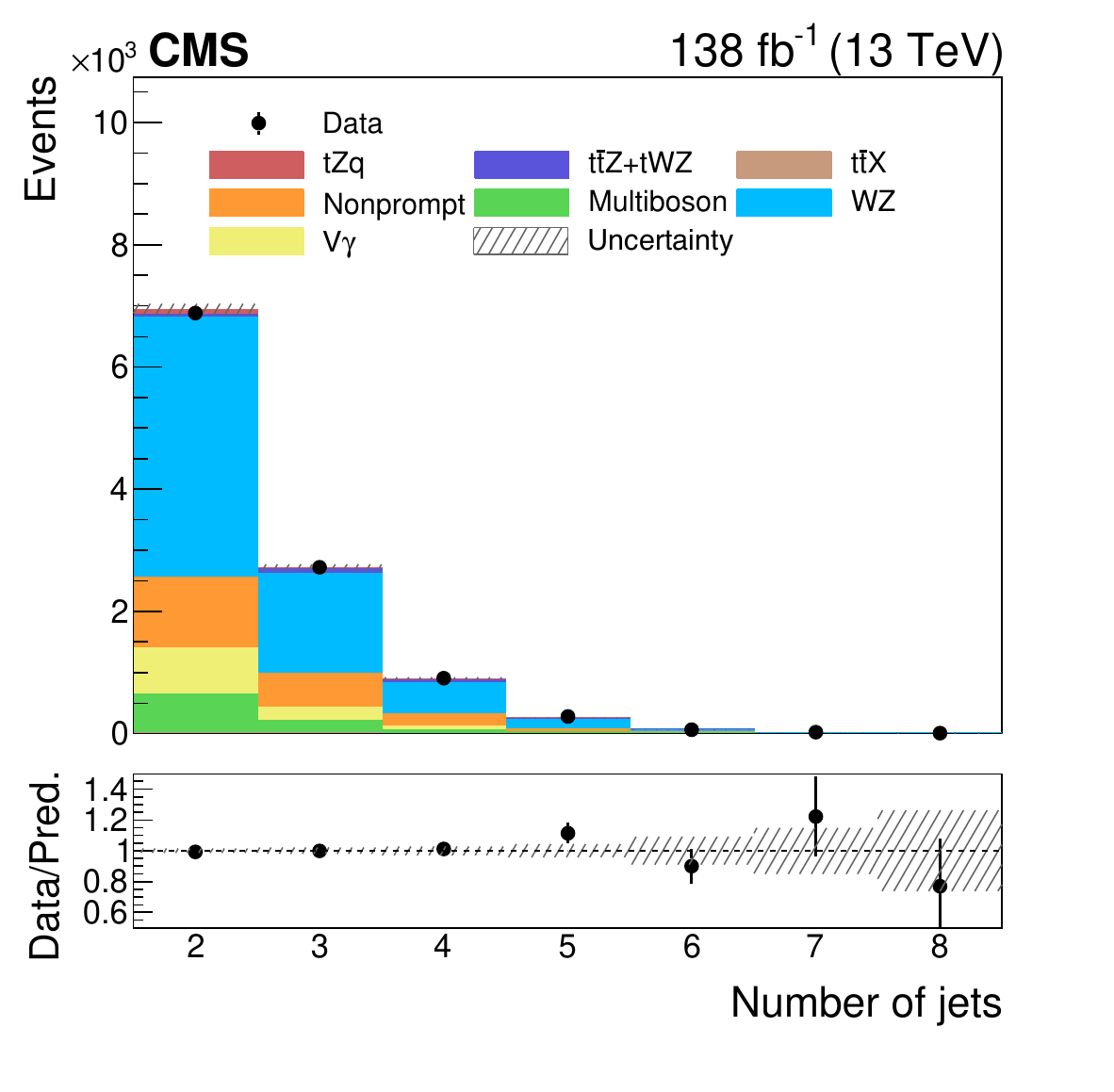} \\
\includegraphics[width=0.45\textwidth]{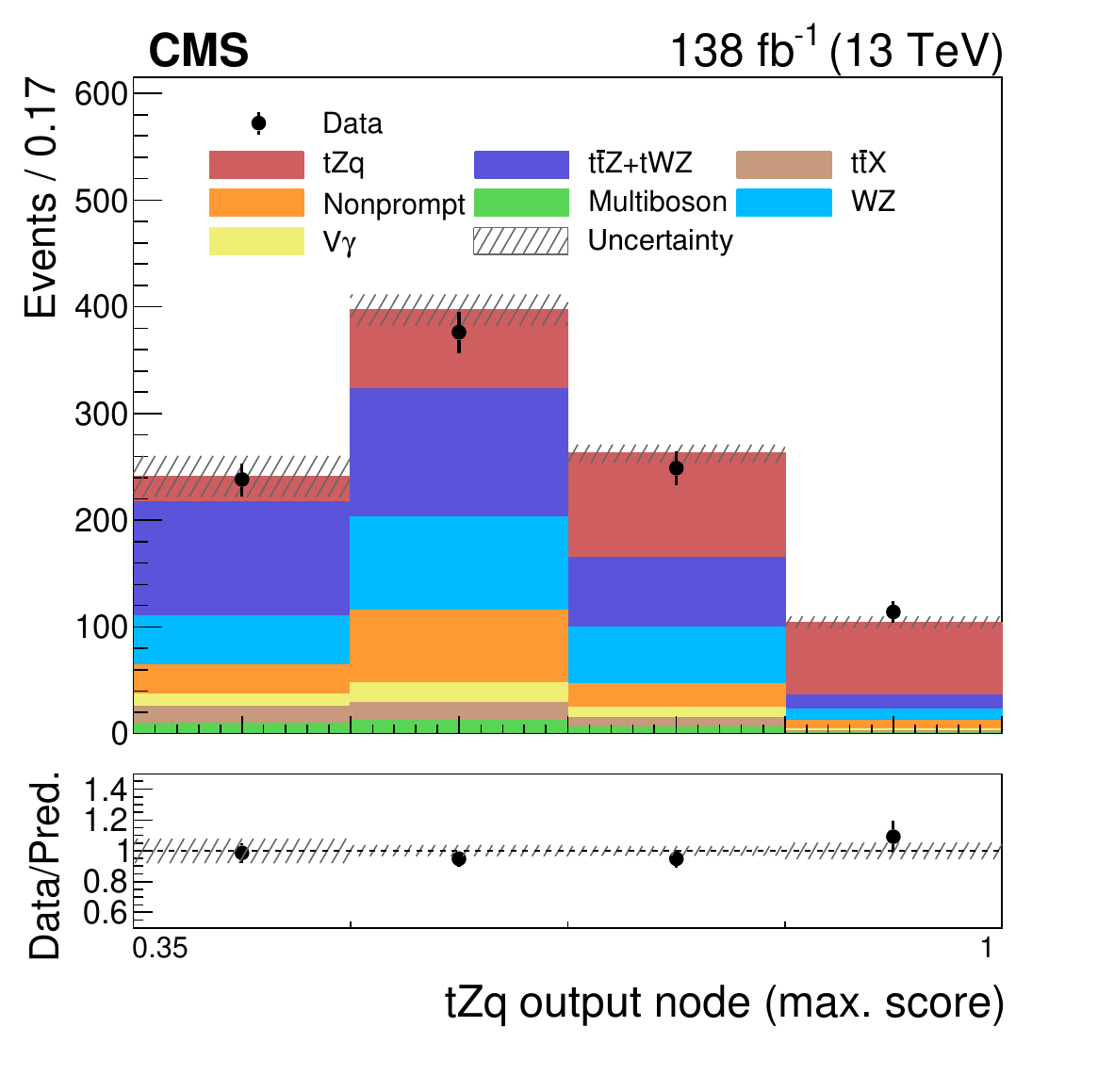}
\includegraphics[width=0.45\textwidth]{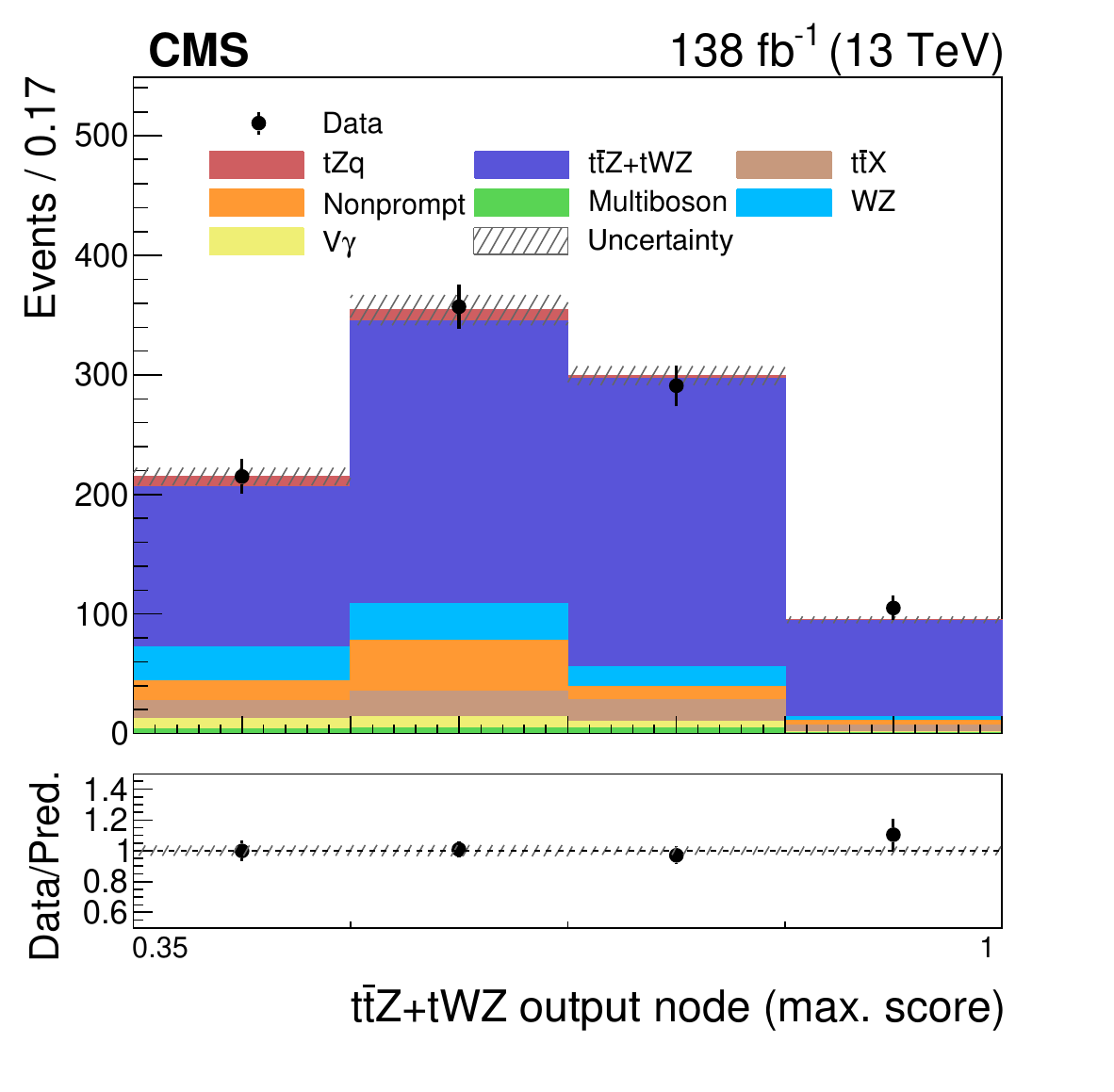} \\
\includegraphics[width=0.45\textwidth]{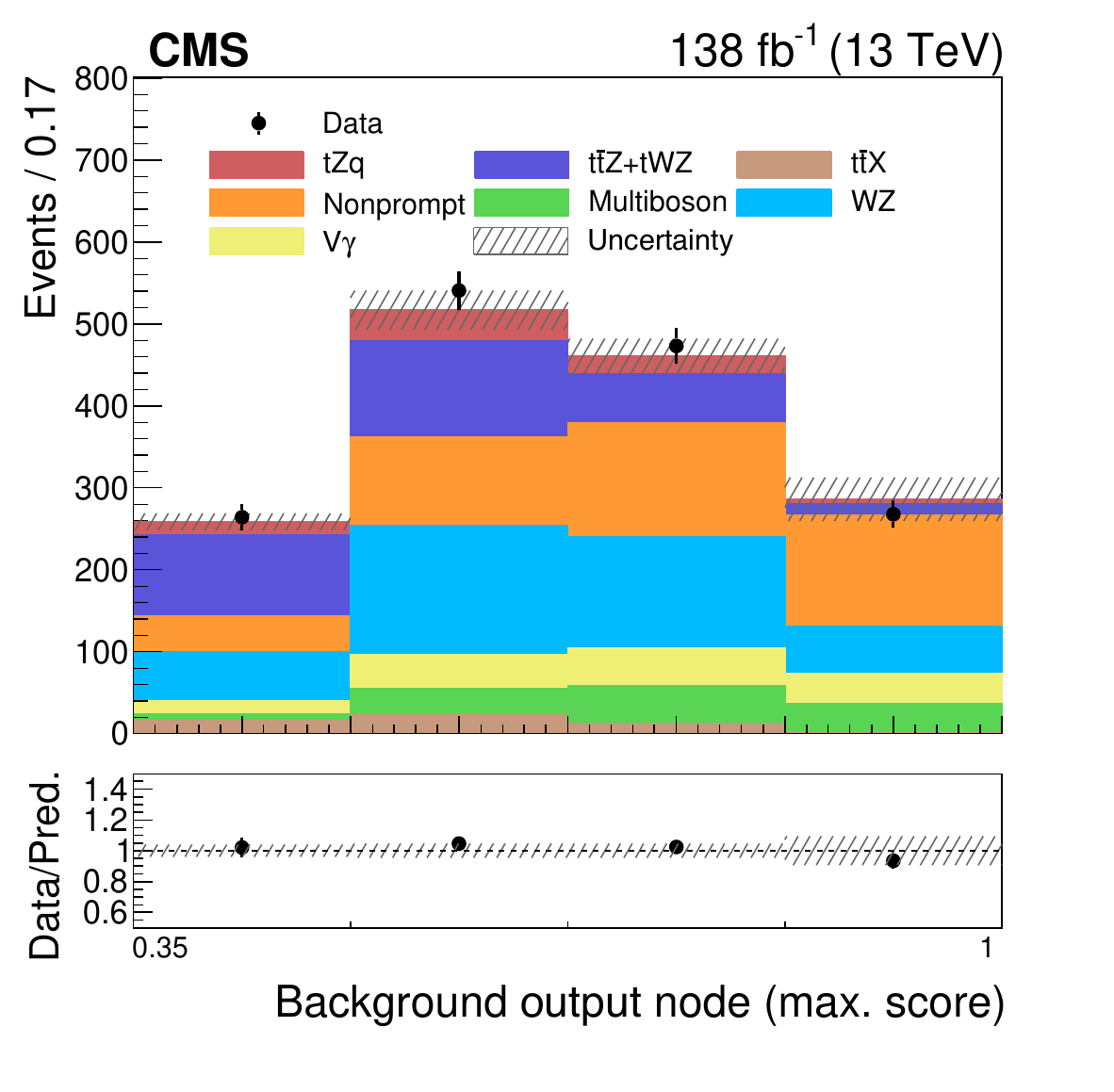}
\caption{Postfit distributions of the \PQb jet multiplicity in events with four leptons (upper left) and the jet multiplicity distribution in events with zero \PQb jets (upper right). Postfit distributions in the output nodes for \tZq (middle left), \ttZplustWZ (middle right), and the background (lower). The data are displayed as points with statistical error bars and the expectation is shown with a histogram with the systematic uncertainty given by the hatched area.}
\label{fig:postfit}
\end{figure}

The fit yields cross section ratios $\mu$ to the central value of the SM predictions of $\mttZtWZ=1.17\pm0.07$ for the sum of the \ttZ and \tWZ processes and $\mtZq=0.99\pm0.13$ for \tZq production.
The measured inclusive cross sections are:
\begin{align*}
    \sttZtWZ & = 1.14\pm0.05\stat\pm0.04\syst\unit{pb}, \\
    \stZq & = 0.81\pm0.07\stat\pm0.06\syst\unit{pb}.
\end{align*}
The distributions resulting from the fit, in the following referred to as postfit distributions, are shown in Fig.~\ref{fig:postfit}.
The corresponding prefit distributions, \ie, before the scaling of the expectations to the result of the fit, are presented in the right column of Fig.~\ref{fig:dnn} and in Fig.~\ref{fig:inclusive}.

\begin{figure}[!ht]
\centering
\includegraphics[width=0.55\textwidth]{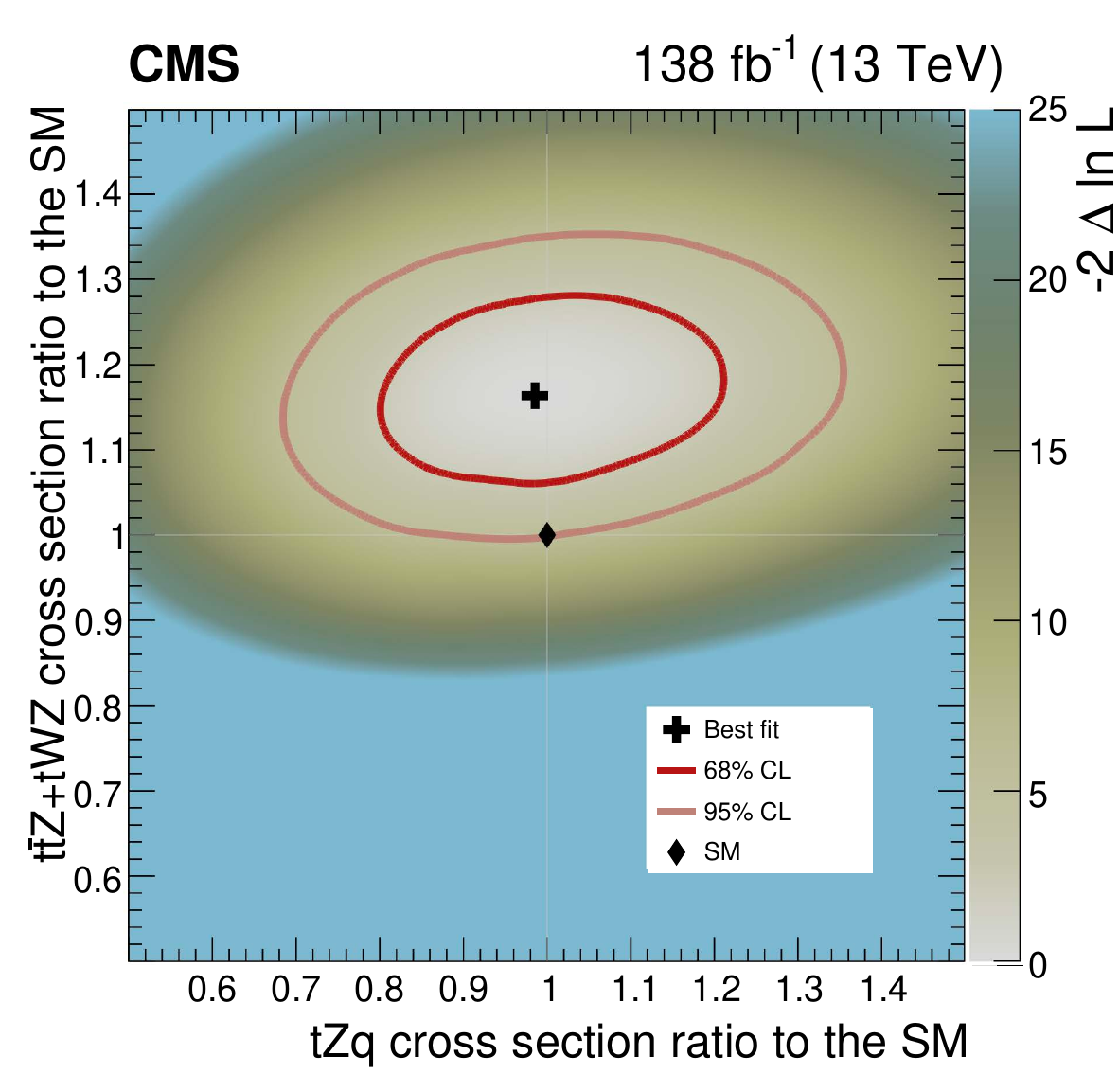}
\caption{Scan of twice the negative log-likelihood difference $-2\Delta\ln{L}$ (color axis) of the two measured inclusive cross sections normalized to the SM predictions \mttZtWZ and \mtZq. The measurement is indicated by a cross, and the SM prediction by a diamond. The 68\% and 95\% confidence levels (\CL) ($-2\Delta\ln{L}=2.30$ and 5.99, respectively) are indicated by the red lines.}
\label{fig:2DScan}
\end{figure}

While the measured \tZq cross section agrees very well with the prediction, the sum of the \ttZplustWZ cross sections is about two standard deviations higher than the central value of the prediction.
The two-dimensional profile likelihood ratio for \mttZtWZ and \mtZq is displayed in Fig.~\ref{fig:2DScan}. A small positive correlation can be seen, which arises from the correlated sources, such as the experimental uncertainties in JES, lepton identification, luminosity, and trigger.

\noindent

The previous CMS measurement of the \tZq cross section~\cite{CMS:2021ugv} was performed using the same data set as this analysis, but the results were reported in the phase space in which the \PZ boson decays into two charged leptons (electrons, muons, or tau leptons) with an invariant mass greater than 30\GeV. In this phase space, the current fit yields $93\,^{+12}_{-11}\unit{fb}$ for the \tZq process when fixing the \ttZ and \tWZ cross sections to their SM predictions, consistent with the previous measurement of $87.9\,^{+10.5}_{-9.5}\unit{fb}$~\cite{CMS:2021ugv}, and with the prediction $94.2\pm3.1\unit{fb}$.

As a cross-check with the previous CMS measurements of the \ttZ~\cite{CMS:2019too} and \tWZ~\cite{CMS:2023krq} cross sections, we repeat the likelihood fit using two alternative configurations. First, fixing the \tZq and \tWZ processes to their SM prediction, with normalization uncertainties of 15 and 40\%, respectively, returns a value of $0.99\pm0.07\unit{pb}$, in agreement with the previous result of $0.95\pm0.08\unit{pb}$~\cite{CMS:2019too}.
Second, constraining the \ttZ and \tZq processes to the SM prediction with normalization uncertainties of 15\% each, the inclusive \tWZ cross section is found to be $0.39\pm0.16\unit{pb}$, in agreement with the CMS measurement of  $0.35\pm0.11\unit{pb}$~\cite{CMS:2023krq}.

\subsection{Differential measurement}
\label{subsec:diff}

The differential cross sections for the sum of \ttZ and \tWZ, and for \tZq are measured as functions of: the transverse momentum \pTZ of the \PZ boson; the transverse momentum \ptlW of the lepton coming from the \PW boson decay; the azimuthal angle \Dphill between the two leptons coming from the \PZ boson; the \DR between the \PZ boson and the lepton coming from the \PW boson, \DRZlW; and the cosine of the angle \cosThetaStar between the \PZ boson and the negatively charged lepton coming from its decay, in the \PZ boson rest frame.
The measured and expected event distributions for these observables have been shown in Fig.~\ref{fig:controlplots_unfolding}.

The chosen observables are expected to be sensitive to QCD modeling and/or beyond-SM scenarios and useful for EFT interpretations. For example, \pTZ and \ptlW are sensitive to the $\mathcal{O}_{\PQt\PW}$ and $\mathcal{O}_{\phi\PQt\PQb}$ operators that describe modified \PW--\PQt--\PQb couplings~\cite{CMS:2021aly,Degrande:2018fog}. In case of new physics, discrepancies between the data and the SM prediction are expected in \Dphill, as the leptons from the \PZ boson decay will be more or less collimated depending on the boost of the \PZ boson. The \DRZlW variable is sensitive to the modeling of the processes, as well as to the coupling between top quark and \PZ boson. The observable \cosThetaStar is an interesting observable to constrain EFT parameters, as discussed in Refs.~\cite{CMS:2019too,spitzbart}.

The reconstructed observables are affected by the detector resolutions and efficiencies, as well as by limitations in the detector acceptance.
The binning of the different observables is chosen by computing a response matrix for both signal samples, maintaining good purity and stability and a sizable number of signal events in each bin. The purity is defined as the fraction of events in a detector-level bin that belong to the corresponding generator-level bin, while the stability is the fraction of events in a generator-level bin that are observed in the corresponding detector-level bin. For all of the observables, the purity and stability are above 85\% in every bin, and no regularization condition is applied in the unfolding. To facilitate comparisons with theory calculations for top quarks in the final state, the data are unfolded to the parton level, after ISR and FSR, and before hadronization.

As described in Section~\ref{sec:evtsel}, each event is assigned to the process category for which the corresponding output node is maximal.
The differential cross sections for \ttZ and \tWZ combined and for \tZq are extracted simultaneously using a profile likelihood fit to the data.
In each bin of the measurement, templates of the expected output score distributions for signals and backgrounds are created. The signal templates are split into subsamples corresponding to the generator-level bins of the observable to unfold, and the events in the signal output nodes are further split into the different categories corresponding to the detector-level bins. In the fit, for each bin, two cross section parameters of interest are used to determine the likelihood ratio in that bin. The postfit results in each bin are the values of the cross section and nuisance parameters for which the profile likelihood ratio is maximized.

\begin{figure}[!ht]
\centering
\includegraphics[width=0.9\textwidth]{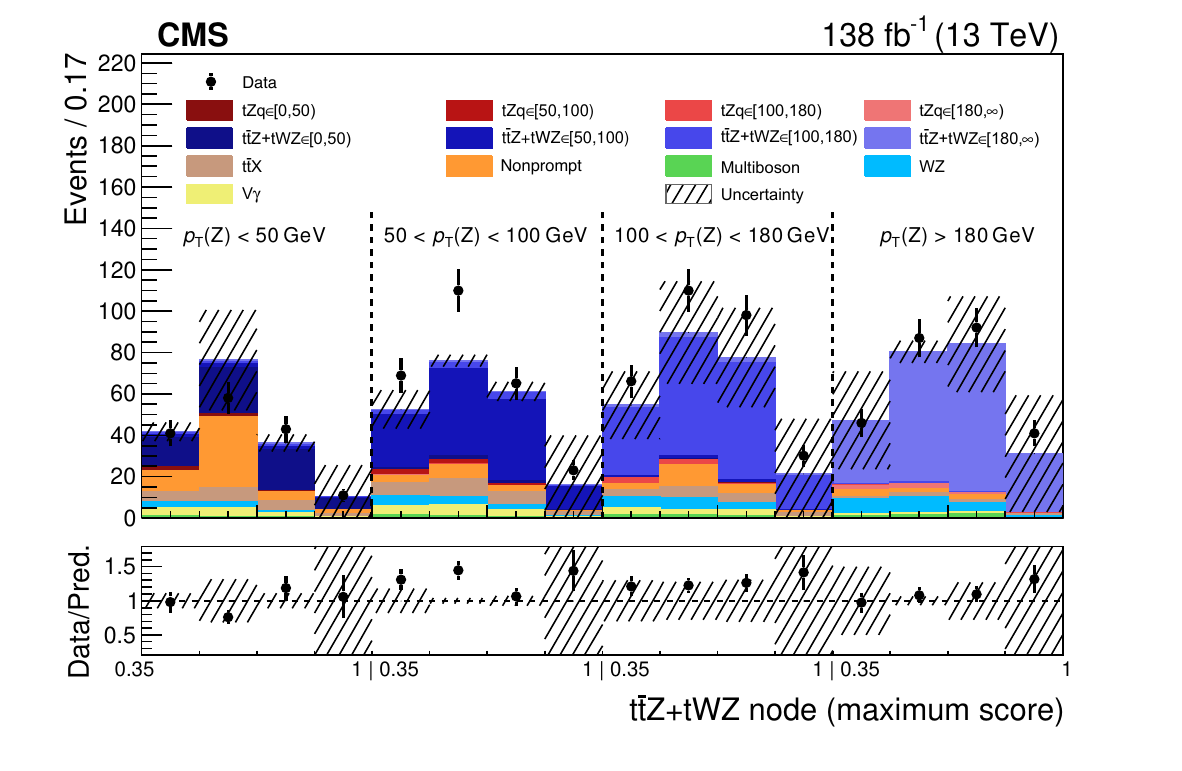} \\
\includegraphics[width=0.9\textwidth]{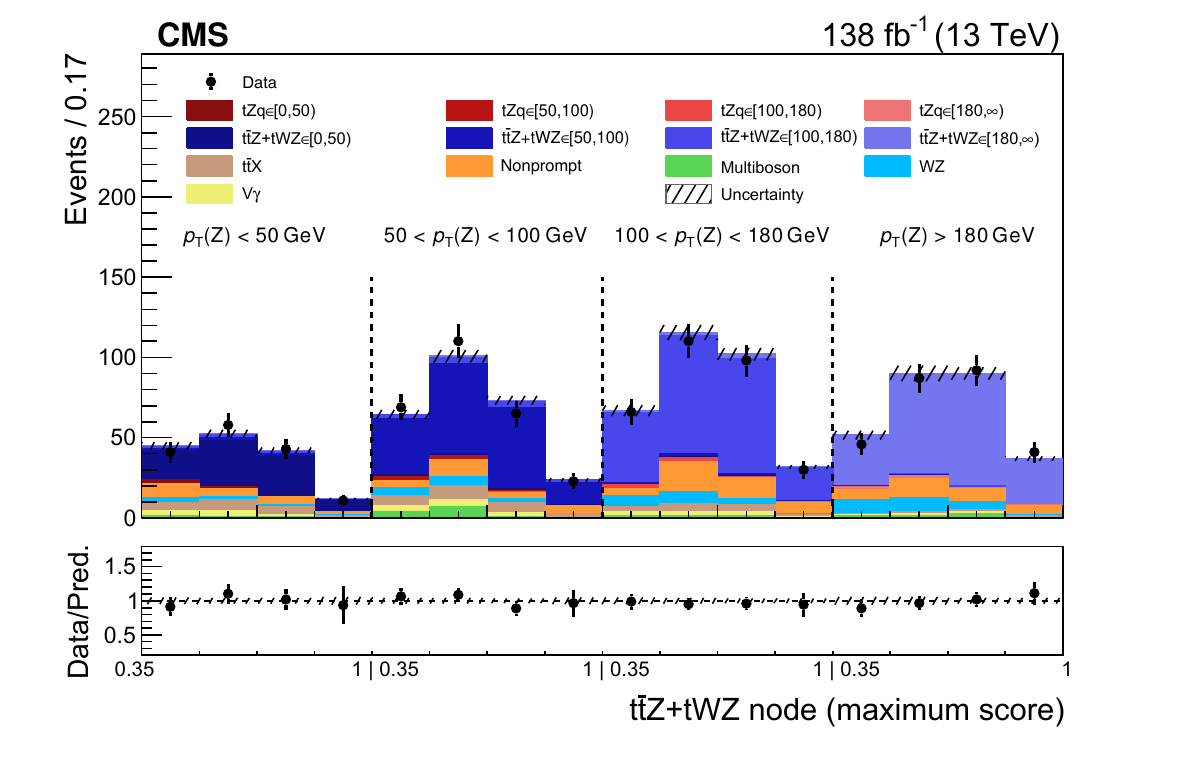}
\caption{Prefit (upper) and postfit (lower) output node distributions for the sum of the \ttZ and \tWZ processes.
Separate templates are shown for each bin of reconstructed \pTZ.
The signal samples are further split into four components each, shown by different colors, according to the generator-level bins of \pTZ.
The fit is performed simultaneously on this distribution and that in Fig.~\ref{fig:fit_ptZ_tZq}.
The data are displayed as points with statistical error bars and the expectation is shown with a histogram with the systematic uncertainty given by the hatched area.}
\label{fig:fit_ptZ_ttZ}
\end{figure}

\begin{figure}[!t]
\centering
\includegraphics[width=0.9\textwidth]{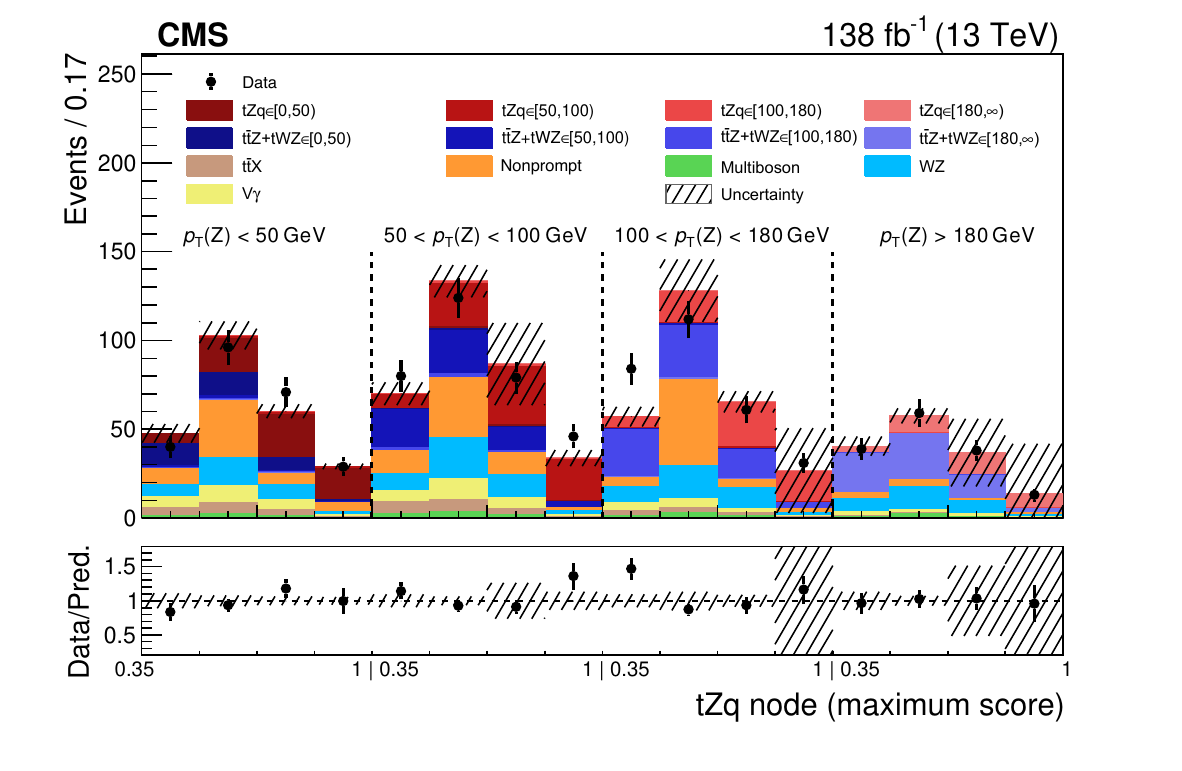} \\
\includegraphics[width=0.9\textwidth]{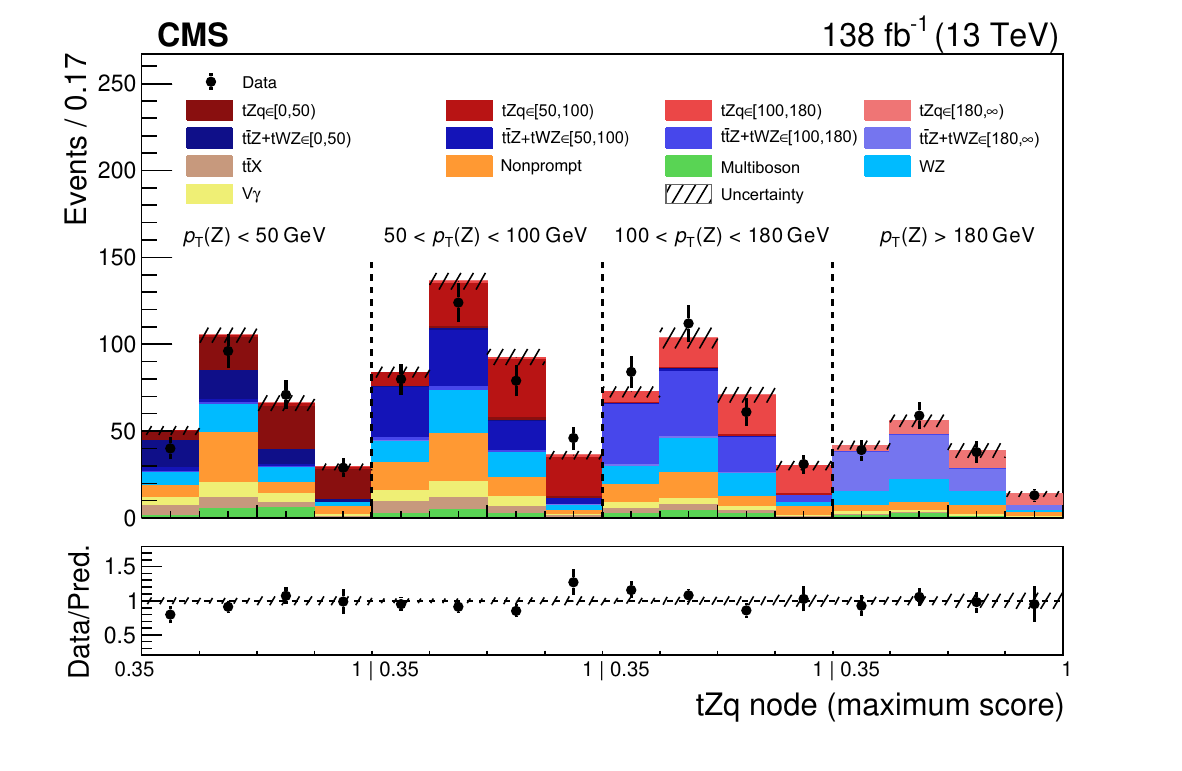}
\caption{Prefit (upper) and postfit (lower) output node distributions for the \tZq process.
Separate templates are shown for each bin of reconstructed \pTZ.
The signal samples are further split into four components each, shown by different colors, according to the generator-level bins of \pTZ.
The fit is performed simultaneously on this distribution and that in Fig.~\ref{fig:fit_ptZ_ttZ}.
The data are displayed as points with statistical error bars and the expectation is shown with a histogram with the systematic uncertainty given by the hatched area.}
\label{fig:fit_ptZ_tZq}
\end{figure}

The inputs and results of the fit are shown in Figs.~\ref{fig:fit_ptZ_ttZ} and \ref{fig:fit_ptZ_tZq} for the example of the differential cross sections as a function of \pTZ. The components \ttZplustWZ and \tZq are measured simultaneously, \ie, eight parameters of interest are determined at the same time.

\begin{figure}[!p]
\centering
\includegraphics[width=0.45\textwidth]{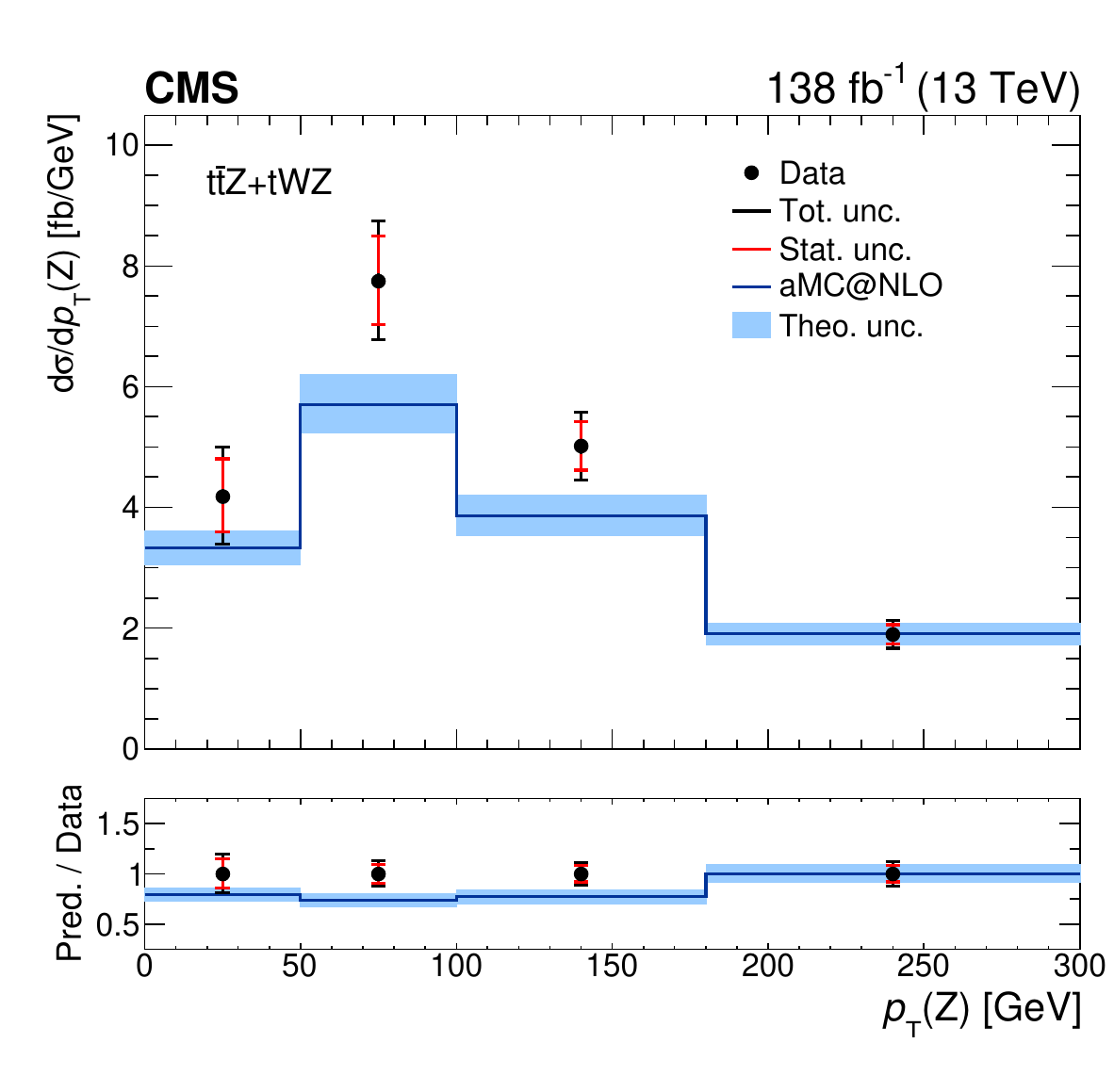}
\includegraphics[width=0.45\textwidth]{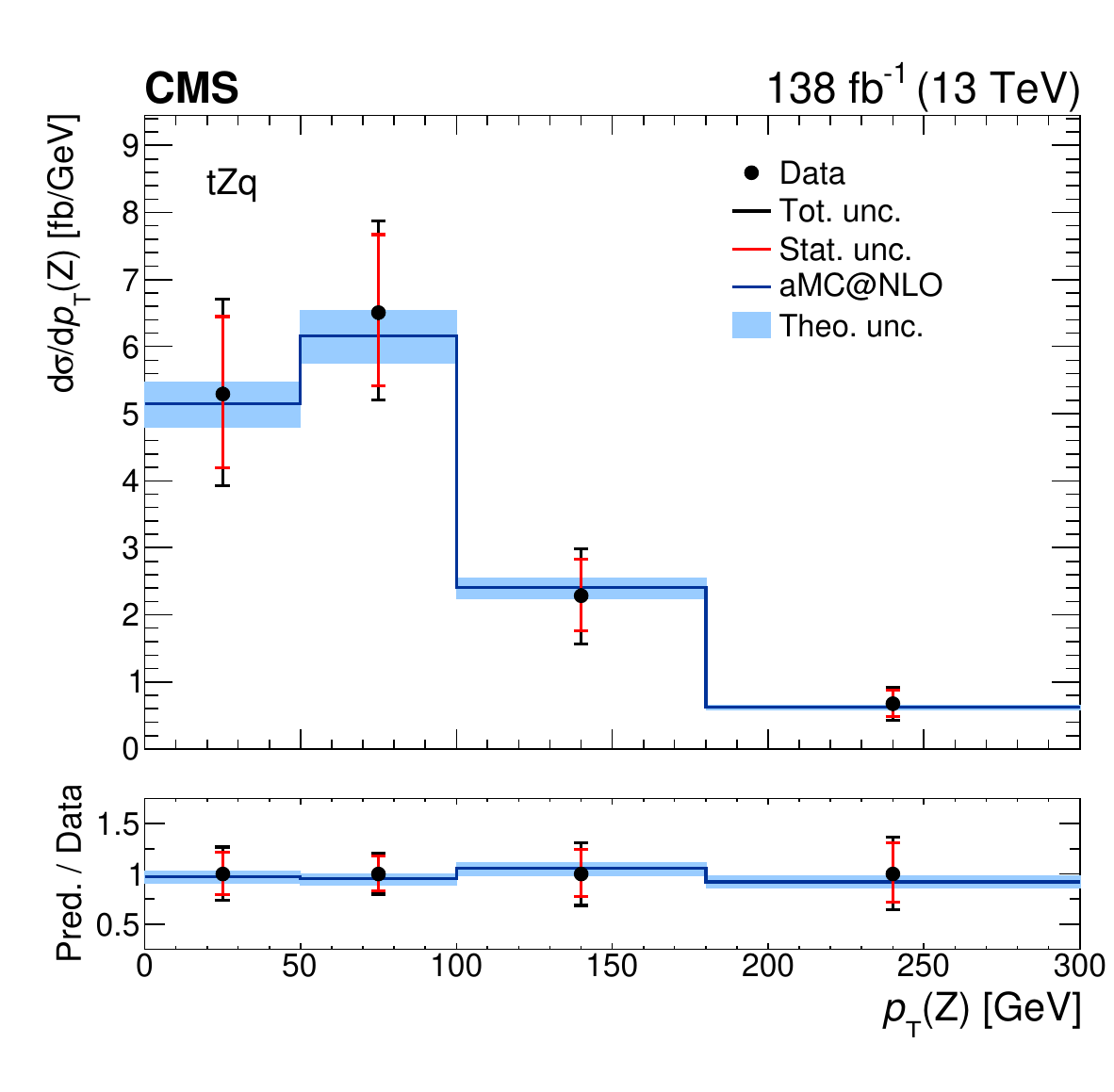} \\
\includegraphics[width=0.45\textwidth]{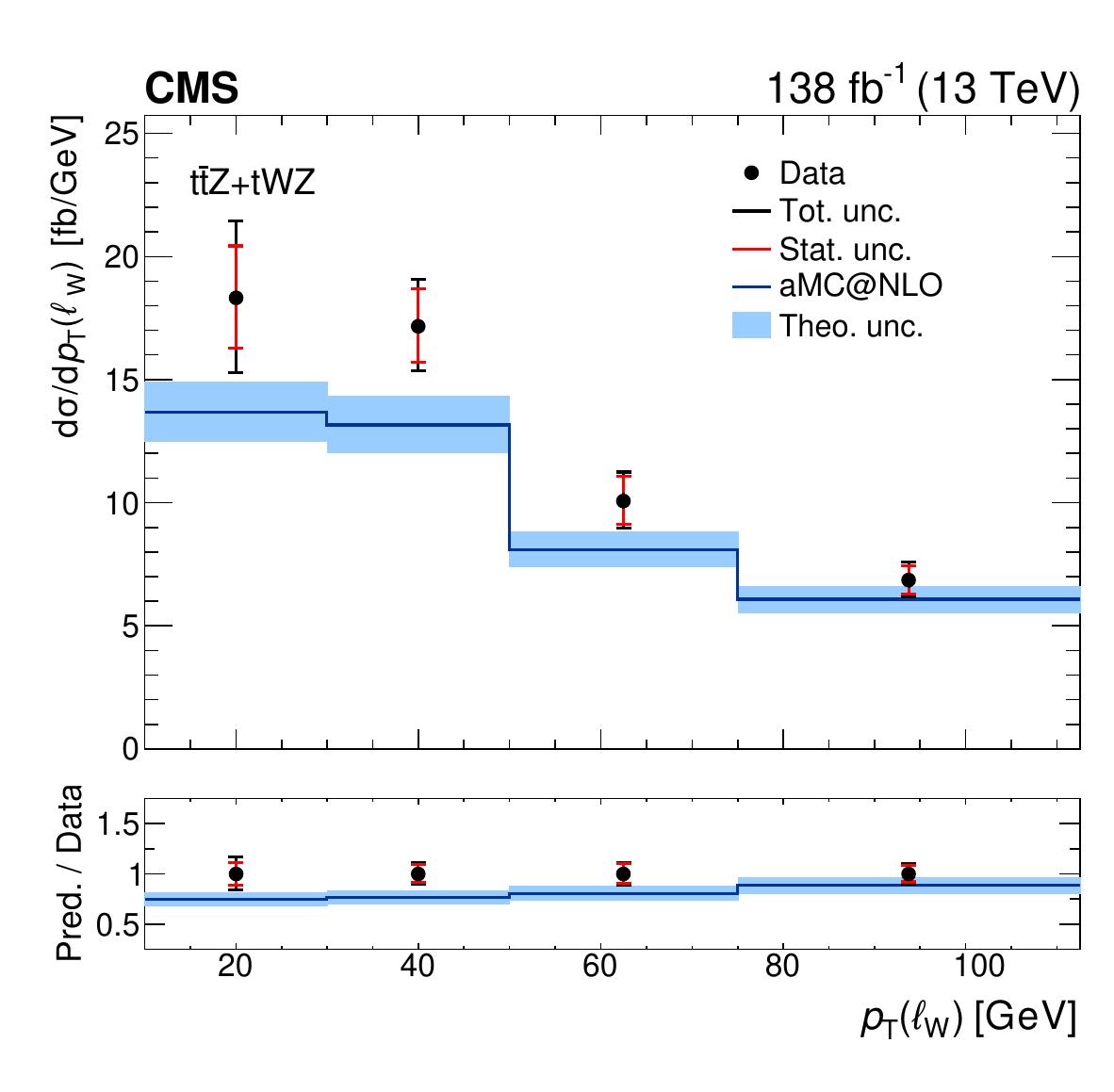}
\includegraphics[width=0.45\textwidth]{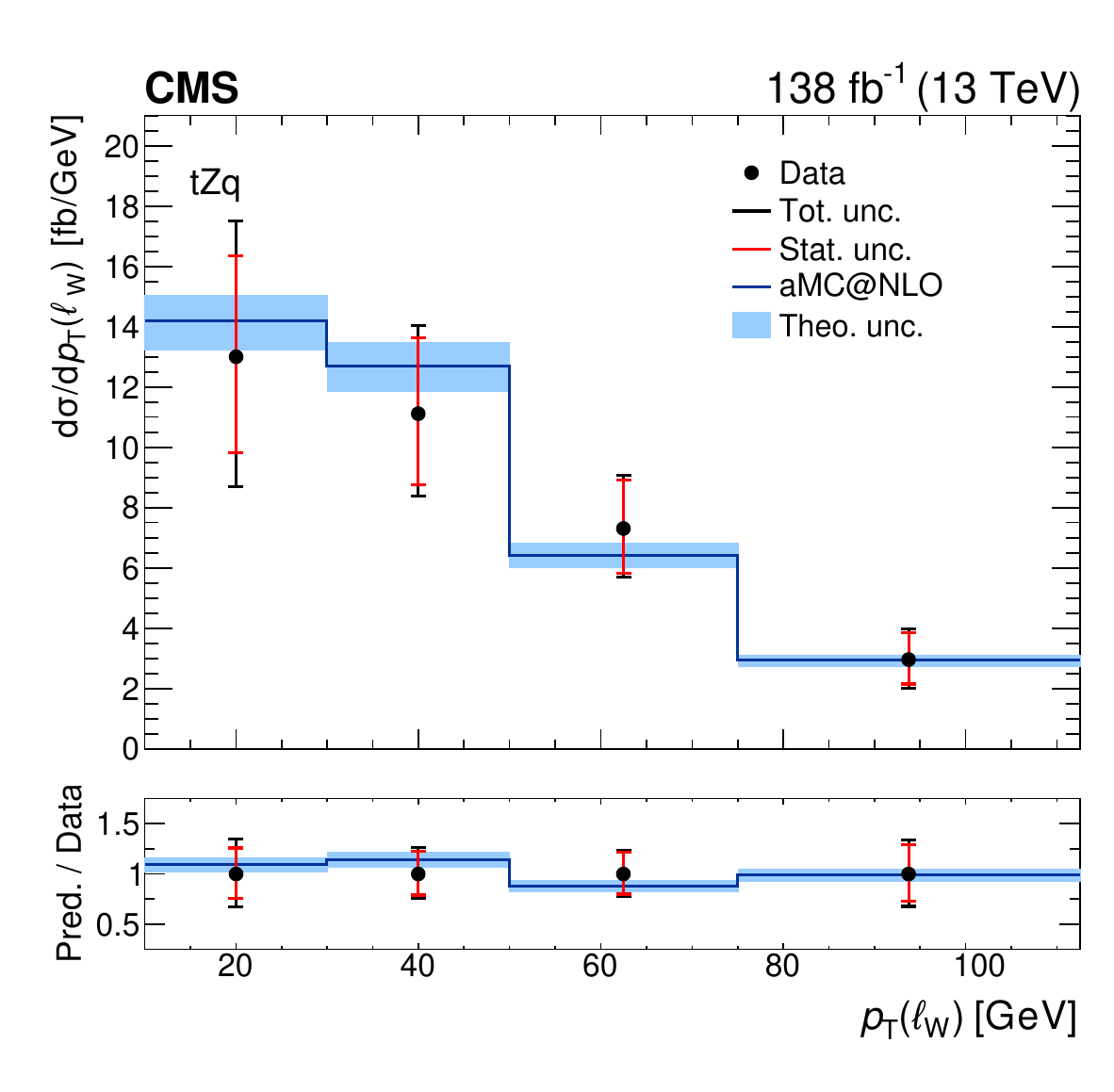} \\
\includegraphics[width=0.45\textwidth]{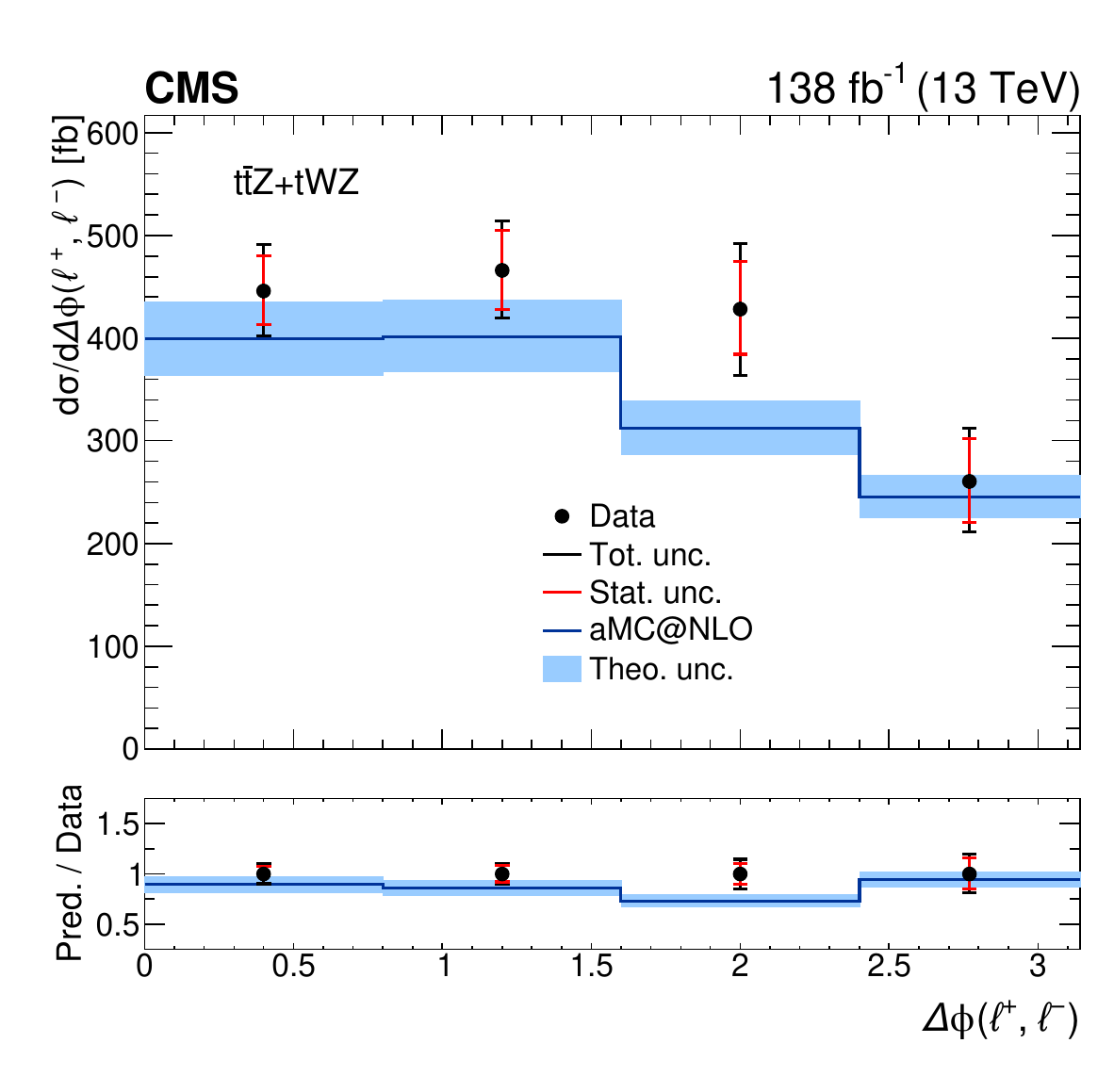}
\includegraphics[width=0.45\textwidth]{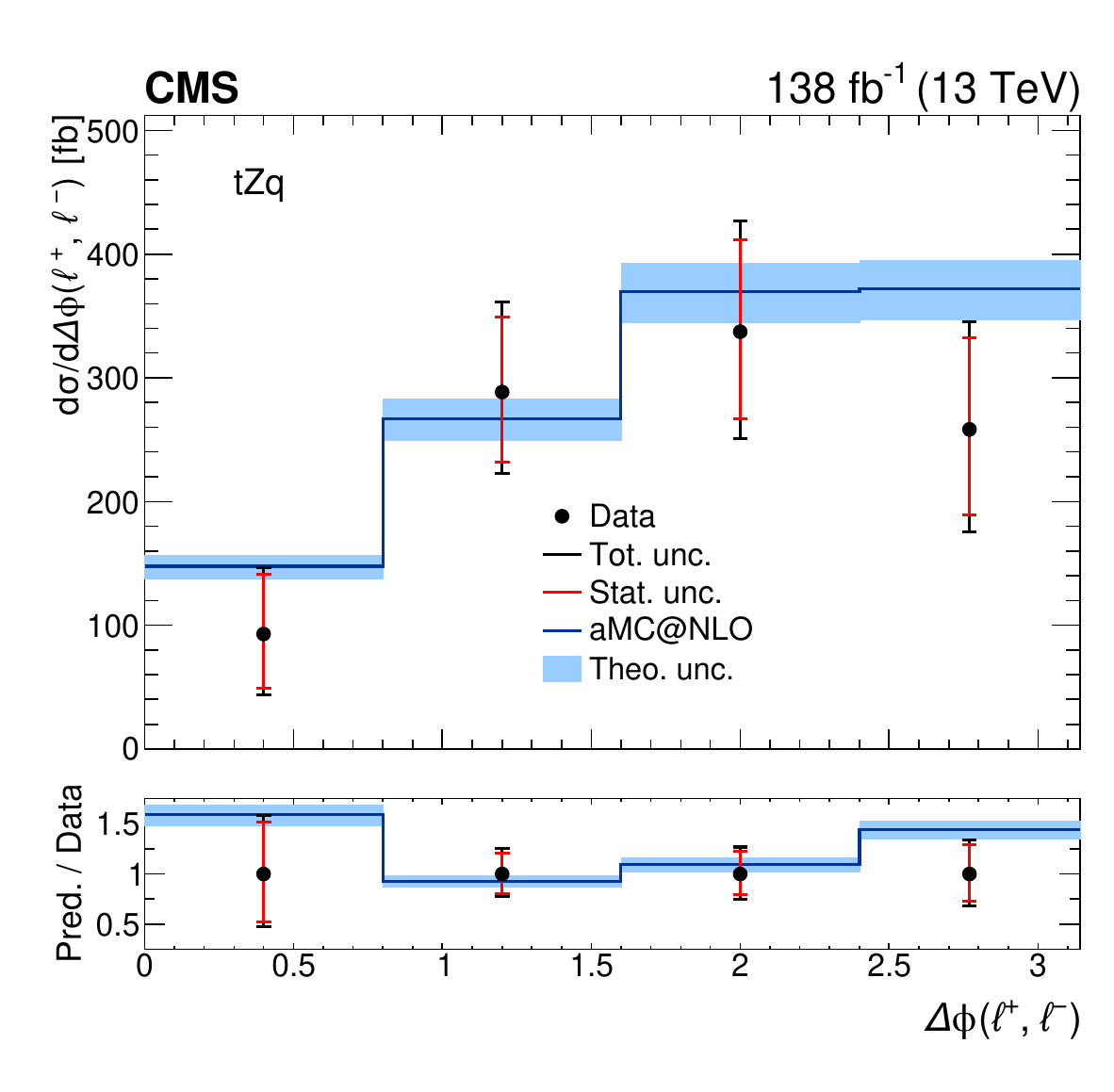}
\caption{Differential cross sections for the sum of \ttZ and \tWZ production (left column) and \tZq production (right column) as a function of \pTZ (upper),
\ptlW (middle), and \Dphill (lower). The inner (outer) error bars indicate the statistical (total) uncertainty, while the shaded area refers to the uncertainty in the theory prediction. The lower panel shows the ratio of the prediction with the data. The points at unity show the uncertainty of the data.}
\label{fig:unfolded1}
\end{figure}

\begin{figure}[!t]
\centering
\includegraphics[width=0.45\textwidth]{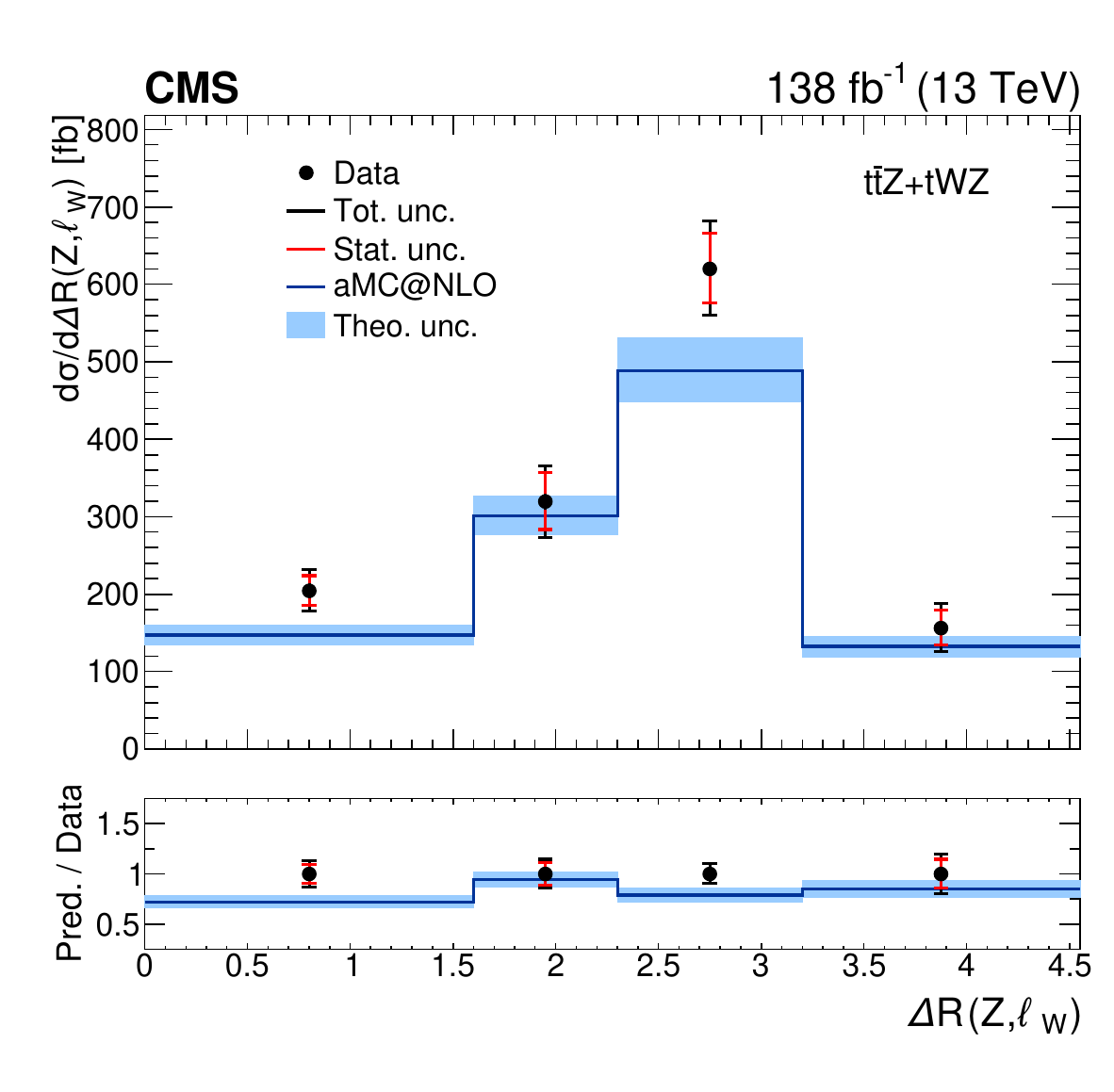}
\includegraphics[width=0.45\textwidth]{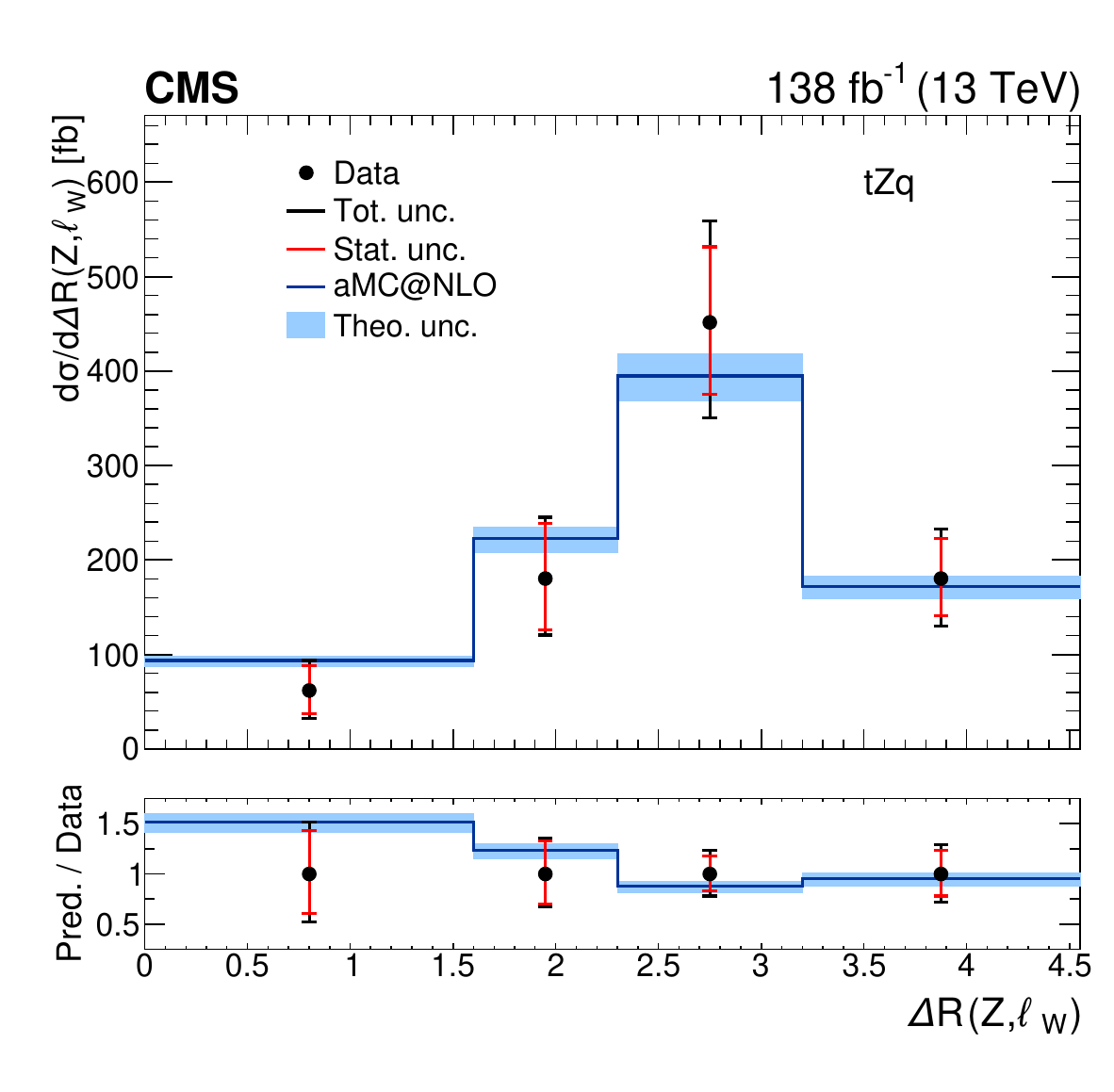}
\includegraphics[width=0.45\textwidth]{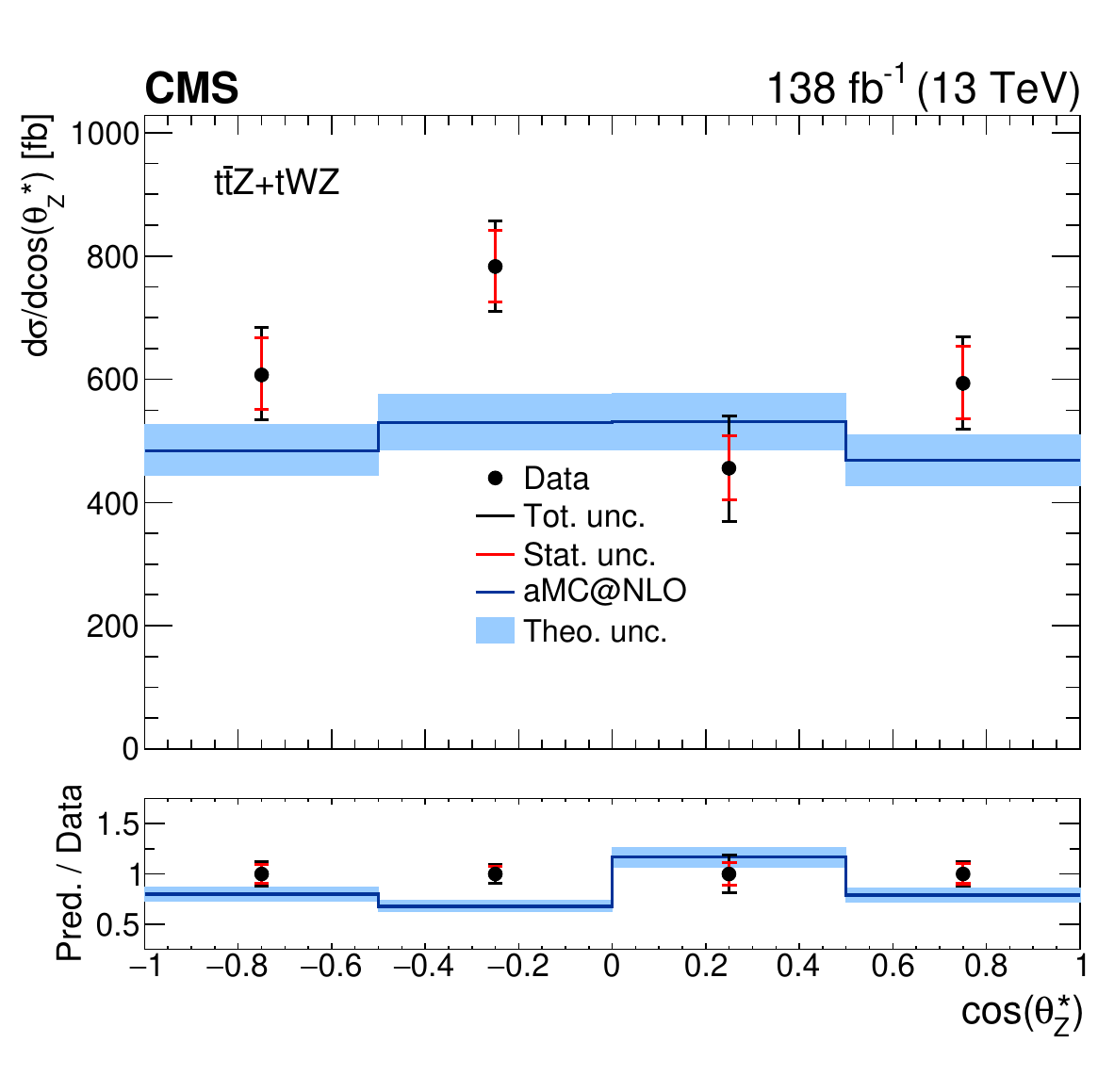}
\includegraphics[width=0.45\textwidth]{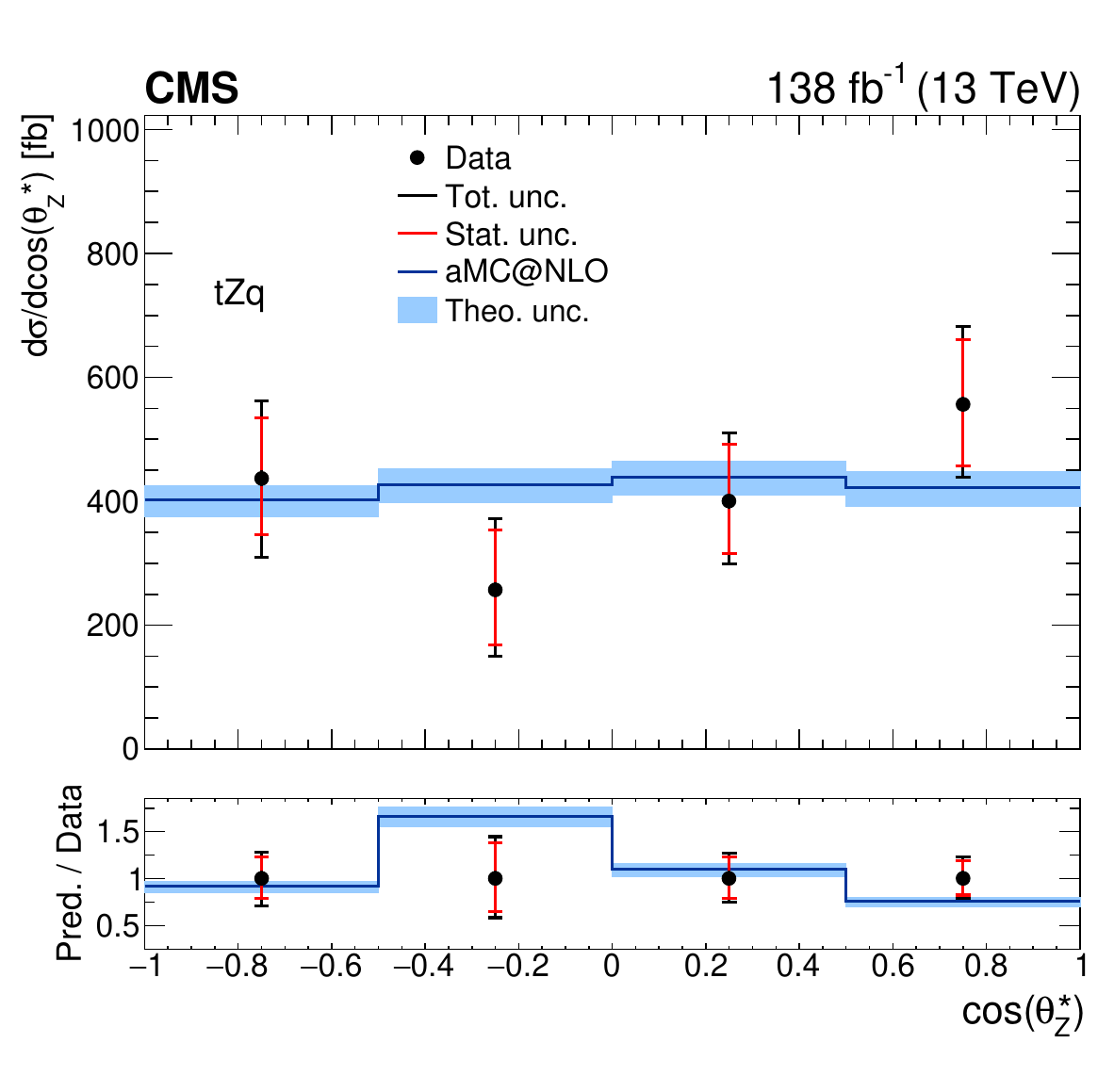}
\caption{Differential cross sections for the sum of \ttZ and \tWZ production (left column) and \tZq production (right column) as a function of \DRZlW (upper), and \cosThetaStar (lower). The inner (outer) error bars indicate the statistical (total) uncertainty, while the shaded area refers to the uncertainty in the theory prediction. The lower panel shows the ratio of the prediction with the data. The points at unity show the uncertainty of the data.}
\label{fig:unfolded2}
\end{figure}

\begin{figure}[!p]
\centering
\includegraphics[width=0.45\textwidth]{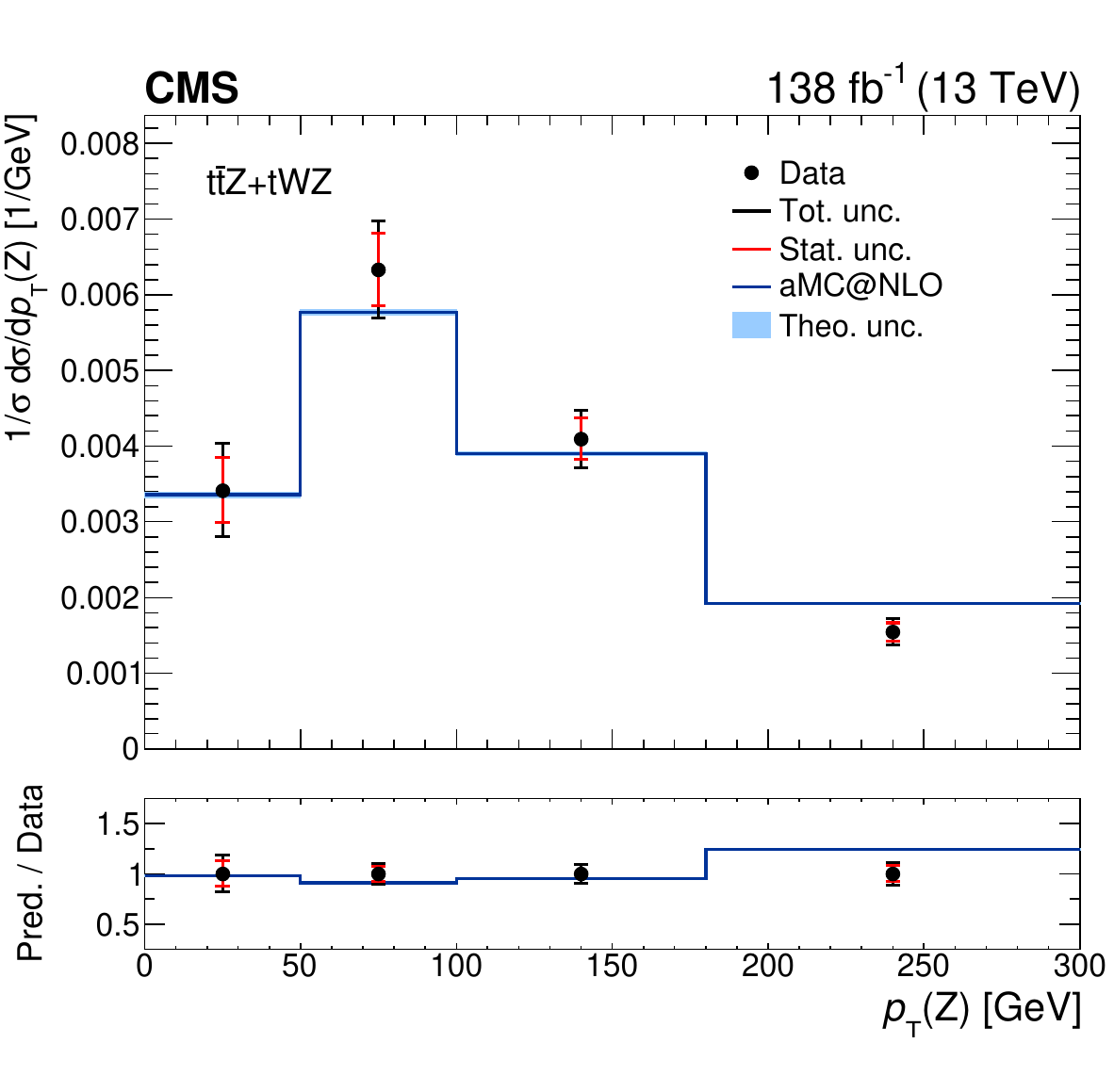}
\includegraphics[width=0.45\textwidth]{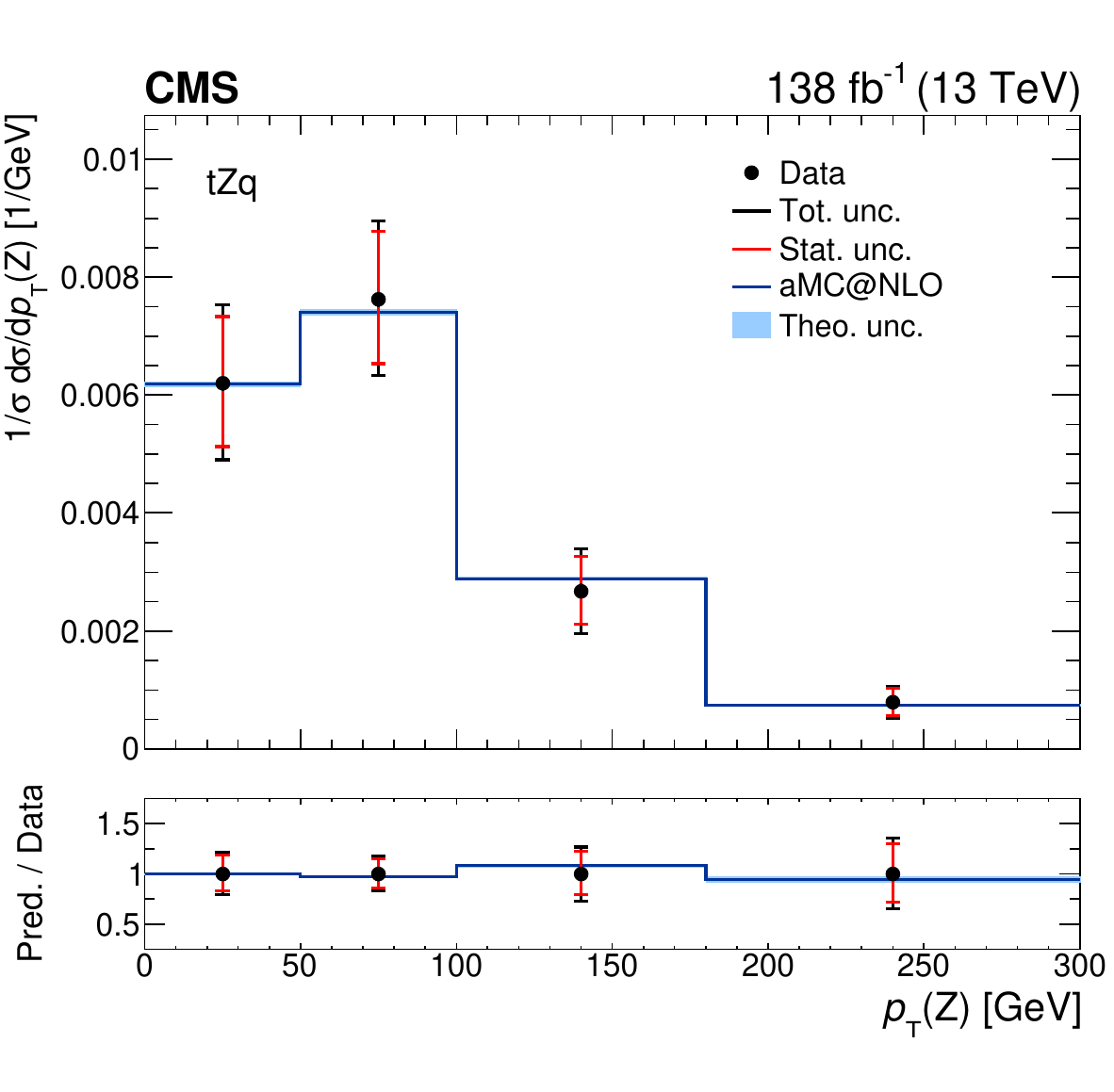} \\
\includegraphics[width=0.45\textwidth]{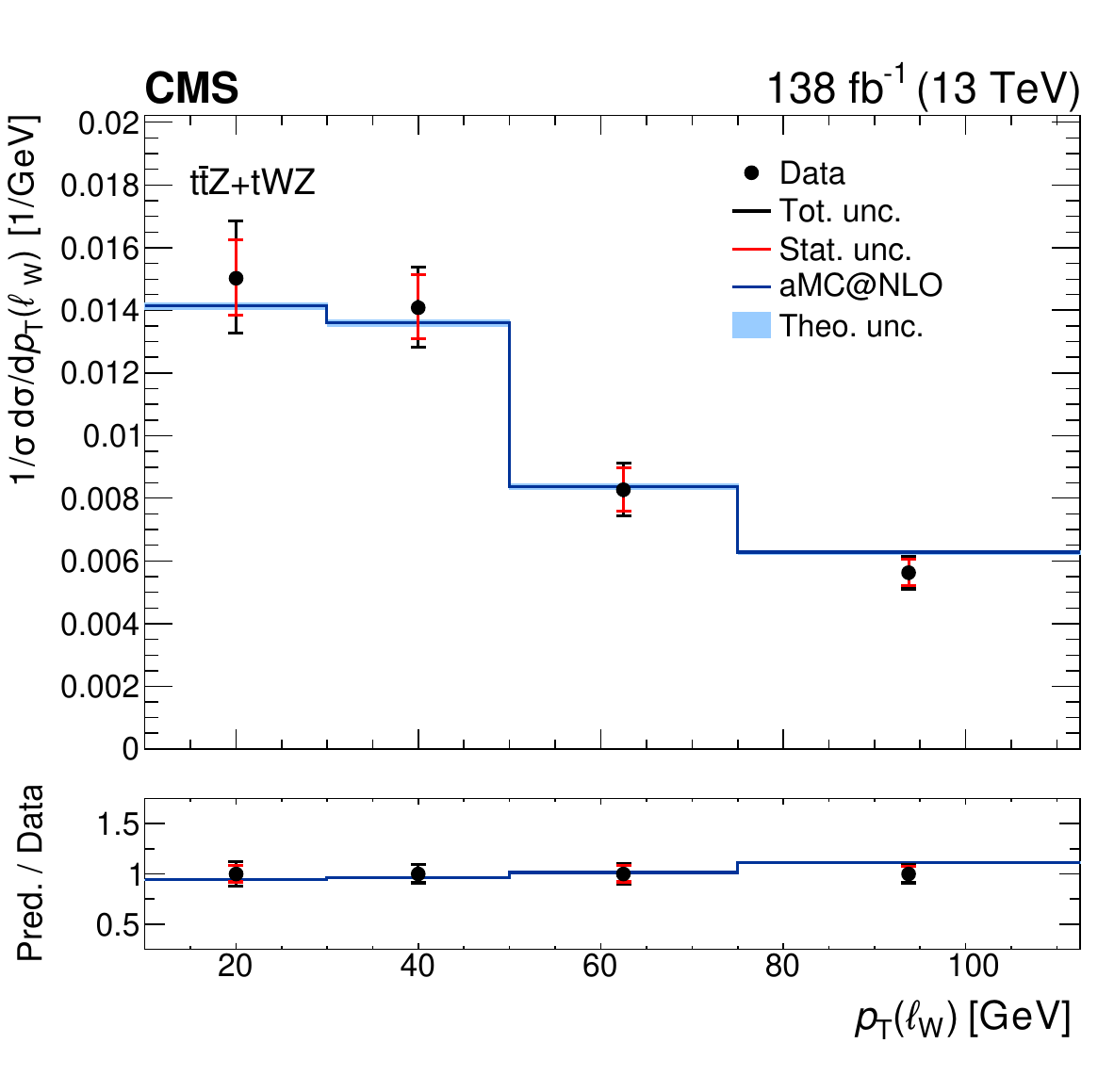}
\includegraphics[width=0.45\textwidth]{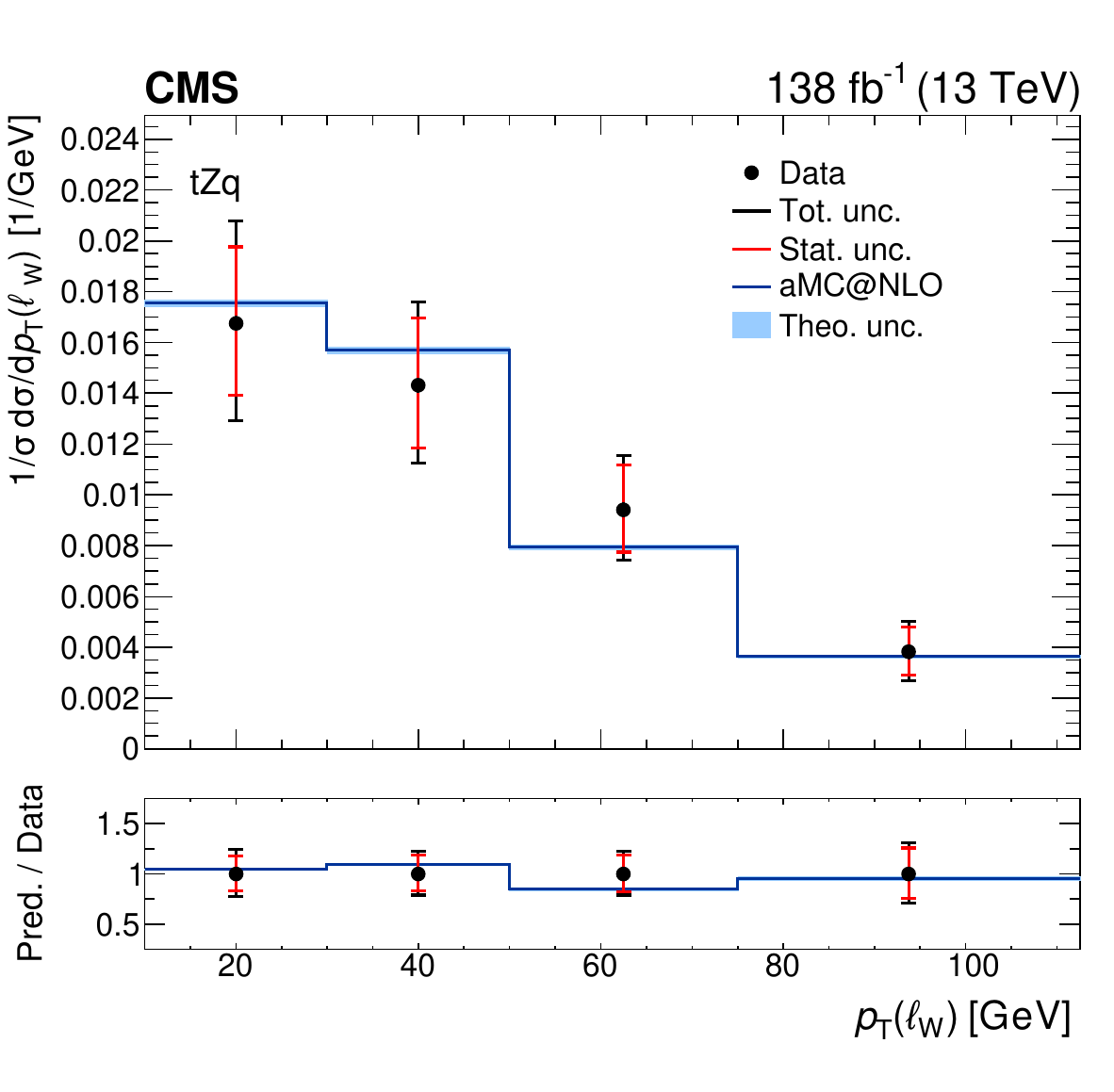} \\
\includegraphics[width=0.45\textwidth]{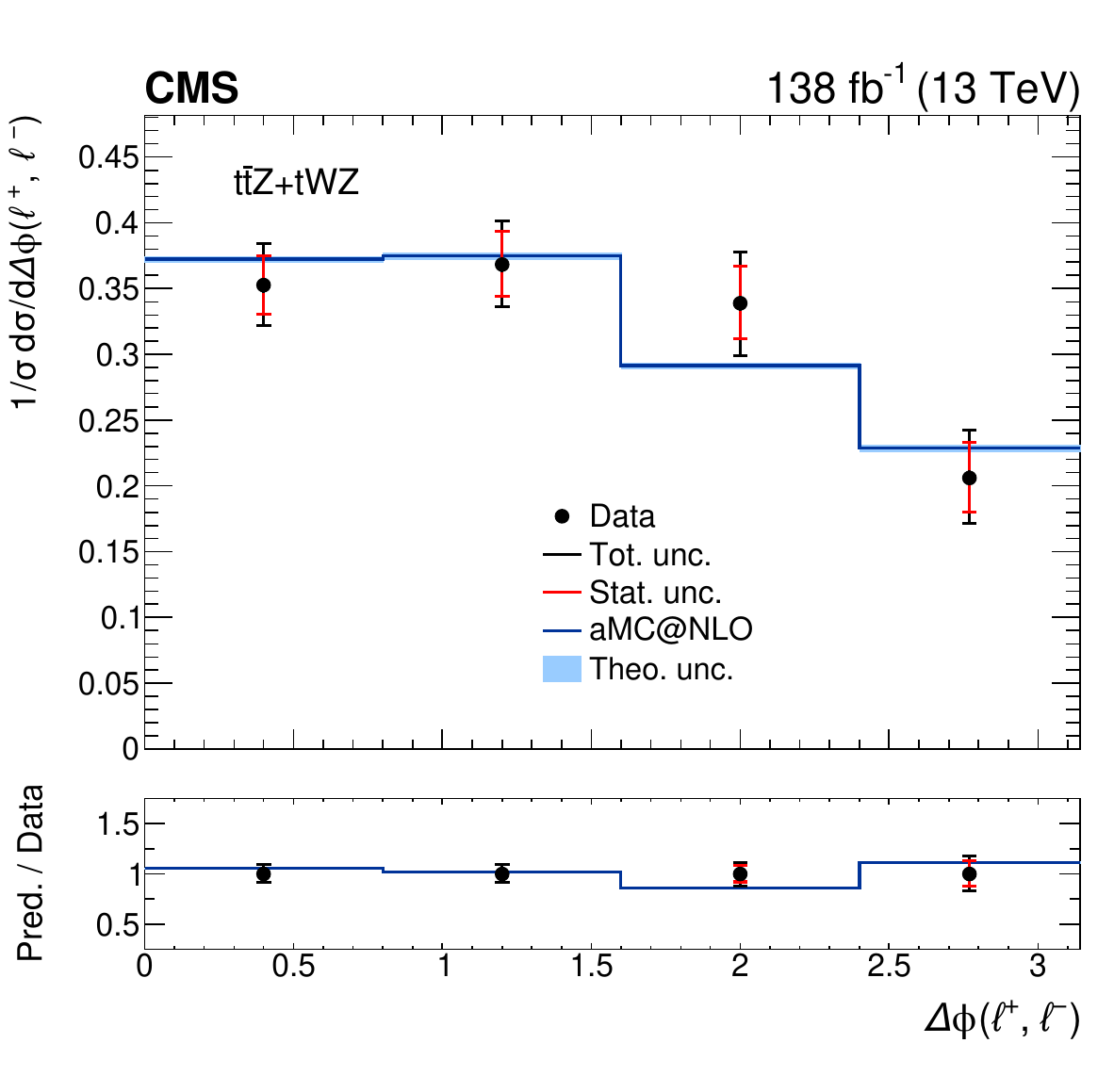}
\includegraphics[width=0.45\textwidth]{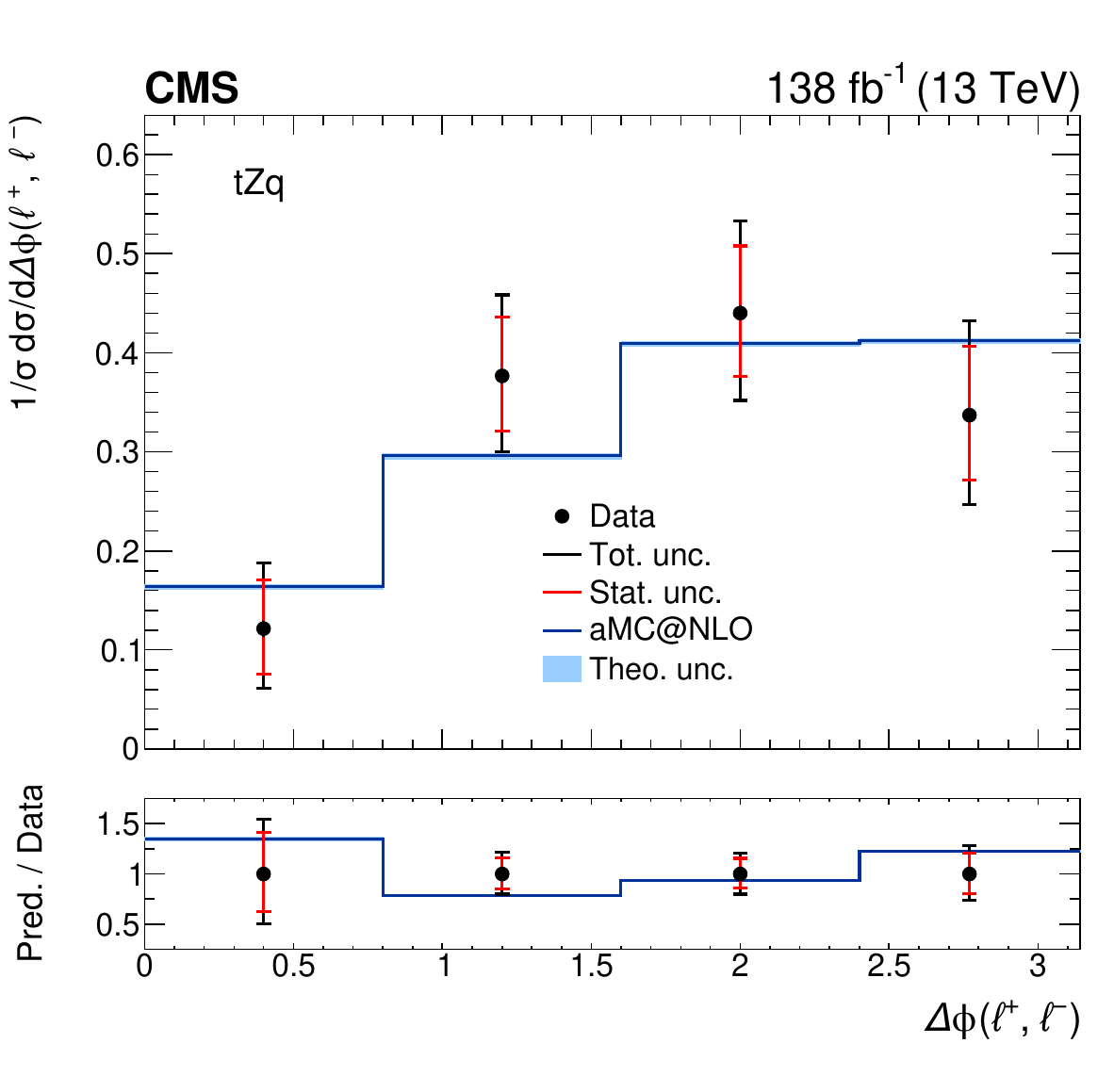}
\caption{Normalized differential cross sections for the sum of \ttZ and \tWZ production (left column) and \tZq production (right column) as a function of \pTZ (upper),  \ptlW (middle), and \Dphill (lower). The inner (outer) error bars indicate the statistical (total) uncertainty, while the shaded area refers to the uncertainty in the theory prediction. The lower panel shows the ratio of the prediction with the data. The points at unity show the uncertainty of the data.}
\label{fig:unfolded1_normalized}
\end{figure}

\begin{figure}[!p]
\centering
\includegraphics[width=0.45\textwidth]{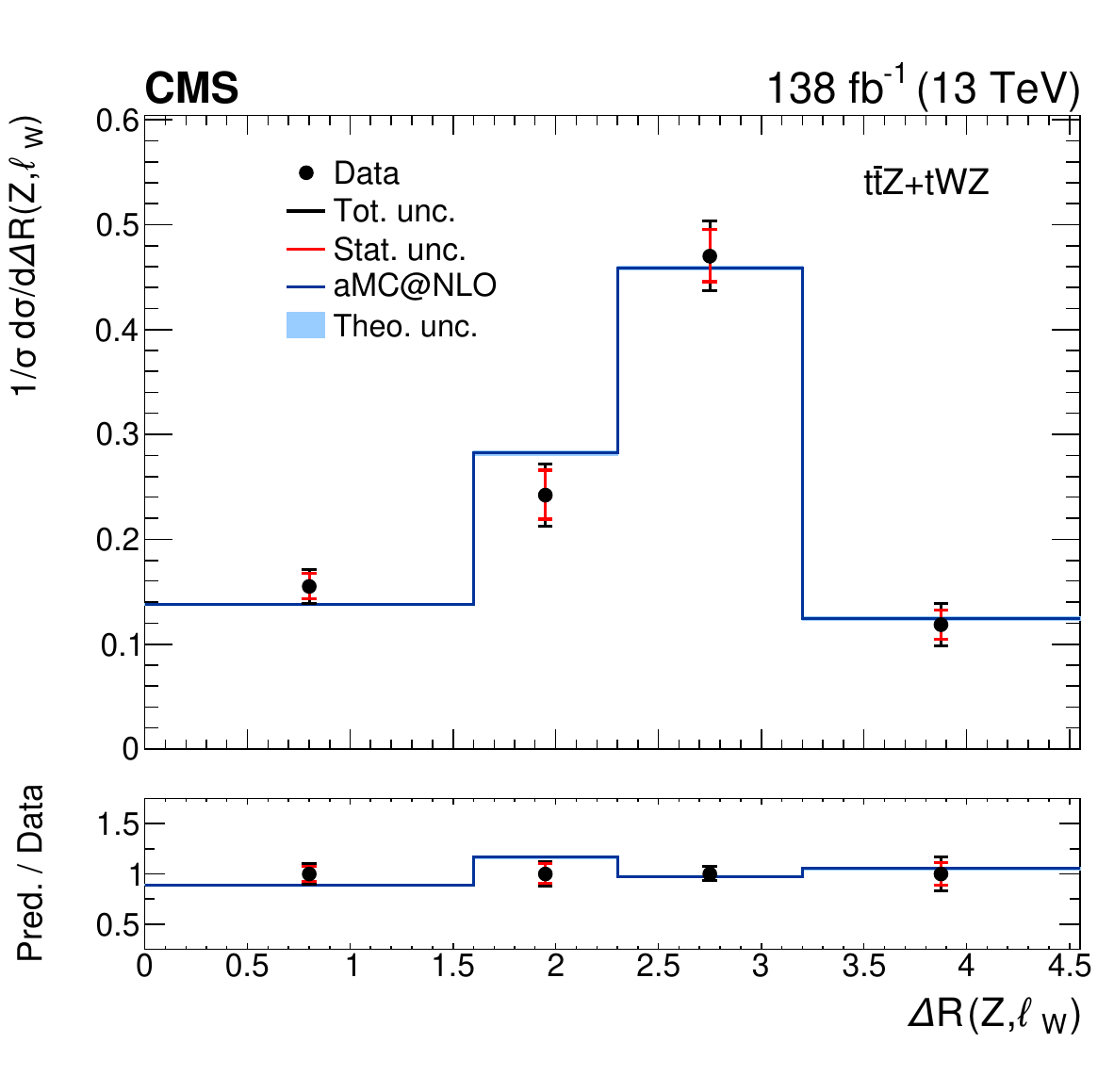}
\includegraphics[width=0.45\textwidth]{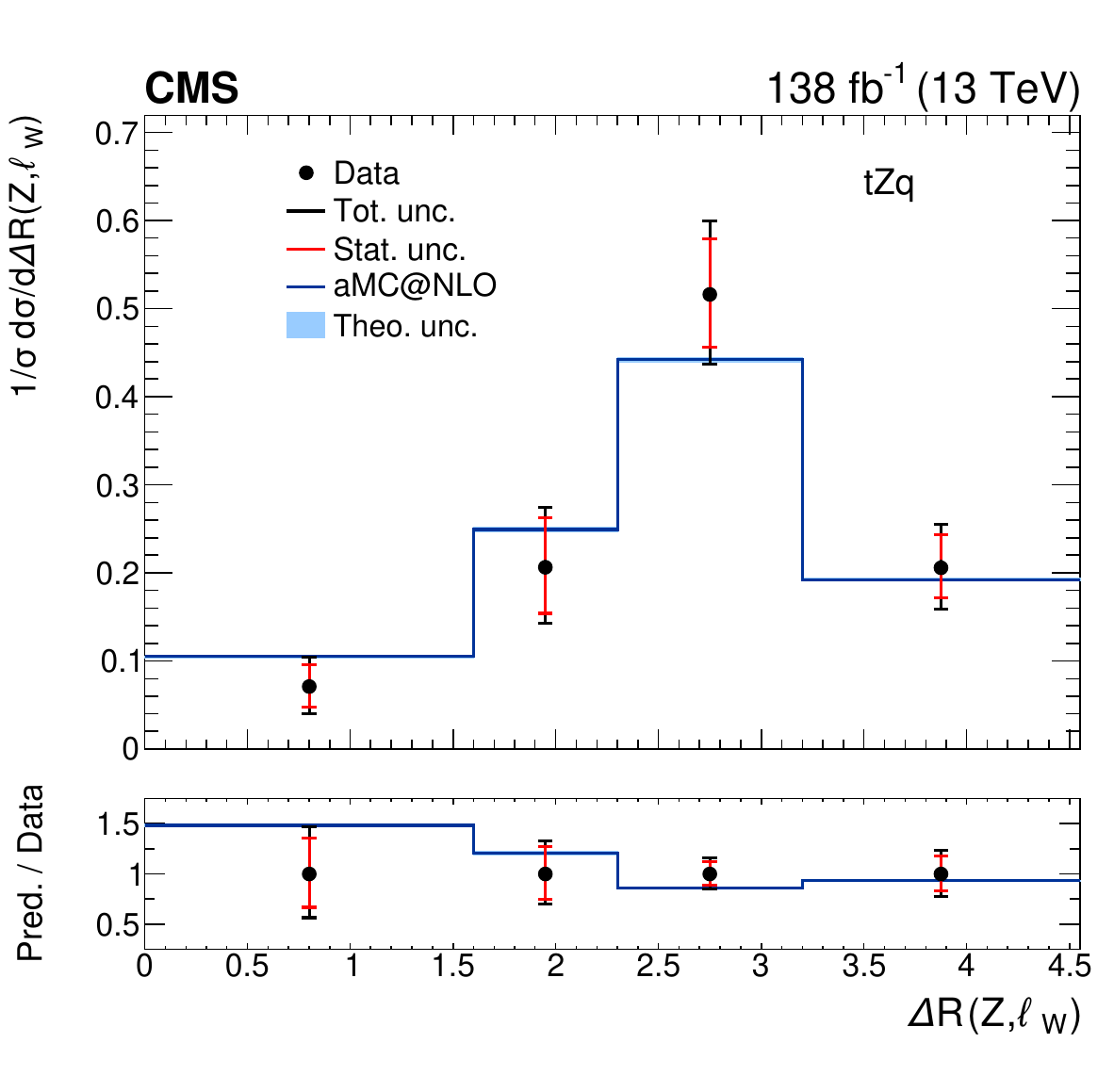} \\
\includegraphics[width=0.45\textwidth]{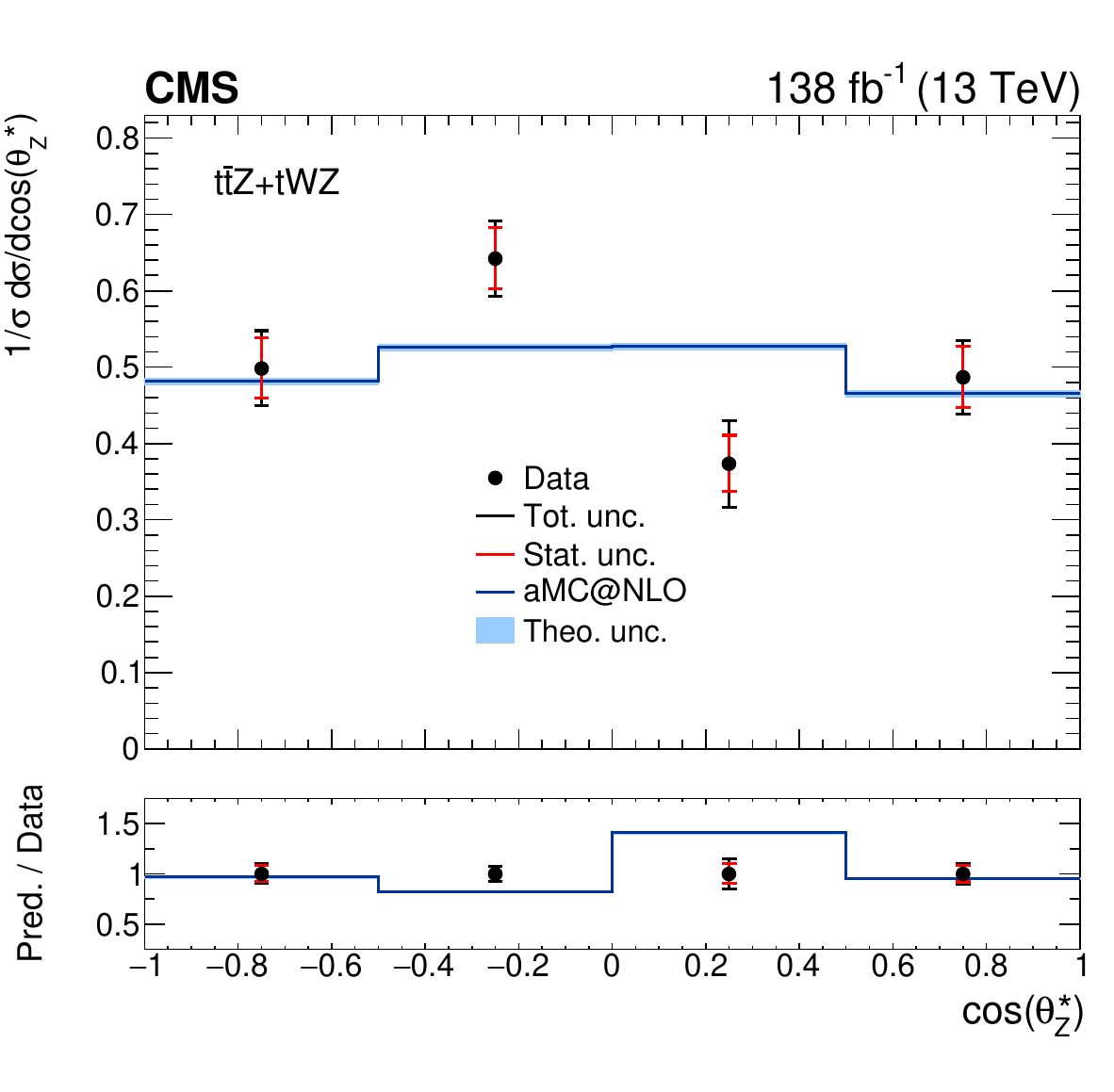}
\includegraphics[width=0.45\textwidth]{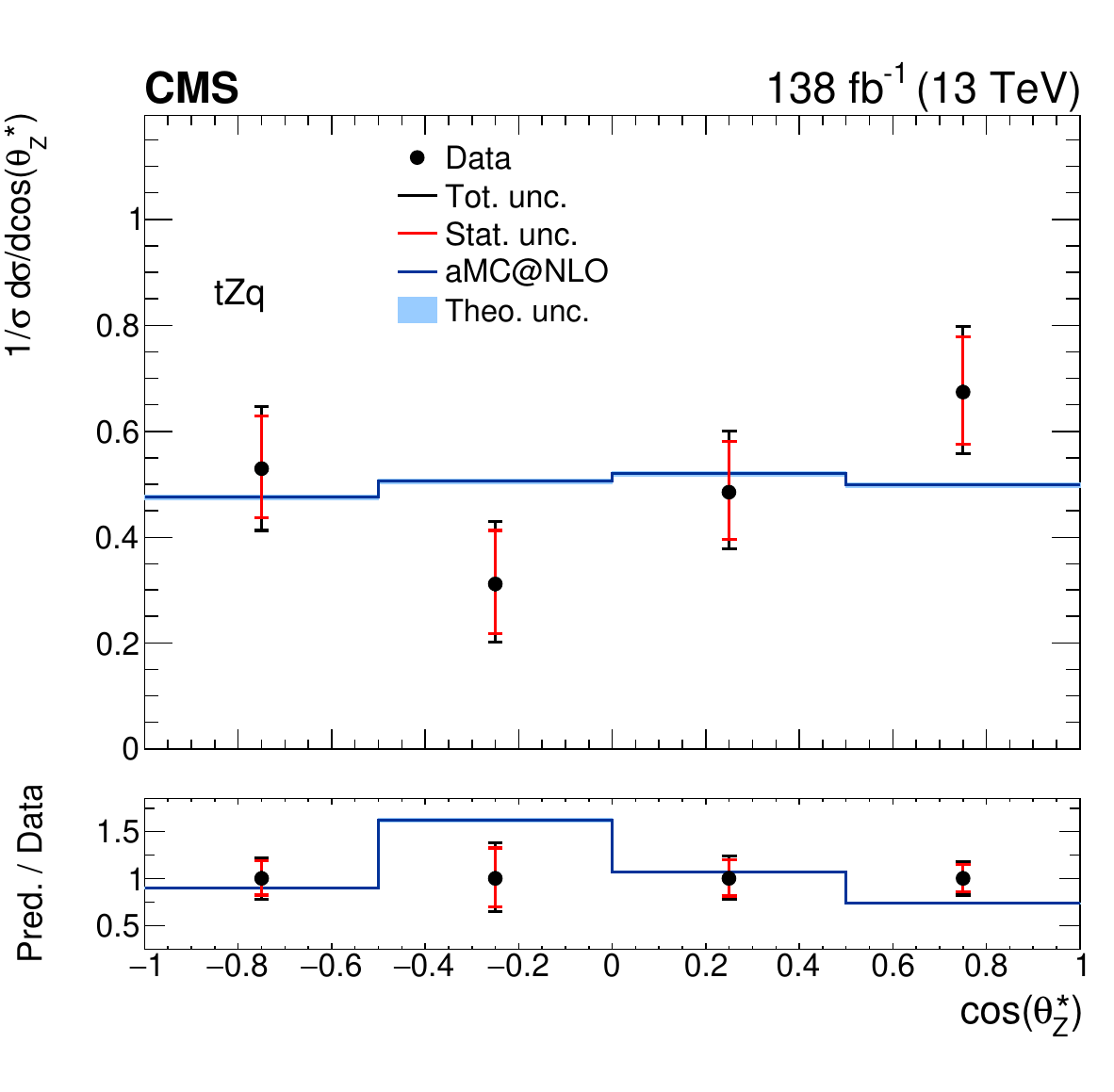}
\caption{Normalized differential cross sections for the sum of \ttZ and \tWZ production (left column) and \tZq production (right column) as a function of \DRZlW (upper), and \cosThetaStar (lower). The inner (outer) error bars indicate the statistical (total) uncertainty, while the shaded area refers to the uncertainty in the theory prediction. The lower panel shows the ratio of the prediction with the data. The points at unity show the uncertainty of the data.}
\label{fig:unfolded2_normalized}
\end{figure}

\begin{figure}[!p]
\centering
\includegraphics[width=0.45\textwidth]{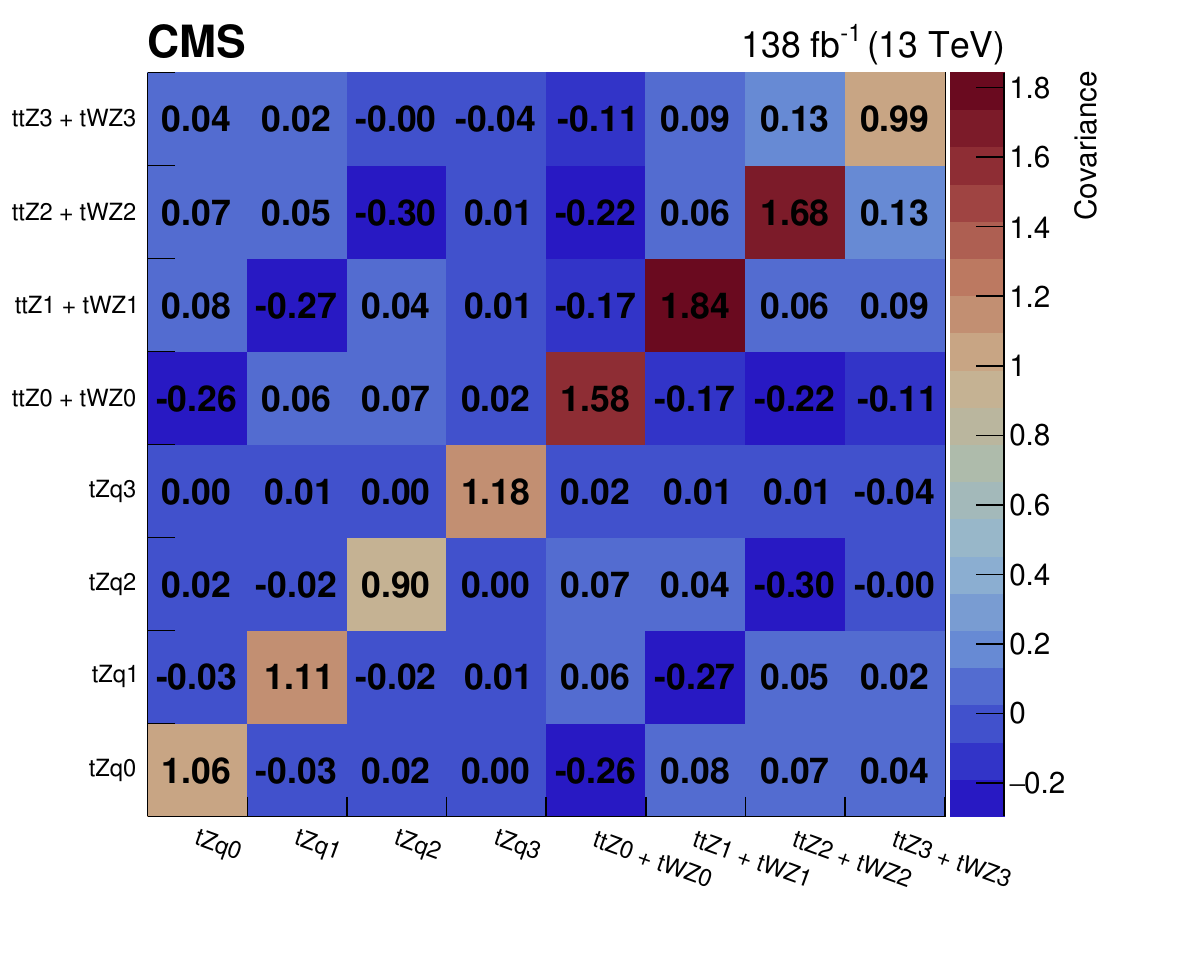}
\includegraphics[width=0.45\textwidth]{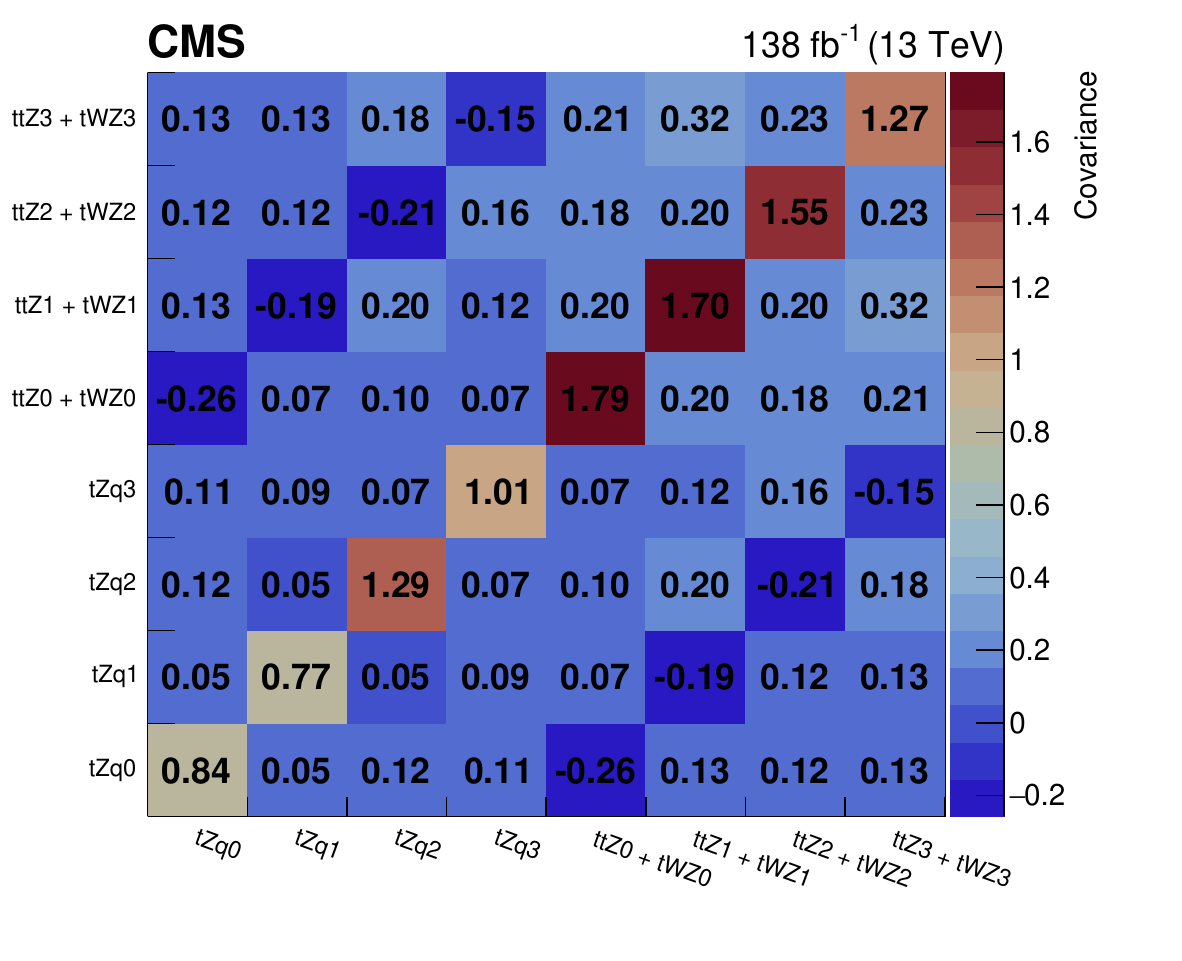} \\
\includegraphics[width=0.45\textwidth]{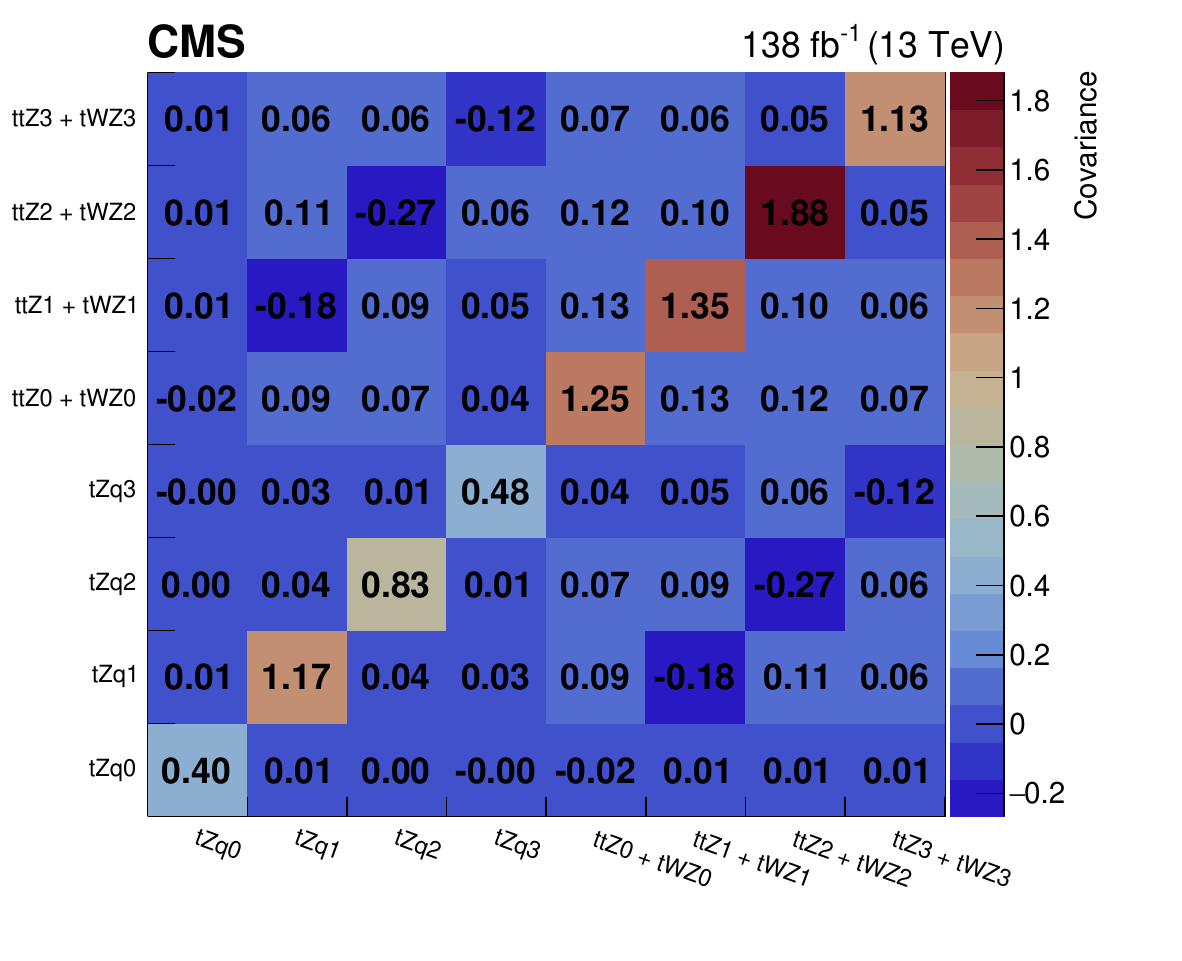}
\includegraphics[width=0.45\textwidth]{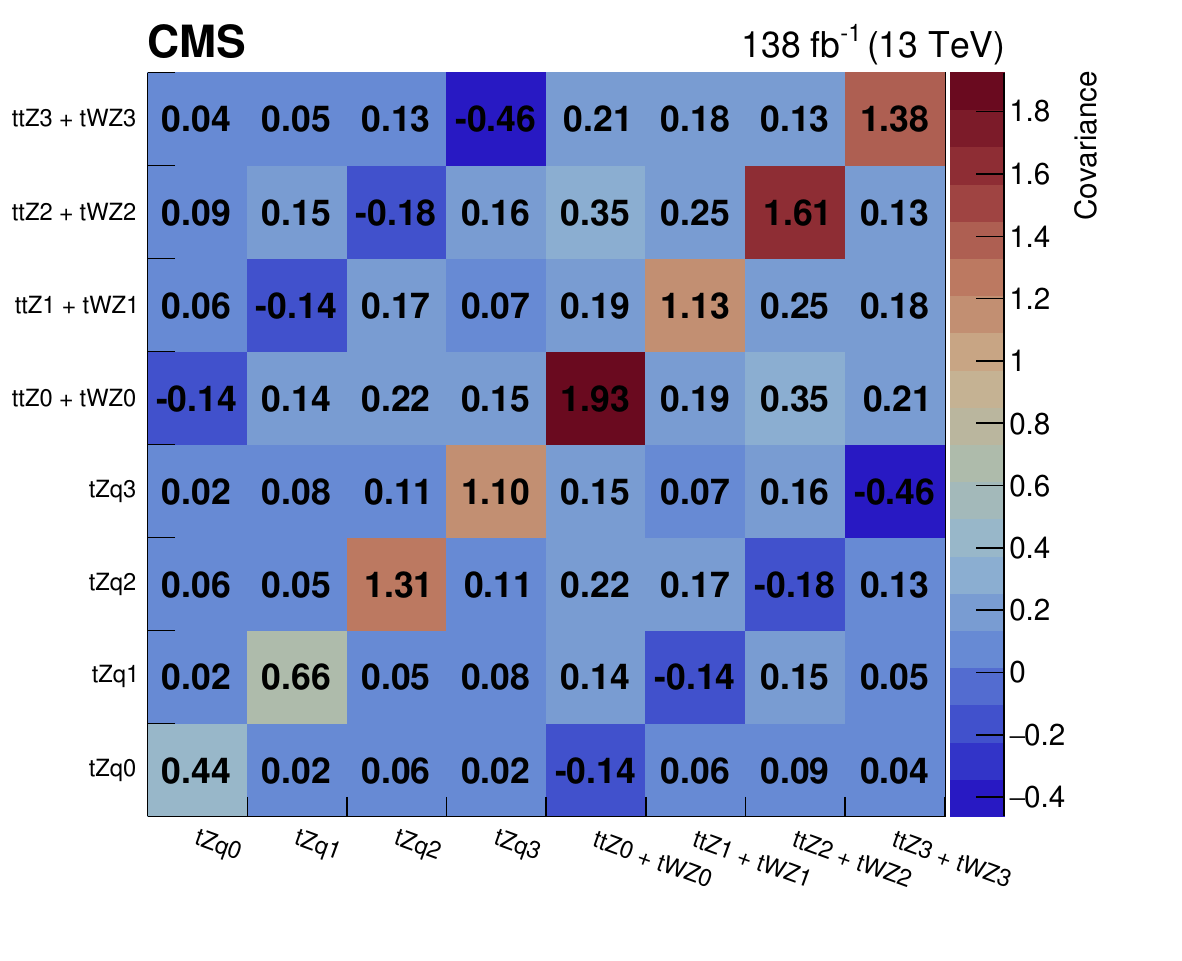} \\
\includegraphics[width=0.45\textwidth]{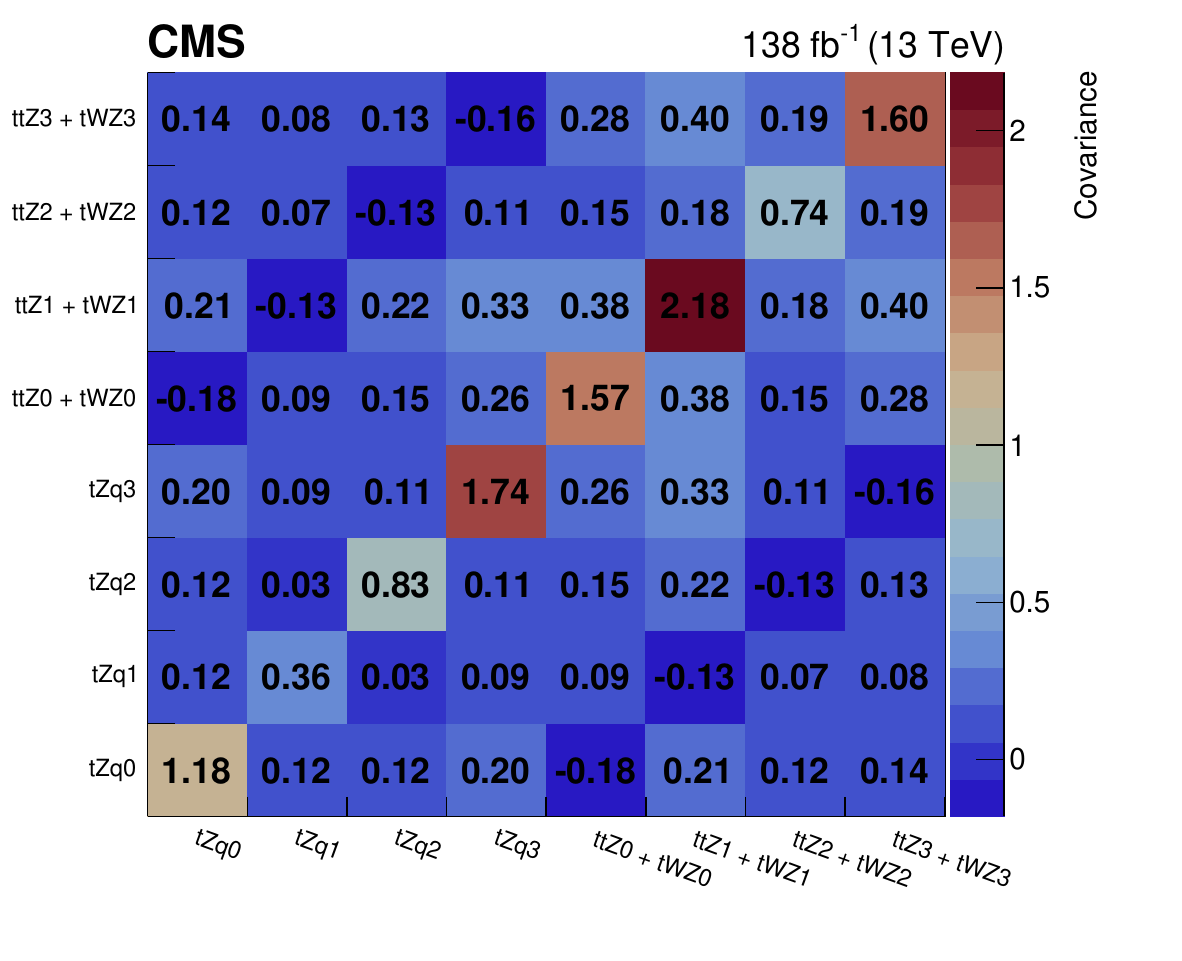}
\caption{Covariance matrices for the simultaneous measurement of the differential cross section as a function of \pTZ, (upper left), \ptlW (upper right), \DRZlW (middle left), \Dphill (middle right), and \cosThetaStar (lower). The last digits in the axis labels refer to the respective bin of the corresponding differential cross section measurement.}
\label{fig:covMat}
\end{figure}

The differential cross sections for the sum of \ttZ and \tWZ, and for \tZq are shown in Figs.~\ref{fig:unfolded1} and \ref{fig:unfolded2}. These figures also include the predictions as obtained from the MC simulation, together with their uncertainties. The uncertainties in the predictions are estimated by summing in quadrature the systematic uncertainties coming from the renormalization and factorization scales, PDFs, and parton showers. For the sum of the \ttZ and \tWZ cross sections, the uncertainties related to the overlap removal are also included.
The predicted cross sections refer to the same phase space as the inclusive measurement. To simplify shape comparisons and to remove bin-to-bin correlated uncertainties, Figs.~\ref{fig:unfolded1_normalized} and \ref{fig:unfolded2_normalized} show the normalized cross sections.
Covariance matrices for the differential cross sections are provided in Fig.~\ref{fig:covMat}.

As for the inclusive results, the \tZq differential cross sections are in good agreement with the theory prediction.
For the sum of the \ttZ and \tWZ cross sections, an excess over expectation is observed (left columns of Figs.~\ref{fig:unfolded1} and \ref{fig:unfolded2}).
A possible trend is observed as a function of \ptlW, as presented in Figs.~\ref{fig:unfolded1} and \ref{fig:unfolded1_normalized} (middle left), which leads to a discrepancy in the region of low \ptlW. This trend is reminiscent of similar trends in inclusive \ttbar production between MC simulations at NLO and data~\cite{CMS:2019too, CMS:2023krq}.
For \ttbar production, it has been shown that calculations at next-to-next-to-leading order give a significantly improved description of the data~\cite{CMS:2024ybg,CMS:2018adi,CMS:2018htd}.

\section{Summary}
\label{sec:summary}

A first simultaneous measurement of single and pair production of top quarks in association with a \PZ boson is presented. The data were recorded by the CMS experiment in proton-proton collisions at a center-of-mass energy of 13\TeV, corresponding to an integrated luminosity of 138\fbinv.
Events with three or more leptons (electrons or muons) are analysed. The separation between the signals is achieved using a deep neural network classifier with three output nodes for the combined \ttZ and \tWZ processes, the \tZq process, and the backgrounds.
The inclusive cross sections are measured to be $\sttZtWZ=1.14\pm0.07\unit{pb}$ for the sum of the \tWZ and \ttZ processes, and $\stZq=0.81\pm0.10\unit{pb}$ for \tZq production. Both results are evaluated for a dilepton invariant mass between 70 and 110\GeV. The cross sections, and their full correlation matrices, are measured differentially as functions of several observables that are expected to be sensitive to effects arising from QCD modeling and beyond the SM contributions, making them also useful for EFT interpretations.
Good agreement with theoretical predictions is found for the \tZq process, while for \ttZplustWZ production, the slope in the transverse momentum distribution of the lepton that originates from the \PW boson is found to be somewhat steeper than predicted.

\begin{acknowledgments}
We congratulate our colleagues in the CERN accelerator departments for the excellent performance of the LHC and thank the technical and administrative staffs at CERN and at other CMS institutes for their contributions to the success of the CMS effort. In addition, we gratefully acknowledge the computing centers and personnel of the Worldwide LHC Computing Grid and other centers for delivering so effectively the computing infrastructure essential to our analyses. Finally, we acknowledge the enduring support for the construction and operation of the LHC, the CMS detector, and the supporting computing infrastructure provided by the following funding agencies: SC (Armenia), BMBWF and FWF (Austria); FNRS and FWO (Belgium); CNPq, CAPES, FAPERJ, FAPERGS, and FAPESP (Brazil); MES and BNSF (Bulgaria); CERN; CAS, MoST, and NSFC (China); MINCIENCIAS (Colombia); MSES and CSF (Croatia); RIF (Cyprus); SENESCYT (Ecuador); ERC PRG, RVTT3 and MoER TK202 (Estonia); Academy of Finland, MEC, and HIP (Finland); CEA and CNRS/IN2P3 (France); SRNSF (Georgia); BMBF, DFG, and HGF (Germany); GSRI (Greece); NKFIH (Hungary); DAE and DST (India); IPM (Iran); SFI (Ireland); INFN (Italy); MSIP and NRF (Republic of Korea); MES (Latvia); LMTLT (Lithuania); MOE and UM (Malaysia); BUAP, CINVESTAV, CONACYT, LNS, SEP, and UASLP-FAI (Mexico); MOS (Montenegro); MBIE (New Zealand); PAEC (Pakistan); MES and NSC (Poland); FCT (Portugal); MESTD (Serbia); MCIN/AEI and PCTI (Spain); MOSTR (Sri Lanka); Swiss Funding Agencies (Switzerland); MST (Taipei); MHESI and NSTDA (Thailand); TUBITAK and TENMAK (Turkey); NASU (Ukraine); STFC (United Kingdom); DOE and NSF (USA).

\hyphenation{Rachada-pisek} Individuals have received support from the Marie-Curie program and the European Research Council and Horizon 2020 Grant, contract Nos.\ 675440, 724704, 752730, 758316, 765710, 824093, 101115353, 101002207, and COST Action CA16108 (European Union); the Leventis Foundation; the Alfred P.\ Sloan Foundation; the Alexander von Humboldt Foundation; the Science Committee, project no. 22rl-037 (Armenia); the Belgian Federal Science Policy Office; the Fonds pour la Formation \`a la Recherche dans l'Industrie et dans l'Agriculture (FRIA-Belgium); the F.R.S.-FNRS and FWO (Belgium) under the ``Excellence of Science -- EOS" -- be.h project n.\ 30820817; the Beijing Municipal Science \& Technology Commission, No. Z191100007219010 and Fundamental Research Funds for the Central Universities (China); the Ministry of Education, Youth and Sports (MEYS) of the Czech Republic; the Shota Rustaveli National Science Foundation, grant FR-22-985 (Georgia); the Deutsche Forschungsgemeinschaft (DFG), among others, under Germany's Excellence Strategy -- EXC 2121 ``Quantum Universe" -- 390833306, and under project number 400140256 - GRK2497; the Hellenic Foundation for Research and Innovation (HFRI), Project Number 2288 (Greece); the Hungarian Academy of Sciences, the New National Excellence Program - \'UNKP, the NKFIH research grants K 131991, K 133046, K 138136, K 143460, K 143477, K 146913, K 146914, K 147048, 2020-2.2.1-ED-2021-00181, and TKP2021-NKTA-64 (Hungary); the Council of Science and Industrial Research, India; ICSC -- National Research Center for High Performance Computing, Big Data and Quantum Computing and FAIR -- Future Artificial Intelligence Research, funded by the NextGenerationEU program (Italy); the Latvian Council of Science; the Ministry of Education and Science, project no. 2022/WK/14, and the National Science Center, contracts Opus 2021/41/B/ST2/01369 and 2021/43/B/ST2/01552 (Poland); the Funda\c{c}\~ao para a Ci\^encia e a Tecnologia, grant CEECIND/01334/2018 (Portugal); the National Priorities Research Program by Qatar National Research Fund; MCIN/AEI/10.13039/501100011033, ERDF ``a way of making Europe", and the Programa Estatal de Fomento de la Investigaci{\'o}n Cient{\'i}fica y T{\'e}cnica de Excelencia Mar\'{\i}a de Maeztu, grant MDM-2017-0765 and Programa Severo Ochoa del Principado de Asturias (Spain); the Chulalongkorn Academic into Its 2nd Century Project Advancement Project, and the National Science, Research and Innovation Fund via the Program Management Unit for Human Resources \& Institutional Development, Research and Innovation, grant B39G670016 (Thailand); the Kavli Foundation; the Nvidia Corporation; the SuperMicro Corporation; the Welch Foundation, contract C-1845; and the Weston Havens Foundation (USA).
\end{acknowledgments}
\bibliography{auto_generated}
\cleardoublepage \appendix\section{The CMS Collaboration \label{app:collab}}\begin{sloppypar}\hyphenpenalty=5000\widowpenalty=500\clubpenalty=5000
\cmsinstitute{Yerevan Physics Institute, Yerevan, Armenia}
{\tolerance=6000
A.~Hayrapetyan, A.~Tumasyan\cmsAuthorMark{1}\cmsorcid{0009-0000-0684-6742}
\par}
\cmsinstitute{Institut f\"{u}r Hochenergiephysik, Vienna, Austria}
{\tolerance=6000
W.~Adam\cmsorcid{0000-0001-9099-4341}, J.W.~Andrejkovic, T.~Bergauer\cmsorcid{0000-0002-5786-0293}, S.~Chatterjee\cmsorcid{0000-0003-2660-0349}, K.~Damanakis\cmsorcid{0000-0001-5389-2872}, M.~Dragicevic\cmsorcid{0000-0003-1967-6783}, P.S.~Hussain\cmsorcid{0000-0002-4825-5278}, M.~Jeitler\cmsAuthorMark{2}\cmsorcid{0000-0002-5141-9560}, N.~Krammer\cmsorcid{0000-0002-0548-0985}, A.~Li\cmsorcid{0000-0002-4547-116X}, D.~Liko\cmsorcid{0000-0002-3380-473X}, I.~Mikulec\cmsorcid{0000-0003-0385-2746}, J.~Schieck\cmsAuthorMark{2}\cmsorcid{0000-0002-1058-8093}, R.~Sch\"{o}fbeck\cmsorcid{0000-0002-2332-8784}, D.~Schwarz\cmsorcid{0000-0002-3821-7331}, M.~Sonawane\cmsorcid{0000-0003-0510-7010}, W.~Waltenberger\cmsorcid{0000-0002-6215-7228}, C.-E.~Wulz\cmsAuthorMark{2}\cmsorcid{0000-0001-9226-5812}
% ...
\par}
\vskip\cmsinstskip
\dag:~Deceased\\
$^{1}$Also at Yerevan State University, Yerevan, Armenia\\
$^{2}$Also at TU Wien, Vienna, Austria\\
% ...
\end{sloppypar}
\end{document}